\documentclass[11pt,fleqn,twoside]{article}
\usepackage{amsmath}
\usepackage{amssymb}
\usepackage{xcolor}
\usepackage{soul}
\usepackage{colortbl} 
\usepackage{booktabs}
\usepackage{pstricks,multido,pst-plot}
\usepackage{tikz}	
\usepackage{pgfplots}
\usepackage{paralist}
\usepackage{wrapfig}
\usepackage{graphicx,epsfig}
\usepackage{marginnote}
\usepackage{todonotes}
\usepackage[margin=2cm,top=1.4cm,bottom=2.1cm,centering]{geometry}   
\usepackage{tcolorbox,pgfkeys}
\usepackage{hyperref}                      
\usepackage{microtype}
\usepackage{mathtools}

\definecolor{lightgray}{gray}{0.95}
\definecolor{lightyellow}{rgb}{.97,.96,.57}
\definecolor{myblue}{rgb}{.39,.54,0.82}
\definecolor{midblue}{rgb}{.7,.7,1}
\definecolor{vertbleu}{rgb}{.73,.88,.93}
\definecolor{lightblue}{rgb}{0.93,0.95,1.0}
\definecolor{mygray}{rgb}{.80,.80,.80}
\definecolor{lightred}{rgb}{0.96,0.84,0.84}
\definecolor{mygreen}{rgb}{0.40,0.75,0.16}
\definecolor{mybrown}{rgb}{0.69,0.49,0.30}
\definecolor{mypink}{rgb}{1.00,0.22,0.57}
\definecolor{mygrey}{rgb}{.90,.90,.90}
\definecolor{cerulean}{cmyk}{0.94,0.11,0,0}

\newtcolorbox{mybox}{colback=yellow!20!white,colframe=gray,boxrule=0.3mm,arc=3mm,outer arc=1mm}

\newcommand\SmallMatrix[1]{{%
  \scriptsize\arraycolsep=0.3\arraycolsep\ensuremath{\begin{pmatrix}#1\end{pmatrix}}}}

\def\Red{} 
\def\Black{} 

\usepackage{enumitem}

\catcode`\@=11
\def\numberbysection{\@addtoreset{equation}{section}
     \def\theequation{\thesection.\arabic{equation}}}
\numberbysection

\def\be{\begin{equation}}
\def\ee{\end{equation}}
\newcommand\bea{\begin{eqnarray}}
\newcommand\eea{\end{eqnarray}}
\renewcommand\phi{\varphi}

\newcommand\egal{&\!\!\!=\!\!\!&}

\renewcommand{\ge}{\geqslant}
\renewcommand{\le}{\leqslant}
\def\benn{\begin{eqnarray*}}
\def\eenn{\end{eqnarray*}}

\def\ldet{$\lambda$-determinant }
\def\ldets{$\lambda$-determinants }
\def\detl{\textstyle \det_\lambda}

\def\Z{{\mathbb Z}}

\def\R{{\mathbb R}}

\def\C{{\mathbb C}}

\def\cqfd{\Red \hskip 1truemm \vrule height2.3mm depth0mm width2.3mm \Black}

\def\vdimerb{\pspolygon[linewidth=0.4pt,linecolor=black,fillstyle=solid,fillcolor=myblue](0,0)(0,2)(1,2)(1,0)}

\def\hdimerg{\pspolygon[linewidth=0.4pt,linecolor=black,fillstyle=solid,fillcolor=green](0,0)(2,0)(2,1)(0,1)}

\def\urban{
\pspolygon[linewidth=0.8pt,linecolor=black](1,0)(2,1)(3,0)(2,-1)(1,0)
\psline[linewidth=1.2pt,linecolor=mygreen](0,0)(1,0) 
\psline[linewidth=1.2pt,linecolor=mygreen](2,1)(2,2) 
\psline[linewidth=1.2pt,linecolor=mygreen](3,0)(4,0) 
\psline[linewidth=1.2pt,linecolor=mygreen](2,-1)(2,-2) 
\pscircle[linecolor=black,fillstyle=solid,fillcolor=black](0,0){0.15}
\pscircle[linecolor=black,fillstyle=solid,fillcolor=black](2,2){0.15}
\pscircle[linecolor=black,fillstyle=solid,fillcolor=black](4,0){0.15}
\pscircle[linecolor=black,fillstyle=solid,fillcolor=black](2,-2){0.15}
\pscircle[linecolor=black,fillstyle=solid,fillcolor=black](1,0){0.15}
\pscircle[linecolor=black,fillstyle=solid,fillcolor=black](2,1){0.15}
\pscircle[linecolor=black,fillstyle=solid,fillcolor=black](3,0){0.15}
\pscircle[linecolor=black,fillstyle=solid,fillcolor=black](2,-1){0.15}
}

\newtheorem{theo}{Theorem}[section]
\newtheorem{coro}[theo]{Corollary}
\newtheorem{prob}[theo]{Problem}

\newcommand*\xbar[1]{%
  \hbox{%
    \vbox{%
      \hrule height 0.5pt 
      \kern0.3ex
      \hbox{%
        \kern 0.0em
        \ensuremath{#1}%
        \kern-0.0em
      }%
    }%
  }%
} 

\hypersetup{
colorlinks=true,
citecolor=red,
linkcolor=blue,
urlcolor=black
}

\begin{document}
\renewcommand{\baselinestretch}{1.1}

\title{\vspace{-2truemm}\textbf{\Large On $\lambda$-determinants and tiling problems}}

\date{}
\maketitle

\begin{center}
{\vspace{-14mm}\large \textsc{Jean-Fran\c cois de Kemmeter$^{1,2}$, Nicolas Robert$^3$ and Philippe Ruelle$^3$}}
\\[.5cm]
{\em $^1$ Department of Mathematics and Namur Institute for Complex Systems (naXys)\\
University of Namur, Rue de Bruxelles 61, Namur, B-5000, Belgium}

{\em $^2$ Department of Mathematics, Florida State University\\ 
1017 Academic Way, Tallahassee, FL 32306, United States of America}

{\em $^3$ Institut de Recherche en Math\'ematique et Physique\\ 
Universit\'e catholique de Louvain, Louvain-la-Neuve, B-1348, Belgium}
\\[.2cm]

\end{center}

\vspace{3mm}
\noindent
Email: jean-francois.dekemmeter@unamur.be, nicolas.robert@uclouvain.be, philippe.ruelle@uclouvain.be

\vspace{0.4cm} 

\begin{abstract}
We review the connections between the octahedral recurrence, $\lambda$-determinants and tiling problems. This provides in particular a direct combinatorial interpretation of the $\lambda$-determinant (and generalizations thereof) of an arbitrary matrix in terms of domino tilings of Aztec diamonds. We also reinterpret  the general Robbins-Rumsey formula for the rational function of consecutive minors, given by a summation over pairs of compatible alternating sign matrices, as the partition function for tilings of Aztec diamonds equipped with a general measure.
\end{abstract}

\vspace{4mm} 
\hrule
\normalsize


\baselineskip=4.5mm

\tableofcontents 

\renewcommand{\baselinestretch}{1.1}
\selectfont
\parskip        0pt


\vskip 1truecm   
\noindent
{\bf \large 1. Introduction}
\addcontentsline{toc}{subsection}{1. Introduction}
\setcounter{section}{1}
\setcounter{equation}{0}

\medskip
\noindent
The \ldets have been introduced by Robbins and Rumsey \cite{RR86} by generalizing the Desnanot-Jacobi formula for ordinary determinants. Let $A$ be a square $n \times n$ matrix and denote by $A_{\mathrm{UL}},A_{\mathrm{LL}},A_{\mathrm{UR}}$, and $A_{\mathrm{LR}}$ the $(n-1) \times (n-1)$ restrictions of $A$ to the upper-left, lower-left, upper-right and lower-right corners of $A$, and denote by $A_\mathrm C$ the $(n-2) \times (n-2)$ central restriction of $A$. The Desnanot-Jacobi identity reads
\be
\det A \cdot \det A_{\mathrm C} = \det A_{\mathrm{UL}} \cdot \det A_{\mathrm{LR}} + \det A_{\mathrm{LL}} \cdot \det A_{\mathrm{UR}}.
\label{dj}
\ee
Setting the determinant of an order zero matrix to 1 and that of an order 1 matrix equal to the matrix itself, this identity allows one to compute recursively the determinant of a matrix of any order (provided the required minors are non-zero). For $n=2$, one easily recovers from (\ref{dj}) the usual form for the determinant of a $2 \times 2$ matrix on \cite{Do66} has used the identity (\ref{dj}) to propose the condensation method, an efficient algorithm that reduces the computation of a determinant to the evaluation of connected $2 \times 2$ minors.

Robbins and Rumsey defined the $\lambda$-determinants, denoted by $\det_\lambda$, by deforming the above identity by the introduction of a complex parameter $\lambda$,
\be
\detl A \cdot \detl A_{\mathrm C} = \detl A_{\mathrm{UL}} \cdot \detl A_{\mathrm{LR}} + \lambda \detl A_{\mathrm{LL}} \cdot \detl A_{\mathrm{UR}}.
\label{rr}
\ee
The parameter $\lambda$ does not affect the determinants of matrices of order 0 and 1, so that $2 \times 2$ $\lambda$-minors are simply given by
\be
\detl \begin{pmatrix} a & b \\ c & d \end{pmatrix} = ad + \lambda \, bc,
\ee
from which \ldets of higher order can be evaluated recursively. Dodgson's condensation method applies equally well to the computation of \ldets provided $\lambda$-minors are used instead of usual minors. Let us briefly recall it \cite{Do66}.

As mentioned above, the method computes the \ldet of a matrix $A$ in a recursive manner, by defining a finite sequence of matrices of decreasing order. If $A$ is $n \times n$, the sequence is initiated by setting $A_0 = (1)_{1 \le i,j \le n+1}$, the all-ones matrix of size $n+1$, and $A_1=A$. Then for $2 \le k \le n$, the matrix $A_k$, of size $n+1-k$, is obtained by computing all connected $2 \times 2$ $\lambda$-minors of $A_{k-1}$ and dividing entrywise by the central submatrix of $A_{k-2}$. Explicitly the entries of $A_k$ are given by 
\be
(A_k)_{i,j} = \big[(A_{k-1})_{i,j} \, (A_{k-1})_{i+1,j+1} + \lambda \, (A_{k-1})_{i+1,j} \, (A_{k-1})_{i,j+1}\big]/(A_{k-2})_{i+1,j+1}.
\label{cond}
\ee
Then $A_k$ is the matrix of all connected $\lambda$-minors of $A$ of size $k$, so that the last term in the sequence yields the result, $A_n = \detl A$. The method proves to be very efficient but can be problematic since the entries by which we divide may vanish. When $\detl A$ is known to exist (like for $\lambda=-1$), the problems can be cured by row or column permutations on $A$ (meaningful only for $\lambda=-1$) or by regularization, see \cite{Pr05} for a short discussion in this direction. As a simple application, the \ldet of a general matrix of order 3 is found to be
\bea
\detl \begin{pmatrix} a_{11} & a_{12} & a_{13} \\ a_{21} & a_{22} & a_{23} \\ a_{31} & a_{32} & a_{33} \end{pmatrix} \egal a_{11}\,a_{22}\,a_{33} + \lambda \, a_{12}\,a_{21}\,a_{33} + \lambda \, a_{11}\,a_{23}\,a_{32} + \lambda^2 \, a_{12}\,a_{23}\,a_{31} \nonumber\\
&& + \: \lambda^2 \, a_{13}\,a_{21}\,a_{32} + \lambda^3 \, a_{13}\,a_{22}\,a_{31} + \lambda(1+\lambda) \,\frac{a_{12}\,a_{21}\,a_{23}\,a_{32}}{a_{22}}.
\label{3det}
\eea

It is a textbook result that the ordinary determinant of an $n \times n$ matrix $A$ can be written explicitly (and non-recursively) in terms of its entries $a_{ij}$ as a sum over the symmetric group $S_n$, $\det A = \sum_{\sigma \in S_n} \, \varepsilon_\sigma \, A^{B(\sigma)}$; here $\varepsilon_\sigma$ is the parity of the permutation $\sigma$, $B(\sigma)$ is the $n \times n$ permutation matrix associated with $\sigma$ and $A^B$ denotes the product $\prod_{1 \le i,j \le n} \: a_{ij}^{b_{ij}}$. 

A natural question is thus whether there exists an analogous formula for $\lambda$-determinants. The result obtained by Robbins and Rumsey is not only that such a formula exists but also that the set of matrices playing the role of the permutation matrices in the case of ordinary determinants is universal, independent of $\lambda \neq -1$. It was in this context that alternating sign matrices made their appearance for the first time \cite{Br99,BP99}. An alternating sign matrix $B$ is a square matrix with entries $-1,0,1$ such that all row and column sums are equal to 1, and such that the non-zero entries $\pm 1$ alternate both in rows and columns. The set of alternating sign matrices of order $n$ will be denoted by ASM$_n$ (it contains in particular all permutation matrices $B(\sigma)$ of size $n$).

The remarkable formula proved in \cite{RR86} reads,
\be
\detl A = \displaystyle \sum_{B \,\in\, {\rm ASM}_n} \; \lambda^{P(B)} \: (1 + \lambda)^{N_-(B)} \; A^B,
\label{RR}
\ee
where $N_-(B)$ is the number of entries of $B$ equal to $-1$ (later on, we will also use $N_+(B)$, the number of entries equal to $+1$), and $P(B) = {\rm Inv}(B) - N_-(B) \ge 0$ with the inversion number given by Inv$(B) = \sum_{i<k} \sum_{j>\ell} \, b_{i,j}b_{k,\ell}$ (for a permutation $\sigma$, Inv($B(\sigma))$ is the minimal number of transpositions of adjacent elements by which $\sigma$ can be obtained from the identity). The number $P(B)$ has been given a more direct interpretation in \cite{DF13}, as the number of zeros in $B$ which have non-zero entries to the right and below, and such that the first non-zero entry in both directions is a 1. 

Comparing (\ref{RR}) with (\ref{3det}), one recognizes the first six terms as given by the six permutation matrices of size 3, while the last term involves the only alternating sign matrix of order 3 with a unique entry $-1$. 
For $\lambda=-1$, the summation reduces to alternating sign matrices with $N_-(B)=0$, that is, to permutation matrices, so one recovers the usual  formula quoted above. One also sees from the expression (\ref{RR}) that, except in the case $\lambda=-1$, the \ldet of a matrix involves positive and negative powers of its entries, and therefore may be undefined if some entries are zero. When $\lambda$ is taken as an indeterminate, the \ldet of a generic matrix of size $n$ is a polynomial in $\lambda$ of degree $\frac{n(n-1)}2$.

The first indication of a relation between \ldets and tiling problems stems from the discovery by Kuo that the number $T_n$ of domino tilings of an Aztec diamond of order $n$ satisfies the following recurrence relation \cite{Ku04},
\be
T_n \, T_{n-2} = 2 T_{n-1}^2,
\label{kuo}
\ee
It is indeed reminiscent of the general form of the $\lambda$-Desnanot-Jacobi identity for a 1-determinant ($\lambda=1$) if $T_n$ can be written as $T_n = \det_1 A$ for a suitable matrix $A$ (such that the two terms in the r.h.s. of (\ref{rr}) are equal). The boundary values $T_0=1$ and $T_1=2$ confirm that $A$ can be taken to be the all-ones constant matrix of order $n+1$, leading to $T_n = 2^{n(n+1)/2}$. Even though the matrix $A$ is very simple in this case, with all entries equal to 1, the summation in (\ref{RR}) is not trivial to evaluate since the alternating sign matrices get weighted according to the number of $-1$ they contain. The Dodgson condensation algorithm however furnishes the result in a straightforward way, as does the recurrence (\ref{kuo}) itself.

If we want to assign different weights (or relative probabilities) to the domino tilings, the formulation in terms of perfect matchings of the dual Aztec graph is more convenient. Kuo showed in \cite{Ku04} that a quadratic recurrence relation similar to the Desnanot-Jacobi identity holds for a general weighting of the edges of the Aztec graph, and applied it in a few examples, among which a particular two-periodic weighting of the Aztec diamond and some other holey domains. Speyer \cite{Sp07} slightly generalized Kuo's recurrence by adding face weights, and, placing it in a wider context, called it the octahedron recurrence (the term has been used earlier, though it is hard to trace its origin). As far as Aztec diamonds are concerned, both Kuo's and Speyer's articles write the quadratic recurrence relation for the most general weighting; the relation they found is a generalized form of the $\lambda$-Desnanot-Jacobi identity and leads to inhomogeneous \ldets \cite{DF13}, see Section 4.

Let us consider weighted perfect matchings of Aztec graphs, for which the weighting is defined in terms of face weights. In this case, the weight of a dimer in a given perfect matching is proportional to the inverse product of the weights of the faces it is adjacent to (see Section 2 for more details), and the weight of a perfect matching is the product of the weights of all dimers. In addition to these face weights, we assign each vertical dimer a bias in the form of an extra weight $\sqrt{\lambda}$. Let us symbolically denote by $W$ the collection of all face weights, and by $T_n(W|\lambda)$ the partition function, that is, the sum of the weights of all perfect matchings of the order $n$ Aztec graph. Then $T_n(W|\lambda)$ satisfies the octahedron recurrence \cite{Ku04,Sp07}
\be
T_n(W|\lambda) \cdot T_{n-2}(W_{\mathrm C}|\lambda) = T_{n-1}(W_{\mathrm{UL}}|\lambda) \cdot T_{n-1}(W_{\mathrm{LR}}|\lambda) + \lambda \, T_{n-1}(W_{\mathrm{LL}}|\lambda) \cdot T_{n-1}(W_{\mathrm{UR}}|\lambda),
\label{TW}
\ee
where the weight systems $W_{\mathrm{UL}},W_{\mathrm{LR}},W_{\mathrm{LL}},W_{\mathrm{UR}}$ and $W_\mathrm{C}$ are the restrictions of $W$ to the Aztec subgraphs in the four corners\footnote{If the central face of the order $n$ graph is centered at the coordinates $(0,0)$, those of the four restrictions UL,\,LR,\,LL,\,UR are centered at $(-1,0),(1,0),(0,-1),(0,1)$ respectively.} and in the center, of order $n-1$ and $n-2$ respectively. When all weights are equal to 1, namely the distribution on perfect matchings is uniform, $T_n(W=1|\lambda=1)$ is just the number of domino tilings, and the previous recurrence reduces to (\ref{kuo}).

The similarity of the octahedron recurrence and the Desnanot-Jacobi identity for \ldets is striking, and has been noticed by many authors. One should not expect that the solutions of the former can always be cast as $\lambda$-determinants (we will indeed see that this is not the case), but one can hope that \ldets can be given a combinatorial interpretation as partition functions of weighted perfect matchings. The simplest example has been mentioned above, namely the 1-determinant of the all-ones matrix, $\det_1(a_{i,j}=1)_{1 \le i,j \le n+1}$, is the number of perfect matchings of the order $n$ Aztec graph. One can readily generalize it to any $\lambda$ and get $\detl(a_{i,j}=1)_{1 \le i,j \le n+1} = (1+\lambda)^{n(n+1)/2}$, the partition function for perfect matchings with bias $\sqrt{\lambda}$ for each vertical dimer. The latter has been noted by Propp in \cite{Pr05}, who went further on to propose another example: if a number of 1's are replaced by 0's in the four corners of the all-ones matrix, the 1-determinant of the resulting matrix yields the number of domino tilings of a square of side $n$. This last example is a strong hint of a more general structure. To the best of our knowledge, no systematic combinatorial interpretation has been given to general \ldets in the context of domino tilings. 

The main goal of this article is precisely to fill this gap, by giving several combinatorial interpretations of general $\lambda$-determinants in relation to perfect matchings of Aztec graphs. This will give us the opportunity to review basic and well-known material about alternating sign matrices in the context of tiling problems. Formulating the counting of perfect matchings in terms of $\lambda$-determinants, and more generally the partition functions when non-trivial weightings are used, does not always make their explicit evaluation easy. However this formulation offers an important conceptual understanding and at the same time provides an extremely versatile method, applicable to many cases that would otherwise be hard to tackle by standard methods. Finally it gives a straight, computationally efficient and quick mean to obtain numerical results. Many illustrative examples will be given. 

As a last remark, and because asymptotic values are in general more useful than exact values at finite size, the use of \ldets also prompts the intriguing question of whether Szeg\"o limit theorems could be formulated for them, when applied to Toeplitz matrices.


\vskip .5truecm
\noindent
{\bf \large 2. General octahedron recurrence}
\addcontentsline{toc}{subsection}{2. General octahedron recurrence}
\setcounter{section}{2}
\setcounter{equation}{0}

\begin{figure}[t]
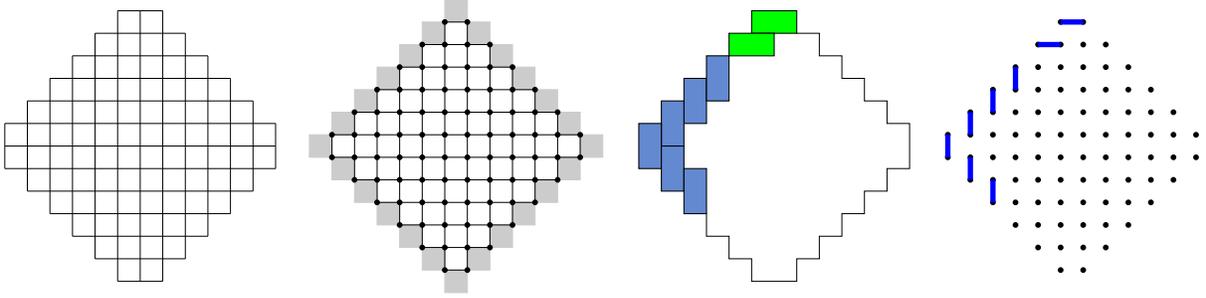

\begin{center}
\psset{unit=.3cm}
\pspicture(2,0)(14,12)
\psline[linewidth=0.2pt,linecolor=black](5,-1)(7,-1)
\psline[linewidth=0.2pt,linecolor=black](4,0)(8,-0)
\psline[linewidth=0.2pt,linecolor=black](3,1)(9,1)
\psline[linewidth=0.2pt,linecolor=black](2,2)(10,2)
\psline[linewidth=0.2pt,linecolor=black](1,3)(11,3)
\psline[linewidth=0.2pt,linecolor=black](0,4)(12,4)
\psline[linewidth=0.2pt,linecolor=black](0,5)(12,5)
\psline[linewidth=0.2pt,linecolor=black](0,6)(12,6)
\psline[linewidth=0.2pt,linecolor=black](1,7)(11,7)
\psline[linewidth=0.2pt,linecolor=black](2,8)(10,8)
\psline[linewidth=0.2pt,linecolor=black](3,9)(9,9)
\psline[linewidth=0.2pt,linecolor=black](4,10)(8,10)
\psline[linewidth=0.2pt,linecolor=black](5,11)(7,11)
\psline[linewidth=0.2pt,linecolor=black](0,4)(0,6)
\psline[linewidth=0.2pt,linecolor=black](1,3)(1,7)
\psline[linewidth=0.2pt,linecolor=black](2,2)(2,8)
\psline[linewidth=0.2pt,linecolor=black](3,1)(3,9)
\psline[linewidth=0.2pt,linecolor=black](4,0)(4,10)
\psline[linewidth=0.2pt,linecolor=black](5,-1)(5,11)
\psline[linewidth=0.2pt,linecolor=black](6,-1)(6,11)
\psline[linewidth=0.2pt,linecolor=black](7,-1)(7,11)
\psline[linewidth=0.2pt,linecolor=black](8,0)(8,10)
\psline[linewidth=0.2pt,linecolor=black](9,1)(9,9)
\psline[linewidth=0.2pt,linecolor=black](10,2)(10,8)
\psline[linewidth=0.2pt,linecolor=black](11,3)(11,7)
\psline[linewidth=0.2pt,linecolor=black](12,4)(12,6)
\endpspicture
\pspicture(0,0)(10,12)
\rput(-0.5,4.5){\multido{\nx=0+1}{7}{\rput(\nx,\nx){\pspolygon[linewidth=0pt,linecolor=mygray,fillstyle=solid,fillcolor=mygray](0,0)(0,1)(1,1)(1,0)}}}
\rput(5.5,-1.5){\multido{\nx=0+1}{7}{\rput(\nx,\nx){\pspolygon[linewidth=0pt,linecolor=mygray,fillstyle=solid,fillcolor=mygray](0,0)(0,1)(1,1)(1,0)}}}
\rput(0.5,3.5){\multido{\nx=0+1}{5}{\rput(\nx,-\nx){\pspolygon[linewidth=0pt,linecolor=mygray,fillstyle=solid,fillcolor=mygray](0,0)(0,1)(1,1)(1,0)}}}
\rput(6.5,9.5){\multido{\nx=0+1}{5}{\rput(\nx,-\nx){\pspolygon[linewidth=0pt,linecolor=mygray,fillstyle=solid,fillcolor=mygray](0,0)(0,1)(1,1)(1,0)}}}
\multido{\nt=4+1}{2}{\rput(0,\nt){\pscircle[linecolor=black,fillstyle=solid,fillcolor=black](0.5,0.5){0.08}}}
\multido{\nt=3+1}{4}{\rput(1,\nt){\pscircle[linecolor=black,fillstyle=solid,fillcolor=black](0.5,0.5){0.08}}}
\multido{\nt=2+1}{6}{\rput(2,\nt){\pscircle[linecolor=black,fillstyle=solid,fillcolor=black](0.5,0.5){0.08}}}
\multido{\nt=1+1}{8}{\rput(3,\nt){\pscircle[linecolor=black,fillstyle=solid,fillcolor=black](0.5,0.5){0.08}}}
\multido{\nt=0+1}{10}{\rput(4,\nt){\pscircle[linecolor=black,fillstyle=solid,fillcolor=black](0.5,0.5){0.08}}}
\multido{\nt=-1+1}{12}{\rput(5,\nt){\pscircle[linecolor=black,fillstyle=solid,fillcolor=black](0.5,0.5){0.08}}}
\multido{\nt=-1+1}{12}{\rput(6,\nt){\pscircle[linecolor=black,fillstyle=solid,fillcolor=black](0.5,0.5){0.08}}}
\multido{\nt=0+1}{10}{\rput(7,\nt){\pscircle[linecolor=black,fillstyle=solid,fillcolor=black](0.5,0.5){0.08}}}
\multido{\nt=1+1}{8}{\rput(8,\nt){\pscircle[linecolor=black,fillstyle=solid,fillcolor=black](0.5,0.5){0.08}}}
\multido{\nt=2+1}{6}{\rput(9,\nt){\pscircle[linecolor=black,fillstyle=solid,fillcolor=black](0.5,0.5){0.08}}}
\multido{\nt=3+1}{4}{\rput(10,\nt){\pscircle[linecolor=black,fillstyle=solid,fillcolor=black](0.5,0.5){0.08}}}
\multido{\nt=4+1}{2}{\rput(11,\nt){\pscircle[linecolor=black,fillstyle=solid,fillcolor=black](0.5,0.5){0.08}}}
\rput(0.5,-0.5){
\psline[linewidth=0.2pt,linecolor=black](5,0)(6,0)
\psline[linewidth=0.2pt,linecolor=black](4,1)(7,1)
\psline[linewidth=0.2pt,linecolor=black](3,2)(8,2)
\psline[linewidth=0.2pt,linecolor=black](2,3)(9,3)
\psline[linewidth=0.2pt,linecolor=black](1,4)(10,4)
\psline[linewidth=0.2pt,linecolor=black](0,5)(11,5)
\psline[linewidth=0.2pt,linecolor=black](0,6)(11,6)
\psline[linewidth=0.2pt,linecolor=black](1,7)(10,7)
\psline[linewidth=0.2pt,linecolor=black](2,8)(9,8)
\psline[linewidth=0.2pt,linecolor=black](3,9)(8,9)
\psline[linewidth=0.2pt,linecolor=black](4,10)(7,10)
\psline[linewidth=0.2pt,linecolor=black](5,11)(6,11)
\psline[linewidth=0.2pt,linecolor=black](0,5)(0,6)
\psline[linewidth=0.2pt,linecolor=black](1,4)(1,7)
\psline[linewidth=0.2pt,linecolor=black](2,3)(2,8)
\psline[linewidth=0.2pt,linecolor=black](3,2)(3,9)
\psline[linewidth=0.2pt,linecolor=black](4,1)(4,10)
\psline[linewidth=0.2pt,linecolor=black](5,0)(5,11)
\psline[linewidth=0.2pt,linecolor=black](6,0)(6,11)
\psline[linewidth=0.2pt,linecolor=black](7,1)(7,10)
\psline[linewidth=0.2pt,linecolor=black](8,2)(8,9)
\psline[linewidth=0.2pt,linecolor=black](9,3)(9,8)
\psline[linewidth=0.2pt,linecolor=black](10,4)(10,7)
\psline[linewidth=0.2pt,linecolor=black](11,5)(11,6)}
\endpspicture
\hspace{2cm}
\pspicture(3,0)(14,12)
\rput(5,10){\hdimerg}
\rput(4,9){\hdimerg}
\rput(2,6){\vdimerb}
\rput(1,5){\vdimerb}
\rput(0,4){\vdimerb}
\rput(1,3){\vdimerb}
\rput(2,2){\vdimerb}
\rput(3,7){\vdimerb}
\psline[linewidth=0.2pt,linecolor=black](7,10)(8,10)(8,9)(9,9)(9,8)(10,8)(10,7)(11,7)(11,6)(12,6)(12,4)(11,4)(11,3)(10,3)(10,2)(9,2)(9,1)(8,1)(8,0)(7,0)(7,-1)(5,-1)(5,0)(4,0)(4,1)(3,1)(3,2)
\endpspicture
\pspicture(0.8,0)(10,12)
\multido{\nt=4+1}{2}{\rput(0,\nt){\pscircle[linecolor=black,fillstyle=solid,fillcolor=black](0.5,0.5){0.08}}}
\multido{\nt=3+1}{4}{\rput(1,\nt){\pscircle[linecolor=black,fillstyle=solid,fillcolor=black](0.5,0.5){0.08}}}
\multido{\nt=2+1}{6}{\rput(2,\nt){\pscircle[linecolor=black,fillstyle=solid,fillcolor=black](0.5,0.5){0.08}}}
\multido{\nt=1+1}{8}{\rput(3,\nt){\pscircle[linecolor=black,fillstyle=solid,fillcolor=black](0.5,0.5){0.08}}}
\multido{\nt=0+1}{10}{\rput(4,\nt){\pscircle[linecolor=black,fillstyle=solid,fillcolor=black](0.5,0.5){0.08}}}
\multido{\nt=-1+1}{12}{\rput(5,\nt){\pscircle[linecolor=black,fillstyle=solid,fillcolor=black](0.5,0.5){0.08}}}
\multido{\nt=-1+1}{12}{\rput(6,\nt){\pscircle[linecolor=black,fillstyle=solid,fillcolor=black](0.5,0.5){0.08}}}
\multido{\nt=0+1}{10}{\rput(7,\nt){\pscircle[linecolor=black,fillstyle=solid,fillcolor=black](0.5,0.5){0.08}}}
\multido{\nt=1+1}{8}{\rput(8,\nt){\pscircle[linecolor=black,fillstyle=solid,fillcolor=black](0.5,0.5){0.08}}}
\multido{\nt=2+1}{6}{\rput(9,\nt){\pscircle[linecolor=black,fillstyle=solid,fillcolor=black](0.5,0.5){0.08}}}
\multido{\nt=3+1}{4}{\rput(10,\nt){\pscircle[linecolor=black,fillstyle=solid,fillcolor=black](0.5,0.5){0.08}}}
\multido{\nt=4+1}{2}{\rput(11,\nt){\pscircle[linecolor=black,fillstyle=solid,fillcolor=black](0.5,0.5){0.08}}}
\psline[linewidth=2.0pt,linecolor=blue](5.5,10.5)(6.5,10.5)
\psline[linewidth=2.0pt,linecolor=blue](4.5,9.5)(5.5,9.5)
\psline[linewidth=2.0pt,linecolor=blue](2.5,6.5)(2.5,7.5)
\psline[linewidth=2.0pt,linecolor=blue](1.5,5.5)(1.5,6.5)
\psline[linewidth=2.0pt,linecolor=blue](0.5,4.5)(0.5,5.5)
\psline[linewidth=2.0pt,linecolor=blue](3.5,7.5)(3.5,8.5)
\psline[linewidth=2.0pt,linecolor=blue](1.5,3.5)(1.5,4.5)
\psline[linewidth=2.0pt,linecolor=blue](2.5,2.5)(2.5,3.5)
\endpspicture
\end{center}
\caption{The two figures on the left show the Aztec diamond of order 6, and next to it, its extended dual graph $\widehat{\cal A}_n$ with the boundary faces (shaded). The two right figures show a partial domino tiling and the corresponding partial perfect matching of ${\cal A}_n$.}
\label{fig1}
\end{figure}

\medskip
\noindent
The Aztec diamond of order $n$ is a planar domain formed of unit cells with staircase boundaries; the left panel of Figure~\ref{fig1} depicts an order 6 Aztec diamond. The number of unit cells on the rows varies from 2 to $2n$. Domino tilings of Aztec diamonds can equally be described as perfect matchings of the dual graph ${\cal A}_n$: a perfect matching $M$ of ${\cal A}_n$ is a subset of edges such that every vertex of the graph is the endpoint of one and only one edge in $M$. The edges of a perfect matching are also called dimers, see Figure~\ref{fig1} where a partial tiling and its dimer content are shown. Finally the graph ${\cal A}_n$ may be extended to $\widehat{\cal A}_n$ by including all faces around the boundary of ${\cal A}_n$, shown in shaded gray in Figure~\ref{fig1}. The faces of $\widehat{\cal A}_n$ contained in ${\cal A}_n$ are called inner faces; the others are called boundary faces. Note that the extension $\widehat{\cal A}_n$ does not contain more edges or more vertices than ${\cal A}_n$.

Perfect matchings of Aztec graphs (or any graph) can be given different weights, which depend on their dimer content. We will consider here two types of weightings, $w_F(M)$ and $w_E(M)$, defined in terms of weights attached to the faces of the graph or attached to its edges. In what follows, we choose a coordinate system such that the faces of $\widehat{\cal A}_n$ are centered at integer coordinates $(k,\ell)$ with $|k|+|\ell| \le n$; the central face is centered at $(0,0)$.

\begin{figure}[t]
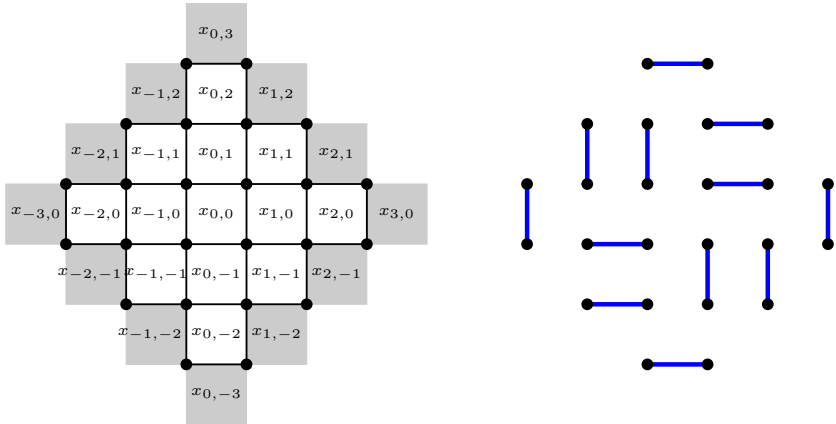

\begin{center}
\psset{unit=.8cm}
\pspicture(0,2)(6,8.5)
\rput(-0.5,4.5){\multido{\nx=0+1}{4}{\rput(\nx,\nx){\pspolygon[linewidth=0pt,linecolor=mygray,fillstyle=solid,fillcolor=mygray](0,0)(0,1)(1,1)(1,0)}}}
\rput(2.5,1.5){\multido{\nx=0+1}{4}{\rput(\nx,\nx){\pspolygon[linewidth=0pt,linecolor=mygray,fillstyle=solid,fillcolor=mygray](0,0)(0,1)(1,1)(1,0)}}}
\rput(0.5,3.5){\multido{\nx=0+1}{2}{\rput(\nx,-\nx){\pspolygon[linewidth=0pt,linecolor=mygray,fillstyle=solid,fillcolor=mygray](0,0)(0,1)(1,1)(1,0)}}}
\rput(3.5,6.5){\multido{\nx=0+1}{2}{\rput(\nx,-\nx){\pspolygon[linewidth=0pt,linecolor=mygray,fillstyle=solid,fillcolor=mygray](0,0)(0,1)(1,1)(1,0)}}}
\multido{\nt=4+1}{2}{\rput(0,\nt){\pscircle[linecolor=black,fillstyle=solid,fillcolor=black](0.5,0.5){0.08}}}
\multido{\nt=3+1}{4}{\rput(1,\nt){\pscircle[linecolor=black,fillstyle=solid,fillcolor=black](0.5,0.5){0.08}}}
\multido{\nt=2+1}{6}{\rput(2,\nt){\pscircle[linecolor=black,fillstyle=solid,fillcolor=black](0.5,0.5){0.08}}}
\multido{\nt=2+1}{6}{\rput(3,\nt){\pscircle[linecolor=black,fillstyle=solid,fillcolor=black](0.5,0.5){0.08}}}
\multido{\nt=3+1}{4}{\rput(4,\nt){\pscircle[linecolor=black,fillstyle=solid,fillcolor=black](0.5,0.5){0.08}}}
\multido{\nt=4+1}{2}{\rput(5,\nt){\pscircle[linecolor=black,fillstyle=solid,fillcolor=black](0.5,0.5){0.08}}}
\psline[linewidth=.7pt,linecolor=black](2.5,7.5)(3.5,7.5)
\psline[linewidth=.7pt,linecolor=black](1.5,6.5)(4.5,6.5)
\psline[linewidth=.7pt,linecolor=black](0.5,5.5)(5.5,5.5)
\psline[linewidth=.7pt,linecolor=black](0.5,4.5)(5.5,4.5)
\psline[linewidth=.7pt,linecolor=black](1.5,3.5)(4.5,3.5)
\psline[linewidth=.7pt,linecolor=black](2.5,2.5)(3.5,2.5)
\psline[linewidth=.7pt,linecolor=black](0.5,5.5)(0.5,4.5)
\psline[linewidth=.7pt,linecolor=black](1.5,6.5)(1.5,3.5)
\psline[linewidth=.7pt,linecolor=black](2.5,7.5)(2.5,2.5)
\psline[linewidth=.7pt,linecolor=black](3.5,7.5)(3.5,2.5)
\psline[linewidth=.7pt,linecolor=black](4.5,6.5)(4.5,3.5)
\psline[linewidth=.7pt,linecolor=black](5.5,5.5)(5.5,4.5)
\rput(3,5){\tiny $x_{0,0}$}
\rput(4,5){\tiny $x_{1,0}$}
\rput(5,5){\tiny $x_{2,0}$}
\rput(6,5){\tiny $x_{3,0}$}
\rput(2,5){\tiny $x_{-1,0}$}
\rput(1,5){\tiny $x_{-2,0}$}
\rput(0,5){\tiny $x_{-3,0}$}
\rput(1,6){\tiny $x_{-2,1}$}
\rput(2,6){\tiny $x_{-1,1}$}
\rput(3,6){\tiny $x_{0,1}$}
\rput(4,6){\tiny $x_{1,1}$}
\rput(5,6){\tiny $x_{2,1}$}
\rput(2,7){\tiny $x_{-1,2}$}
\rput(3,7){\tiny $x_{0,2}$}
\rput(4,7){\tiny $x_{1,2}$}
\rput(3,8){\tiny $x_{0,3}$}
\rput(0.9,4){\tiny $x_{-2,-1}$}
\rput(2,4){\tiny $x_{-1,-1}$}
\rput(3,4){\tiny $x_{0,-1}$}
\rput(4,4){\tiny $x_{1,-1}$}
\rput(5,4){\tiny $x_{2,-1}$}
\rput(1.9,3){\tiny $x_{-1,-2}$}
\rput(3,3){\tiny $x_{0,-2}$}
\rput(4,3){\tiny $x_{1,-2}$}
\rput(3,2){\tiny $x_{0,-3}$}
\endpspicture
\hspace{1.2cm}
\pspicture(0,2)(6,8)
%
\psline[linewidth=1.7pt,linecolor=blue](5.5,5.5)(5.5,4.5)
\psline[linewidth=1.7pt,linecolor=blue](3.5,5.5)(4.5,5.5)
\psline[linewidth=1.7pt,linecolor=blue](3.5,6.5)(4.5,6.5)
\psline[linewidth=1.7pt,linecolor=blue](3.5,7.5)(2.5,7.5)
\psline[linewidth=1.7pt,linecolor=blue](2.5,6.5)(2.5,5.5)
\psline[linewidth=1.7pt,linecolor=blue](1.5,6.5)(1.5,5.5)
\psline[linewidth=1.7pt,linecolor=blue](0.5,5.5)(0.5,4.5)
\psline[linewidth=1.7pt,linecolor=blue](2.5,4.5)(1.5,4.5)
\psline[linewidth=1.7pt,linecolor=blue](2.5,3.5)(1.5,3.5)
\psline[linewidth=1.7pt,linecolor=blue](2.5,2.5)(3.5,2.5)
\psline[linewidth=1.7pt,linecolor=blue](3.5,3.5)(3.5,4.5)
\psline[linewidth=1.7pt,linecolor=blue](4.5,3.5)(4.5,4.5)
\multido{\nt=4+1}{2}{\rput(0,\nt){\pscircle[linecolor=black,fillstyle=solid,fillcolor=black](0.5,0.5){0.08}}}
\multido{\nt=3+1}{4}{\rput(1,\nt){\pscircle[linecolor=black,fillstyle=solid,fillcolor=black](0.5,0.5){0.08}}}
\multido{\nt=2+1}{6}{\rput(2,\nt){\pscircle[linecolor=black,fillstyle=solid,fillcolor=black](0.5,0.5){0.08}}}
\multido{\nt=2+1}{6}{\rput(3,\nt){\pscircle[linecolor=black,fillstyle=solid,fillcolor=black](0.5,0.5){0.08}}}
\multido{\nt=3+1}{4}{\rput(4,\nt){\pscircle[linecolor=black,fillstyle=solid,fillcolor=black](0.5,0.5){0.08}}}
\multido{\nt=4+1}{2}{\rput(5,\nt){\pscircle[linecolor=black,fillstyle=solid,fillcolor=black](0.5,0.5){0.08}}}
\endpspicture
\end{center}
\caption{{\it Left}: the extended graph $\hat{\cal A}_3$ of order 3 with the face variables. {\it Right}: example of  perfect matching with a face weight $w_F(M) = x_{1,-2} \, x_{-2,-1} \, x_{-1,-1}^{-1} \, x_{1,-1}^{-1} \, x_{0,0} \, x_{-1,1}^{-1} \, x_{1,1}^{-1} \, x_{2,1} \, x_{-1,2}$.}
\label{fig2}
\end{figure}

In the first weighting, we assign each face (including the boundary faces) a weight $x_{k,\ell}$, see Figure~\ref{fig2}. For each face $(k,\ell)$, we count the number $N_{k,\ell}$ of dimers which are adjacent to that face and let it contribute a factor $x_{k,\ell}^{1-N_{k,\ell}}$ to the weight of a perfect matching. The total weight of $M$ is the product of these factors, $w_F(M) = \prod_{(k,\ell) \in \widehat{\cal A}_n} x_{k,\ell}^{1-N_{k,\ell}}$. We note that $1-N_{k,\ell}$ is equal to $0,+1$ or $-1$ for an inner face and equal to 0 or $+1$ for a boundary face.

For the second weighting, all edges $e$ of the graph are assigned a weight $w(e)$. The weight of a perfect matching is then the product of the weights of the edges contained in the matching, $w_E(M) = \prod_{e \in M} w(e)$. In practice, and because every edge is adjacent to an even face ($k+\ell$ even), it is enough to assign weights $\alpha_{k,\ell},\beta_{k,\ell}, \gamma_{k,\ell}$ and $\delta_{k,\ell}$ to the east, north, west and south edges of the even faces. It will be convenient to assign edge weights to odd faces too. Since the east edge of an odd face is the west edge of the right neighbouring even face, we set $\alpha_{k,\ell} = \gamma_{k+1,\ell}$ for $k+\ell$ odd; similarly we set $\beta_{k,\ell} = \delta_{k,\ell+1}, \, \gamma_{k,\ell} = \alpha_{k-1,\ell}$ and $\delta_{k,\ell} = \beta_{k,\ell-1}$ for $k+\ell$ odd.

Both weightings $w_F(M)$ and $w_E(M)$ are complete in the sense that either weight uniquely determines $M$, and so are mutually redundant. We nevertheless keep them both since, depending on the specialization of weights we are interested in, one may be more convenient than the other. Let 
\be
T_{n;(0,0)} = \sum_{M {\rm of } {\cal A}_n} w_F(M) \, w_E(M) = \sum_{M {\rm of } {\cal A}_n} \; \prod_{(k,\ell) \in \widehat{\cal A}_n} x_{k,\ell}^{1-N_{k,\ell}} \times \prod_{e \in M} w(e)
\ee
be the partition function for perfect matchings of the Aztec graph of order $n$, with face and edge weights as defined above. The subscript $(0,0)$ indicates that the central face of the Aztec graph has weight $x_{0,0}$. Then the partition functions satisfy the following quadratic recurrence relation \cite{Ku04,Sp07}
\be
T_{n;(0,0)} \, T_{n-2;(0,0)} = \beta_{0,n-1} \, \delta_{0,1-n} \: T_{n-1;(-1,0)} \, T_{n-1;(1,0)} + \alpha_{n-1,0} \, \gamma_{1-n,0} \: T_{n-1;(0,-1)} \, T_{n-1;(0,1)},
\label{octa}
\ee
where $T_{k;(i,j)}$ is the partition function for the perfect matchings of the subgraph ${\cal A}_k \subset {\cal A}_n$ of order $k$ whose central face has weight $x_{(i,j)}$, and computed with respect to the face and edge weights inherited from $\widehat{\cal A}_n$, so that $T_{k;(i,j)}$ depends on a subset of the weights used for $T_{n;(0,0)}$. As noted before, the order $n-1$ subgraphs centered at $(-1,0),\, (1,0),\,(0,-1)$ and $(0,1)$ correspond, after the rotation of the order $n$ Aztec graph by 45 degrees clockwise so that the graph roughly looks like a square, to the UL,\,LR,\,LL and UR restrictions alluded to above in (\ref{TW}).

Together with the initial conditions for $n=0$ and $n=1$, namely,
\be
T_{0;(i,j)} = x_{i,j}, \qquad T_{1;(i,j)} = x_{i,j}^{-1} \, \big(\alpha_{i,j}\,\gamma_{i,j}\,x^{}_{i,j-1} \, x^{}_{i,j+1} + \beta_{i,j}\,\delta_{i,j}\,x^{}_{i-1,j} \, x^{}_{i+1,j}\big),
\label{gen}
\ee
the partition function for all higher values of $n$ are uniquely determined. As $n$ increases, the full expressions of $T_{n;(0,0)}$ become rapidly large and awkward. 

At this level of generality, there does not seem to be any relation with \ldets because the coefficients of the two terms in the r.h.s. of the recurrence (\ref{octa}) both depend on $n$. In order to establish a relation, an easy start is to set the horizontal edge weights $\beta_{i,j},\delta_{i,j}$ equal to 1 and the vertical edge weights $\alpha_{i,j},\gamma_{i,j}$ equal to $\sqrt{\lambda}$. Although it considerably simplifies the expressions of the partition functions, it alone does not guarantee that the latter can be written as $\lambda$-determinants (of what matrix?). In general, they cannot as we will see in the next section, although algorithmically, the general case remains very close to a $\lambda$-determinant.


\vskip .5truecm
\noindent
{\bf \large 3. Face weights}
\addcontentsline{toc}{subsection}{3. Face weights}
\setcounter{section}{3}
\setcounter{equation}{0}

\vspace{5mm}
\begin{figure}[t]
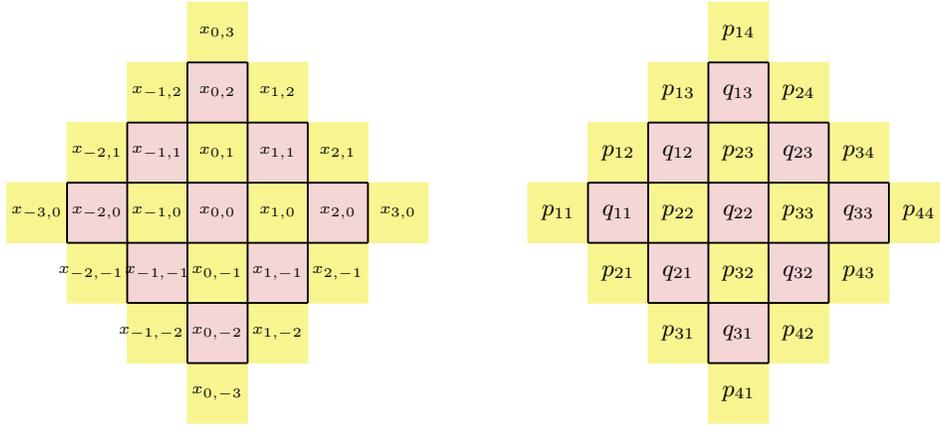

\begin{center}
\psset{unit=.8cm}
\pspicture(0,2)(6,8)
\rput(-0.5,4.5){\multido{\nx=0+1}{4}{\rput(\nx,\nx){\pspolygon[linewidth=0pt,linecolor=lightyellow,fillstyle=solid,fillcolor=lightyellow](0,0)(0,1)(1,1)(1,0)}}}
\rput(0.5,4.5){\multido{\nx=0+1}{3}{\rput(\nx,\nx){\pspolygon[linewidth=0pt,linecolor=lightred,fillstyle=solid,fillcolor=lightred](0,0)(0,1)(1,1)(1,0)}}}
\rput(0.5,3.5){\multido{\nx=0+1}{4}{\rput(\nx,\nx){\pspolygon[linewidth=0pt,linecolor=lightyellow,fillstyle=solid,fillcolor=lightyellow](0,0)(0,1)(1,1)(1,0)}}}
\rput(1.5,3.5){\multido{\nx=0+1}{3}{\rput(\nx,\nx){\pspolygon[linewidth=0pt,linecolor=lightred,fillstyle=solid,fillcolor=lightred](0,0)(0,1)(1,1)(1,0)}}}
\rput(1.5,2.5){\multido{\nx=0+1}{4}{\rput(\nx,\nx){\pspolygon[linewidth=0pt,linecolor=lightyellow,fillstyle=solid,fillcolor=lightyellow](0,0)(0,1)(1,1)(1,0)}}}
\rput(2.5,2.5){\multido{\nx=0+1}{3}{\rput(\nx,\nx){\pspolygon[linewidth=0pt,linecolor=lightred,fillstyle=solid,fillcolor=lightred](0,0)(0,1)(1,1)(1,0)}}}
\rput(2.5,1.5){\multido{\nx=0+1}{4}{\rput(\nx,\nx){\pspolygon[linewidth=0pt,linecolor=lightyellow,fillstyle=solid,fillcolor=lightyellow](0,0)(0,1)(1,1)(1,0)}}}
\psline[linewidth=.7pt,linecolor=black](2.5,7.5)(3.5,7.5)
\psline[linewidth=.7pt,linecolor=black](1.5,6.5)(4.5,6.5)
\psline[linewidth=.7pt,linecolor=black](0.5,5.5)(5.5,5.5)
\psline[linewidth=.7pt,linecolor=black](0.5,4.5)(5.5,4.5)
\psline[linewidth=.7pt,linecolor=black](1.5,3.5)(4.5,3.5)
\psline[linewidth=.7pt,linecolor=black](2.5,2.5)(3.5,2.5)
\psline[linewidth=.7pt,linecolor=black](0.5,5.5)(0.5,4.5)
\psline[linewidth=.7pt,linecolor=black](1.5,6.5)(1.5,3.5)
\psline[linewidth=.7pt,linecolor=black](2.5,7.5)(2.5,2.5)
\psline[linewidth=.7pt,linecolor=black](3.5,7.5)(3.5,2.5)
\psline[linewidth=.7pt,linecolor=black](4.5,6.5)(4.5,3.5)
\psline[linewidth=.7pt,linecolor=black](5.5,5.5)(5.5,4.5)
\rput(3,5){\tiny $x_{0,0}$}
\rput(4,5){\tiny $x_{1,0}$}
\rput(5,5){\tiny $x_{2,0}$}
\rput(6,5){\tiny $x_{3,0}$}
\rput(2,5){\tiny $x_{-1,0}$}
\rput(1,5){\tiny $x_{-2,0}$}
\rput(0,5){\tiny $x_{-3,0}$}
\rput(1,6){\tiny $x_{-2,1}$}
\rput(2,6){\tiny $x_{-1,1}$}
\rput(3,6){\tiny $x_{0,1}$}
\rput(4,6){\tiny $x_{1,1}$}
\rput(5,6){\tiny $x_{2,1}$}
\rput(2,7){\tiny $x_{-1,2}$}
\rput(3,7){\tiny $x_{0,2}$}
\rput(4,7){\tiny $x_{1,2}$}
\rput(3,8){\tiny $x_{0,3}$}
\rput(0.9,4){\tiny $x_{-2,-1}$}
\rput(2,4){\tiny $x_{-1,-1}$}
\rput(3,4){\tiny $x_{0,-1}$}
\rput(4,4){\tiny $x_{1,-1}$}
\rput(5,4){\tiny $x_{2,-1}$}
\rput(1.9,3){\tiny $x_{-1,-2}$}
\rput(3,3){\tiny $x_{0,-2}$}
\rput(4,3){\tiny $x_{1,-2}$}
\rput(3,2){\tiny $x_{0,-3}$}
\endpspicture
\hspace{2cm}
\pspicture(0,2)(6,8)
\rput(-0.5,4.5){\multido{\nx=0+1}{4}{\rput(\nx,\nx){\pspolygon[linewidth=0pt,linecolor=lightyellow,fillstyle=solid,fillcolor=lightyellow](0,0)(0,1)(1,1)(1,0)}}}
\rput(0.5,4.5){\multido{\nx=0+1}{3}{\rput(\nx,\nx){\pspolygon[linewidth=0pt,linecolor=lightred,fillstyle=solid,fillcolor=lightred](0,0)(0,1)(1,1)(1,0)}}}
\rput(0.5,3.5){\multido{\nx=0+1}{4}{\rput(\nx,\nx){\pspolygon[linewidth=0pt,linecolor=lightyellow,fillstyle=solid,fillcolor=lightyellow](0,0)(0,1)(1,1)(1,0)}}}
\rput(1.5,3.5){\multido{\nx=0+1}{3}{\rput(\nx,\nx){\pspolygon[linewidth=0pt,linecolor=lightred,fillstyle=solid,fillcolor=lightred](0,0)(0,1)(1,1)(1,0)}}}
\rput(1.5,2.5){\multido{\nx=0+1}{4}{\rput(\nx,\nx){\pspolygon[linewidth=0pt,linecolor=lightyellow,fillstyle=solid,fillcolor=lightyellow](0,0)(0,1)(1,1)(1,0)}}}
\rput(2.5,2.5){\multido{\nx=0+1}{3}{\rput(\nx,\nx){\pspolygon[linewidth=0pt,linecolor=lightred,fillstyle=solid,fillcolor=lightred](0,0)(0,1)(1,1)(1,0)}}}
\rput(2.5,1.5){\multido{\nx=0+1}{4}{\rput(\nx,\nx){\pspolygon[linewidth=0pt,linecolor=lightyellow,fillstyle=solid,fillcolor=lightyellow](0,0)(0,1)(1,1)(1,0)}}}
\psline[linewidth=.7pt,linecolor=black](2.5,7.5)(3.5,7.5)
\psline[linewidth=.7pt,linecolor=black](1.5,6.5)(4.5,6.5)
\psline[linewidth=.7pt,linecolor=black](0.5,5.5)(5.5,5.5)
\psline[linewidth=.7pt,linecolor=black](0.5,4.5)(5.5,4.5)
\psline[linewidth=.7pt,linecolor=black](1.5,3.5)(4.5,3.5)
\psline[linewidth=.7pt,linecolor=black](2.5,2.5)(3.5,2.5)
\psline[linewidth=.7pt,linecolor=black](0.5,5.5)(0.5,4.5)
\psline[linewidth=.7pt,linecolor=black](1.5,6.5)(1.5,3.5)
\psline[linewidth=.7pt,linecolor=black](2.5,7.5)(2.5,2.5)
\psline[linewidth=.7pt,linecolor=black](3.5,7.5)(3.5,2.5)
\psline[linewidth=.7pt,linecolor=black](4.5,6.5)(4.5,3.5)
\psline[linewidth=.7pt,linecolor=black](5.5,5.5)(5.5,4.5)
\rput(3,5){\footnotesize $q_{22}$}
\rput(4,5){\footnotesize $p_{33}$}
\rput(5,5){\footnotesize $q_{33}$}
\rput(6,5){\footnotesize $p_{44}$}
\rput(2,5){\footnotesize $p_{22}$}
\rput(1,5){\footnotesize $q_{11}$}
\rput(0,5){\footnotesize $p_{11}$}
\rput(1,6){\footnotesize $p_{12}$}
\rput(2,6){\footnotesize $q_{12}$}
\rput(3,6){\footnotesize $p_{23}$}
\rput(4,6){\footnotesize $q_{23}$}
\rput(5,6){\footnotesize $p_{34}$}
\rput(2,7){\footnotesize $p_{13}$}
\rput(3,7){\footnotesize $q_{13}$}
\rput(4,7){\footnotesize $p_{24}$}
\rput(3,8){\footnotesize $p_{14}$}
\rput(1,4){\footnotesize $p_{21}$}
\rput(2,4){\footnotesize $q_{21}$}
\rput(3,4){\footnotesize $p_{32}$}
\rput(4,4){\footnotesize $q_{32}$}
\rput(5,4){\footnotesize $p_{43}$}
\rput(2,3){\footnotesize $p_{31}$}
\rput(3,3){\footnotesize $q_{31}$}
\rput(4,3){\footnotesize $p_{42}$}
\rput(3,2){\footnotesize $p_{41}$}
\endpspicture
\end{center}
\caption{Relabelling of the general weights $x_{k,\ell}$ into $p_{i,j}$ and $q_{i,j}$ forming two matrices of order $n+1$ and $n$, respectively ($n=3$ in the figures).}
\label{fig3}
\end{figure}

\vspace{-3mm}
\noindent
In this section, we keep general face weights, for both inner and boundary faces, and restrict the edge variables to a  constant weight $\sqrt{\lambda}$ on vertical dimers, which we call a vertical bias. A good starting point is to write out the corresponding partition function for $n=2$. Before doing that, it turns out to be more convenient to change coordinates and use different labels for even and odd faces. The new labelling is depicted in Figure~\ref{fig3}. 

The resulting labelling makes two matrices appear: a larger one, $P$, of order $n+1$, whose entries are located on the yellow faces (of the same parity as $n$), and a smaller one, $Q$, of order $n$, whose entries occupy the red faces (of parity opposite to that of $n$). 

We will refer to this as the $(P,Q)$ weighting, and will denote the partition functions by $T_n(P,Q|\lambda)$, in which $P$ and $Q$ have respectively order $n+1$ and $n$. If we want to set all weights $p_{ij}$ or $q_{ij}$ to 1, we will write $P=1$ or $Q=1$ (not to be confused with the identity matrix ${\mathbb I}$); similarly we write $P^{-1}$ or $Q^{-1}$ for the matrices whose entries are $p_{ij}^{-1}$ or $q_{ij}^{-1}$.


\vskip 0.5truecm
\noindent
{\bf 3.1 Restricted weights: $\lambda$-determinants}
\addcontentsline{toc}{subsubsection}{3.1 Restricted weights: $\lambda$-determinants}

\medskip
\noindent
In terms of the $(P,Q)$ weighting and the choice of edge weights $\alpha_{i,j}=\gamma_{i,j}=\sqrt{\lambda}$ and $\beta_{i,j}=\delta_{i,j}=1$, the octahedron recurrence (\ref{octa}) reads
\bea
T_n(P,Q|\lambda) \, T_{n-2}(P_\mathrm{C},Q_\mathrm{C}|\lambda) \egal T_{n-1}(P_{\mathrm{UL}},Q_{\mathrm{UL}}|\lambda) \, T_{n-1}(P_{\mathrm{LR}},Q_{\mathrm{LR}}|\lambda) \nonumber\\
\noalign{\smallskip}
&& \qquad + \: \lambda \, T_{n-1}(P_{\mathrm{LL}},Q_{\mathrm{LL}}|\lambda) \, T_{n-1}(P_{\mathrm{UR}},Q_{\mathrm{UR}}|\lambda),
\label{octaPQ}
\eea
with initial conditions (for $n=0$, the matrix $Q$ has size zero and is represented by $-$)
\be
T_0(p_{11},-|\lambda) = p_{11}, \qquad T_1\left(\SmallMatrix{p_{11} & p_{12} \\ p_{21} & p_{22}},q_{11}\big|\:\lambda\right) = \frac{p_{11}\,p_{22} + \lambda \, p_{12}\,p_{21}}{q_{11}}.
\label{initPQ}
\ee

Coming back to $n=2$, the eight perfect matchings and their weights are listed below; the partition function $T_2(P,Q|\lambda)$ is the sum of the eight contributions. \begin{figure}[h]

\begin{equation*}
\pspicture(0,0)(0.5,0)
\psset{unit=.4cm}
\rput(0,-1.3){
\psline[linewidth=2.0pt,linecolor=blue](0,0)(1,0)
\psline[linewidth=2.0pt,linecolor=blue](-1,2)(0,2)
\psline[linewidth=2.0pt,linecolor=blue](-1,1)(0,1)
\psline[linewidth=2.0pt,linecolor=blue](1,2)(2,2)
\psline[linewidth=2.0pt,linecolor=blue](1,1)(2,1)
\psline[linewidth=2.0pt,linecolor=blue](0,3)(1,3)
\pscircle[linecolor=black,fillstyle=solid,fillcolor=black](0,0){0.12}
\pscircle[linecolor=black,fillstyle=solid,fillcolor=black](1,0){0.12}
\pscircle[linecolor=black,fillstyle=solid,fillcolor=black](-1,1){0.12}
\pscircle[linecolor=black,fillstyle=solid,fillcolor=black](0,1){0.12}
\pscircle[linecolor=black,fillstyle=solid,fillcolor=black](1,1){0.12}
\pscircle[linecolor=black,fillstyle=solid,fillcolor=black](2,1){0.12}
\pscircle[linecolor=black,fillstyle=solid,fillcolor=black](-1,2){0.12}
\pscircle[linecolor=black,fillstyle=solid,fillcolor=black](0,2){0.12}
\pscircle[linecolor=black,fillstyle=solid,fillcolor=black](1,2){0.12}
\pscircle[linecolor=black,fillstyle=solid,fillcolor=black](2,2){0.12}
\pscircle[linecolor=black,fillstyle=solid,fillcolor=black](0,3){0.12}
\pscircle[linecolor=black,fillstyle=solid,fillcolor=black](1,3){0.12}}
\endpspicture
\hspace{8mm} = \;\frac{p_{11}\,p_{22}\,p_{33}}{q_{11}\,q_{22}} 
\hspace{2cm}
\pspicture(0,0)(0.5,0)
\psset{unit=.4cm}
\rput(0,-1.3){
\psline[linewidth=2.0pt,linecolor=blue](0,0)(1,0)
\psline[linewidth=2.0pt,linecolor=blue](0,1)(0,2)
\psline[linewidth=2.0pt,linecolor=blue](-1,1)(-1,2)
\psline[linewidth=2.0pt,linecolor=blue](1,2)(2,2)
\psline[linewidth=2.0pt,linecolor=blue](1,1)(2,1)
\psline[linewidth=2.0pt,linecolor=blue](0,3)(1,3)
\pscircle[linecolor=black,fillstyle=solid,fillcolor=black](0,0){0.12}
\pscircle[linecolor=black,fillstyle=solid,fillcolor=black](1,0){0.12}
\pscircle[linecolor=black,fillstyle=solid,fillcolor=black](-1,1){0.12}
\pscircle[linecolor=black,fillstyle=solid,fillcolor=black](0,1){0.12}
\pscircle[linecolor=black,fillstyle=solid,fillcolor=black](1,1){0.12}
\pscircle[linecolor=black,fillstyle=solid,fillcolor=black](2,1){0.12}
\pscircle[linecolor=black,fillstyle=solid,fillcolor=black](-1,2){0.12}
\pscircle[linecolor=black,fillstyle=solid,fillcolor=black](0,2){0.12}
\pscircle[linecolor=black,fillstyle=solid,fillcolor=black](1,2){0.12}
\pscircle[linecolor=black,fillstyle=solid,fillcolor=black](2,2){0.12}
\pscircle[linecolor=black,fillstyle=solid,fillcolor=black](0,3){0.12}
\pscircle[linecolor=black,fillstyle=solid,fillcolor=black](1,3){0.12}}
\endpspicture
\hspace{8mm} = \; \lambda \: \frac{p_{12}\,p_{21}\,p_{33}}{q_{11}\,q_{22}} 
\hspace{2cm}
\pspicture(0,0)(0.5,0)
\psset{unit=.4cm}
\rput(0,-1.3){
\psline[linewidth=2.0pt,linecolor=blue](0,0)(1,0)
\psline[linewidth=2.0pt,linecolor=blue](-1,2)(0,2)
\psline[linewidth=2.0pt,linecolor=blue](-1,1)(0,1)
\psline[linewidth=2.0pt,linecolor=blue](2,1)(2,2)
\psline[linewidth=2.0pt,linecolor=blue](1,1)(1,2)
\psline[linewidth=2.0pt,linecolor=blue](0,3)(1,3)
\pscircle[linecolor=black,fillstyle=solid,fillcolor=black](0,0){0.12}
\pscircle[linecolor=black,fillstyle=solid,fillcolor=black](1,0){0.12}
\pscircle[linecolor=black,fillstyle=solid,fillcolor=black](-1,1){0.12}
\pscircle[linecolor=black,fillstyle=solid,fillcolor=black](0,1){0.12}
\pscircle[linecolor=black,fillstyle=solid,fillcolor=black](1,1){0.12}
\pscircle[linecolor=black,fillstyle=solid,fillcolor=black](2,1){0.12}
\pscircle[linecolor=black,fillstyle=solid,fillcolor=black](-1,2){0.12}
\pscircle[linecolor=black,fillstyle=solid,fillcolor=black](0,2){0.12}
\pscircle[linecolor=black,fillstyle=solid,fillcolor=black](1,2){0.12}
\pscircle[linecolor=black,fillstyle=solid,fillcolor=black](2,2){0.12}
\pscircle[linecolor=black,fillstyle=solid,fillcolor=black](0,3){0.12}
\pscircle[linecolor=black,fillstyle=solid,fillcolor=black](1,3){0.12}}
\endpspicture
\hspace{8mm} = \; \lambda \: \frac{p_{11}\,p_{23}\,p_{32}}{q_{11}\,q_{22}} 
\end{equation*}

\begin{equation*}
\pspicture(0,0)(0.5,0)
\psset{unit=.4cm}
\rput(0,-1.3){
\psline[linewidth=2.0pt,linecolor=blue](0,0)(0,1)
\psline[linewidth=2.0pt,linecolor=blue](1,0)(1,1)
\psline[linewidth=2.0pt,linecolor=blue](-1,1)(-1,2)
\psline[linewidth=2.0pt,linecolor=blue](2,1)(2,2)
\psline[linewidth=2.0pt,linecolor=blue](0,2)(1,2)
\psline[linewidth=2.0pt,linecolor=blue](0,3)(1,3)
\pscircle[linecolor=black,fillstyle=solid,fillcolor=black](0,0){0.12}
\pscircle[linecolor=black,fillstyle=solid,fillcolor=black](1,0){0.12}
\pscircle[linecolor=black,fillstyle=solid,fillcolor=black](-1,1){0.12}
\pscircle[linecolor=black,fillstyle=solid,fillcolor=black](0,1){0.12}
\pscircle[linecolor=black,fillstyle=solid,fillcolor=black](1,1){0.12}
\pscircle[linecolor=black,fillstyle=solid,fillcolor=black](2,1){0.12}
\pscircle[linecolor=black,fillstyle=solid,fillcolor=black](-1,2){0.12}
\pscircle[linecolor=black,fillstyle=solid,fillcolor=black](0,2){0.12}
\pscircle[linecolor=black,fillstyle=solid,fillcolor=black](1,2){0.12}
\pscircle[linecolor=black,fillstyle=solid,fillcolor=black](2,2){0.12}
\pscircle[linecolor=black,fillstyle=solid,fillcolor=black](0,3){0.12}
\pscircle[linecolor=black,fillstyle=solid,fillcolor=black](1,3){0.12}}
\endpspicture
\hspace{8mm} = \; \lambda^2 \: \frac{p_{12}\,p_{23}\,p_{31}}{q_{12}\,q_{21}} 
\hspace{1.5cm}
\pspicture(0,0)(0.5,0)
\psset{unit=.4cm}
\rput(0,-1.3){
\psline[linewidth=2.0pt,linecolor=blue](0,0)(1,0)
\psline[linewidth=2.0pt,linecolor=blue](0,1)(1,1)
\psline[linewidth=2.0pt,linecolor=blue](-1,1)(-1,2)
\psline[linewidth=2.0pt,linecolor=blue](2,1)(2,2)
\psline[linewidth=2.0pt,linecolor=blue](0,2)(0,3)
\psline[linewidth=2.0pt,linecolor=blue](1,2)(1,3)
\pscircle[linecolor=black,fillstyle=solid,fillcolor=black](0,0){0.12}
\pscircle[linecolor=black,fillstyle=solid,fillcolor=black](1,0){0.12}
\pscircle[linecolor=black,fillstyle=solid,fillcolor=black](-1,1){0.12}
\pscircle[linecolor=black,fillstyle=solid,fillcolor=black](0,1){0.12}
\pscircle[linecolor=black,fillstyle=solid,fillcolor=black](1,1){0.12}
\pscircle[linecolor=black,fillstyle=solid,fillcolor=black](2,1){0.12}
\pscircle[linecolor=black,fillstyle=solid,fillcolor=black](-1,2){0.12}
\pscircle[linecolor=black,fillstyle=solid,fillcolor=black](0,2){0.12}
\pscircle[linecolor=black,fillstyle=solid,fillcolor=black](1,2){0.12}
\pscircle[linecolor=black,fillstyle=solid,fillcolor=black](2,2){0.12}
\pscircle[linecolor=black,fillstyle=solid,fillcolor=black](0,3){0.12}
\pscircle[linecolor=black,fillstyle=solid,fillcolor=black](1,3){0.12}}
\endpspicture
\hspace{8mm} = \; \lambda^2 \: \frac{p_{13}\,p_{21}\,p_{32}}{q_{12}\,q_{21}} 
\hspace{1.8cm}
\pspicture(0,0)(0.5,0)
\psset{unit=.4cm}
\rput(0,-1.3){
\psline[linewidth=2.0pt,linecolor=blue](0,0)(0,1)
\psline[linewidth=2.0pt,linecolor=blue](1,0)(1,1)
\psline[linewidth=2.0pt,linecolor=blue](-1,1)(-1,2)
\psline[linewidth=2.0pt,linecolor=blue](2,1)(2,2)
\psline[linewidth=2.0pt,linecolor=blue](0,2)(0,3)
\psline[linewidth=2.0pt,linecolor=blue](1,2)(1,3)
\pscircle[linecolor=black,fillstyle=solid,fillcolor=black](0,0){0.12}
\pscircle[linecolor=black,fillstyle=solid,fillcolor=black](1,0){0.12}
\pscircle[linecolor=black,fillstyle=solid,fillcolor=black](-1,1){0.12}
\pscircle[linecolor=black,fillstyle=solid,fillcolor=black](0,1){0.12}
\pscircle[linecolor=black,fillstyle=solid,fillcolor=black](1,1){0.12}
\pscircle[linecolor=black,fillstyle=solid,fillcolor=black](2,1){0.12}
\pscircle[linecolor=black,fillstyle=solid,fillcolor=black](-1,2){0.12}
\pscircle[linecolor=black,fillstyle=solid,fillcolor=black](0,2){0.12}
\pscircle[linecolor=black,fillstyle=solid,fillcolor=black](1,2){0.12}
\pscircle[linecolor=black,fillstyle=solid,fillcolor=black](2,2){0.12}
\pscircle[linecolor=black,fillstyle=solid,fillcolor=black](0,3){0.12}
\pscircle[linecolor=black,fillstyle=solid,fillcolor=black](1,3){0.12}}
\endpspicture
\hspace{8mm} = \; \lambda^3 \: \frac{p_{13}\,p_{22}\,p_{31}}{q_{12}\,q_{21}} 
\end{equation*}

\begin{equation*}
\hspace{2.2cm}
\pspicture(0,0)(0.5,0)
\psset{unit=.4cm}
\rput(0,-1.3){
\psline[linewidth=2.0pt,linecolor=blue](0,0)(1,0)
\psline[linewidth=2.0pt,linecolor=blue](0,1)(1,1)
\psline[linewidth=2.0pt,linecolor=blue](-1,1)(-1,2)
\psline[linewidth=2.0pt,linecolor=blue](2,1)(2,2)
\psline[linewidth=2.0pt,linecolor=blue](0,2)(1,2)
\psline[linewidth=2.0pt,linecolor=blue](0,3)(1,3)
\pscircle[linecolor=black,fillstyle=solid,fillcolor=black](0,0){0.12}
\pscircle[linecolor=black,fillstyle=solid,fillcolor=black](1,0){0.12}
\pscircle[linecolor=black,fillstyle=solid,fillcolor=black](-1,1){0.12}
\pscircle[linecolor=black,fillstyle=solid,fillcolor=black](0,1){0.12}
\pscircle[linecolor=black,fillstyle=solid,fillcolor=black](1,1){0.12}
\pscircle[linecolor=black,fillstyle=solid,fillcolor=black](2,1){0.12}
\pscircle[linecolor=black,fillstyle=solid,fillcolor=black](-1,2){0.12}
\pscircle[linecolor=black,fillstyle=solid,fillcolor=black](0,2){0.12}
\pscircle[linecolor=black,fillstyle=solid,fillcolor=black](1,2){0.12}
\pscircle[linecolor=black,fillstyle=solid,fillcolor=black](2,2){0.12}
\pscircle[linecolor=black,fillstyle=solid,fillcolor=black](0,3){0.12}
\pscircle[linecolor=black,fillstyle=solid,fillcolor=black](1,3){0.12}}
\endpspicture
\hspace{8mm} = \; \lambda \: \frac{p_{12}\,p_{21}\,p_{23}\,p_{32}}{p_{22}\,q_{12}\,q_{21}} 
\hspace{2cm}
\pspicture(0,0)(0.5,0)
\psset{unit=.4cm}
\rput(0,-1.3){
\psline[linewidth=2.0pt,linecolor=blue](0,0)(1,0)
\psline[linewidth=2.0pt,linecolor=blue](0,1)(0,2)
\psline[linewidth=2.0pt,linecolor=blue](-1,1)(-1,2)
\psline[linewidth=2.0pt,linecolor=blue](2,1)(2,2)
\psline[linewidth=2.0pt,linecolor=blue](1,1)(1,2)
\psline[linewidth=2.0pt,linecolor=blue](0,3)(1,3)
\pscircle[linecolor=black,fillstyle=solid,fillcolor=black](0,0){0.12}
\pscircle[linecolor=black,fillstyle=solid,fillcolor=black](1,0){0.12}
\pscircle[linecolor=black,fillstyle=solid,fillcolor=black](-1,1){0.12}
\pscircle[linecolor=black,fillstyle=solid,fillcolor=black](0,1){0.12}
\pscircle[linecolor=black,fillstyle=solid,fillcolor=black](1,1){0.12}
\pscircle[linecolor=black,fillstyle=solid,fillcolor=black](2,1){0.12}
\pscircle[linecolor=black,fillstyle=solid,fillcolor=black](-1,2){0.12}
\pscircle[linecolor=black,fillstyle=solid,fillcolor=black](0,2){0.12}
\pscircle[linecolor=black,fillstyle=solid,fillcolor=black](1,2){0.12}
\pscircle[linecolor=black,fillstyle=solid,fillcolor=black](2,2){0.12}
\pscircle[linecolor=black,fillstyle=solid,fillcolor=black](0,3){0.12}
\pscircle[linecolor=black,fillstyle=solid,fillcolor=black](1,3){0.12}}
\endpspicture
\hspace{8mm} = \; \lambda^2 \: \frac{p_{12}\,p_{21}\,p_{23}\,p_{32}}{p_{22}\,q_{11}\,q_{22}} 
\end{equation*}
\end{figure}

In agreement with the recurrence (\ref{octaPQ}), one may indeed check that it is equal to 
\bea
&& \hspace{-1cm} T_2(P,Q|\lambda) = T_2\left(\SmallMatrix{p_{11} & p_{12} & p_{13} \\ p_{21} & p_{22} & p_{23} \\ p_{31} & p_{32} & p_{33}},
\SmallMatrix{q_{11} & q_{12} \\ q_{21} & q_{22}}\Big|\: \lambda\right) \nonumber\\ 
\noalign{\medskip}
\egal \frac{T_1\left(\SmallMatrix{p_{11} & p_{12} \\ p_{21} & p_{22}},q_{11}\big|\:\lambda\right) \cdot T_1\left(\SmallMatrix{p_{22} & p_{23} \\ p_{32} & p_{33}},q_{22}\big|\:\lambda\right) + \lambda \: T_1\left(\SmallMatrix{p_{21} & p_{22} \\ p_{31} & p_{32}},q_{21}\big|\:\lambda\right) \cdot T_1\left(\SmallMatrix{p_{12} & p_{13} \\ p_{22} & p_{23}},q_{12}\big|\:\lambda\right)}{T_0(p_{22},-|\lambda)} \nonumber\\
\noalign{\medskip}
\egal \frac 1{p_{22}} \Big[\frac{p_{11}\,p_{22} + \lambda \, p_{12}\,p_{21}}{q_{11}} \cdot \frac{p_{22}\,p_{33} + \lambda \, p_{23}\,p_{32}}{q_{22}} + \lambda \, \frac{p_{21}\,p_{32} + \lambda \, p_{22}\,p_{31}}{q_{21}} \cdot \frac{p_{12}\,p_{23} + \lambda \, p_{13}\,p_{22}}{q_{12}}\Big].
\label{T2}
\eea

Comparing with the general expression (\ref{3det}) of the \ldet of a general matrix of order 3, one immediately observes that $T_2(P,1|\lambda) = \detl P$. Likewise for $P=1$, the partition function reads 
\be
T_2(1,Q|\lambda) = (1+\lambda)^2 \Big[\frac 1{q_{11}q_{22}} + \frac \lambda{q_{12}q_{21}}\Big] = (1+\lambda)^2 \, \detl (Q^{-1}).
\ee
This gives us a first general result.

\begin{theo} \label{thm1}
The partition functions for perfect matchings  of the order $n$ Aztec graph with respect to the general $(P,Q)$ weighting and vertical bias $\sqrt{\lambda}$ satisfy the following identities,
\be
T_n(P,1|\lambda) = \detl P, \qquad T_n(1,Q|\lambda) = (1+\lambda)^n \, \detl(Q^{-1}).
\label{eqthm1}
\ee
\end{theo}

\noindent
{\it Proof.} From (\ref{initPQ}), the two identities hold for $n=0,1$. (Note that for $n=0$ and $n=1$, $P$ has size $1$ and $2$, whereas $Q$ has size 0 and 1.)  Therefore they hold for any $n$ since in each case, both sides satisfy the same recurrence relations, given by (\ref{octaPQ}) and (\ref{rr}) for the l.h.s. and for the r.h.s.  respectively. \cqfd

\smallskip
In the rest of this section, we consider some applications of these two formulae.


\vskip 0.5truecm
\noindent
{\bf 3.2 Applications to periodic and biased Aztec diamonds}
\addcontentsline{toc}{subsubsection}{3.2 Applications to periodic and biased Aztec diamonds}

\medskip
\noindent
At a basic level, the partition function for all faces equal to 1 easily follows from Theorem \ref{thm1}. 

\begin{coro} \label{bias}
\cite{EKLP92,Pr05} The partition function $T_n(\lambda)$ for perfect matchings of the Aztec graph of order $n$ with vertical bias $\sqrt{\lambda}$ is given by
\be
T_n(\lambda) = (1 + \lambda)^{n(n+1)/2}.
\ee
\end{coro}

\noindent
{\it Proof.} Straightforward from the condensation algorithm since $T_n(\lambda) = T_n(1,1|\lambda)$ is the $\lambda$-determi\-nant of the all-ones matrix of order $n+1$. The recurrence $T_n(\lambda) = (1+\lambda)^n \, T_{n-1}(\lambda)$ also follows by combining the two identities in (\ref{eqthm1}). \cqfd

\medskip
One may refine the previous result by restricting to those perfect matchings having a fixed number of vertical dimers along the NW boundary.

\begin{coro}
The refined partition function $T_{n,\ell}(\lambda)$ for the perfect matchings of the Aztec graph of order $n$ which have exactly $\ell$ vertical dimers along the NW boundary is given by
\be
T_{n,\ell}(\lambda) = {n \choose \ell} \, \lambda^\ell \, (1 + \lambda)^{n(n-1)/2}, \qquad 0 \le \ell \le n.
\ee
\end{coro}

\noindent
{\it Proof.} For any perfect matching, among the $n+1$ outer faces along the NW boundary, $n$ are adjacent to one dimer and one is adjacent to none: the face with weight $p_{1,\ell+1}$ is adjacent to no dimer if the matching has exactly $\ell$ vertical dimers along the NW boundary, see Figure~\ref{fig1}. Taking $P=1$ except the first row set to $(p_{11},p_{12}, \ldots,p_{1,n+1})$ and $Q=1$, we obtain
\be
T_n(P,1|\lambda) = \sum_{\ell=0}^n p_{1,\ell+1} \, T_{n,\ell}(\lambda).
\ee 
It follows that $T_{n,\ell}(\lambda) = \detl P_\ell$ where $P_\ell$ is the all-ones matrix except for the first row, equal to $(0,\ldots,0,1,0,\ldots,0)$ with the unique 1 in position $\ell+1$. The condensation method is easily worked out to compute $\detl P_\ell$ and yields the result. \cqfd

\medskip
According to the Robbins-Rumsey formula (\ref{RR}), the value $T_{n,\ell}(\lambda=1)$ yields the 2-enumeration\footnote{The 2-enumeration of a set $S$ of alternating sign matrices is the weighted enumeration of $S$, where each matrix $B$ of $S$ contributes a factor $2^{N_-(B)}$.}  of the alternating sign matrices of size $n+1$ which have the unique 1 in the first row at position $\ell+1$ \cite{MRR83}, while $T_n(\lambda=1)$ yields the 2-enumeration of all alternating sign matrices of order $n+1$.

\begin{coro} \label{coper}
\cite{DFSG14,Ru22} The partition function $T_n(a,b)$ for perfect matchings of the two-periodic Aztec graph of order $n$ with parameters $a,b$ and no vertical bias ($\lambda=1$) is given by
\be
T_n(a,b) = \Big(\frac2{ab}\Big)^{\lfloor\frac{(n+1)^2}4\rfloor} \big(a^2 + b^2 \big)^{\lfloor\frac{n^2}4\rfloor} \times \begin{cases}
\vspace{-1.5mm}
1 & {\rm if\ }n = 0 \bmod 2,\\
\vspace{-1.5mm}
b & {\rm if\ }n = 1 \bmod 4,\\
a & {\rm if\ }n = 3 \bmod 4.\\
\end{cases}
\ee
\end{coro}

\noindent
{\it Proof.} The two-periodic weighting of the Aztec graph of order $n$ corresponds to $P=1$ and $Q$ the matrix with $q_{11}=a$ and the two parameters $a,b$ alternating on rows and columns. As there is no bias, from Theorem \ref{thm1}, we can write $T_n(a,b)$ as the 1-determinant of $Q^{-1}$, namely,
\medskip
\be
T_n(a,b) = 2^n \: {\textstyle \det_1} \, Q^{-1} = 2^n \: {\textstyle \det_1} \begin{pmatrix}
a^{-1} & b^{-1} & a^{-1} & \ldots \\
b^{-1} & a^{-1} & b^{-1} & \ldots \\
a^{-1} & b^{-1} & a^{-1} & \ldots \\
\ldots & \ldots & \ldots & \ldots 
\end{pmatrix}_{n \times n}\,.
\label{matQ}
\ee

Let us examine the condensation algorithm to compute this 1-determinant. The matrix $Q^{-1}$ is a symmetric Toeplitz matrix, with two alternating quantities on the first row, which completely determine the whole matrix. The algorithm produces a sequence of matrices $A_k$ of decreasing order, starting with $A_0=1$ and $A_1=Q^{-1}$, all of which are symmetric Toeplitz matrices with two alternating quantities on the first row. Let us denote them by $a_k$ and $b_k$, with $a_0=b_0=1$ and $a_1=a^{-1}$, $b_1=b^{-1}$. The condensation algorithm implies that $a_k$ and $b_k$ satisfy the following coupled recurrence relations,
\be
a_k = \frac{a_{k-1}^2 + b_{k-1}^2}{a_{k-2}}, \qquad b_k = \frac{b_{k-1}^2 + a_{k-1}^2}{b_{k-2}},
\label{crec}
\ee
and finishes with $a_n = \det_1 Q^{-1}$.

Defining $r_k \equiv b_k/a_k$, the first recurrence relation yields
\be
\frac{a_k}{a_{k-1}} = \frac{a_{k-1}}{a_{k-2}} (1 + r_{k-1}^2) = a^{-1} \, (1 + r_1^2) \ldots (1 + r_{k-2}^2) \, (1 + r_{k-1}^2), \qquad k \ge 1.
\ee
Forming the telescopic product, we obtain, with $r_0=1$,
\be
T_n(a,b) = 2^n \, a_n = a^{-n} \, (1 + r_0^2)^n \, (1 + r_1^2)^{n-1} \, (1 + r_2^2)^{n-2} \ldots (1 + r_{n-1}^2).
\ee

Taking the ratio of the two relations in (\ref{crec}), we see that $r_k$ satisfies $r^{}_k = r_{k-2}^{-1} = r_{k-4}$ and is therefore 4-periodic. From $r_0=1$ and $r_1=\frac ab$, we obtain the explicit values $r_k = 1,\frac ab,1,\frac ba$ for $k=0,1,2,3 \bmod 4$, from which the product in the previous equation is straightforward to evaluate and yields the result. \cqfd

\begin{prob} \label{problem}
A natural and seemingly innocuous generalization of the previous result would be to add vertical bias and compute $T_n(a,b|\lambda)$. One would similarly find $T_n(a,b|\lambda) = (1+\lambda)^n \: \detl Q^{-1}$, and that likewise, the $A_k$ matrices are symmetric Toeplitz with two alternating quantities in the first row. The bias variable $\lambda$ enters the recurrence relations, which are now given by (the initial conditions remain unchanged and the definition of $r_k = b_k/a_k$ is the same)
\be
a_k = \frac{a_{k-1}^2 + \lambda \, b_{k-1}^2}{a_{k-2}}, \qquad b_k = \frac{b_{k-1}^2 + \lambda \, a_{k-1}^2}{b_{k-2}}, \qquad r_k = \frac{\lambda + r_{k-1}^2}{1 + \lambda \, r_{k-1}^2} \, \frac 1{r_{k-2}}.
\label{akbk}
\ee
Solve these to compute $a_n = \detl Q^{-1}$, and/or find the asymptotic value of $a_n$ for large $n$.
\end{prob}

\smallskip
Following the proof of Corollary \ref{coper}, we readily obtain
\be
\frac{a_k}{a_{k-1}} = \frac{a_{k-1}}{a_{k-2}} (1 + \lambda \, r_{k-1}^2) = a^{-1} \, (1 + \lambda \, r_1^2) \ldots (1 + \lambda \, r_{k-2}^2) \, (1 + \lambda \, r_{k-1}^2), \qquad k \ge 1,
\label{recak}
\ee
from which the partition function for general values of $a,b,\lambda$ follows,
\be
T_n(a,b|\lambda) = (1+\lambda)^n \, a_n = a^{-n} \, \prod_{k=0}^{n-1} \, (1 + \lambda \, r_k^2)^{n-k}.
\label{tnabl}
\ee
We note that under the exchange of $a$ and $b$, the two sequences $(a_k)$ and $(b_k)$ are interchanged, so that 
\be
T_n(b,a|\lambda) = (1 + \lambda)^n \, b_n = r_n \, T_n(a,b|\lambda).
\ee 

The determination of the partition functions for periodic Aztec graphs of order $n$ with vertical bias $\sqrt\lambda$ only requires the knowledge of the $n$ first terms in the sequence $(r_k)$. However these are rational functions of $\lambda$ and $t \equiv r_1 = \frac ab$ and become rapidly untractable, like for instance,
\be
r_4 = \frac{t^8 \lambda^3+t^6 \left(2 \lambda^3+3 \lambda^2-2 \lambda+1\right)+3 t^4 \lambda\left(\lambda^2+1\right)+t^2 \lambda \left(\lambda^3-2 \lambda^2+3 \lambda+2\right)+\lambda}{t^8 \lambda+t^6 \lambda
   \left(\lambda^3-2 \lambda^2+3 \lambda+2\right)+3 t^4 \lambda\left(\lambda^2+1\right)+t^2 \left(2 \lambda^3+3 \lambda^2-2 \lambda+1\right)+\lambda^3}.
\ee

A partial solution to Problem \ref{problem} can be given as follows. Going back to the case $\lambda=1$, one checks that the above expression for $r_4$ collapses to 1, and that $r_5$ equals $t$, implying that the sequence $(r_k)_{k \ge 0}$ is 4-periodic, as noted above, and has a simple expression, $r_k = 1,t,1,t^{-1},1,t,1,t^{-1},\ldots$ for $k \ge 0$. This suggests to see whether other values of $\lambda$ make the sequence periodic. The conditions for the sequence to be $p$-periodic, $r_{k+p}=r_k$ for all $k \ge 0$, read $r_p = 1$ and $r_{p+1}=t$, or equivalently $r_{p-1} = t^{-1}$ and $r_{p}=1$.

For $p=2$, the two conditions yield $t = 1$ and a constant sequence $r_k=1$; this corresponds to the usual, one-periodic measure with a vertical bias (one recovers the result of Corollary \ref{bias} up to a power of $a$). For higher values of $p$, the two conditions imply a polynomial relation between $\lambda$ and $t$, the analysis of which can however be simplified. Indeed, we note that, when extended towards the negative values of $k$, the sequence $(r_k)$ satisfies $r_{-k}^{} = r_k^{-1}$, for all $k \ge 0$; this implies that a sequence which is $p$-periodic on $\Z_+$ is also $p$-periodic on $\Z$, $r_{k+p}=r_k$ for all $k \in \Z$. For even $p$, we have $r^{}_{-p/2} = r^{}_{p/2} = r_{p/2}^{-1}$ and therefore $r_{p/2}^2 = 1$; for odd $p$, we have similarly $r^{}_{-(p+1)/2} = r^{}_{(p-1)/2} = r_{(p+1)/2}^{-1}$. Restricting to positive sequences ($t$ and $\lambda$ are both positive), we see that in both cases, the $p$-periodicity implies a single condition,
\be
r_{\frac{p-1}2} r_{\frac{p+1}2} = 1 \quad \text{for $p$ odd}, \qquad r_{\frac p2}=1  \quad \text{for $p$ even}.
\label{1cond}
\ee
Conversely these relations alone imply a mirror-inversion property, namely $r^{}_{p-k} = r^{-1}_k$, which itself implies $r_{p-1}=t^{-1}, \, r_p=1$ and therefore the $p$-periodicity. 

The resulting polynomial conditions are straightforward to compute. For $p$ up to 12, by imposing (\ref{1cond}), we find the following periodicity conditions, where $\tau_k \equiv t^k + t^{-k}$,
\begin{subequations}
\allowdisplaybreaks
\bea
p=3 \;: && \lambda - (1 + \tau_1) = 0, \\
p=4 \;: && \lambda - 1 = 0, \\
p=5 \;: && \lambda^3 - (3 + 2 \tau_1 + \tau_3) \, \lambda^2 + (1 - 2 \tau_1 - \tau_2) \, \lambda + (1 + \tau_1 + \tau_2) = 0, \\
p=6 \;: && (1 + \tau_1) \, \lambda - 1 = 0, \\
p=7 \;: && \lambda^6 - (6 + 3\tau_1 + 2\tau_3 + \tau_5)\,\lambda^5 + (7 - 7 \tau_1 - 5 \tau_2 - 4 \tau_3 - \tau_4 + \tau_5) \, \lambda^4 \nonumber\\
&& \hspace{5mm} + \; (12 \tau_2 + 5 \tau_3 + 2 \tau_4 - \tau_5)\, \lambda^3 + (17 + 12 \tau_1 + 5 \tau_2 - \tau_3 + 5 \tau_4 + \tau_5 + \tau_6) \, \lambda^2 \nonumber\\
&& \hspace{5mm} + \; (2 - \tau_1 + 4 \tau_2 + 3 \tau_3) \, \lambda - (1 + \tau_1 + \tau_2 + \tau_3) = 0, \\
p=8 \;: && \lambda^2 - (4 + \tau_2) \, \lambda + 1 = 0, \label{per8} \\
p=9 \;: && \lambda^9 - (9 + 3\tau_1 + 3\tau_3 + 2\tau_5 + \tau_7) \, \lambda^8  \nonumber\\
&& \hspace{5mm} + \; (10 - 28 \tau_1 - 20 \tau_2 - 16 \tau_3 - 10 \tau_4 - 6 \tau_5 - 4 \tau_6 + 2 \tau_7 - \tau_8) \, \lambda^7 \nonumber\\
&& \hspace{5mm} - \; (74 + 50 \tau_1 + 4 \tau_2 + 35 \tau_3 + 14 \tau_4 + 5 \tau_5 + 4 \tau_6 + 9 \tau_7 - \tau_8 + \tau_9) \, \lambda^6 \nonumber\\
&& \hspace{5mm} + \; (16 - 40 \tau_1 - 56 \tau_2 - 16 \tau_3 + 8 \tau_4 - 38 \tau_5 + 8 \tau_6 - 2 \tau_7 - \tau_8)\, \lambda^5 \nonumber\\
&& \hspace{5mm} - \; (16 + 5 \tau_1 - 56 \tau_2 + 44 \tau_3 + 8 \tau_4 - 18 \tau_5 + 8 \tau_6 - \tau_7 - \tau_8) \, \lambda^4 \nonumber\\
&& \hspace{5mm} + \; (74 - 22 \tau_1 + 4 \tau_2 + 42 \tau_3 + 14 \tau_4 - 4 \tau_5 + 4 \tau_6 - \tau_8) \, \lambda^3 \nonumber\\
&& \hspace{5mm} - \; (10 - 28 \tau_1 - 20 \tau_2 + 17 \tau_3 - 10 \tau_4 - \tau_5 - 4 \tau_6 - \tau_8)\,\lambda^2 \nonumber\\
&& \hspace{5mm} + \; (9 - 6 \tau_1 + 6 \tau_3) \, \lambda - (1 + \tau_3) = 0, \\
p=10 \;: && (1 + \tau_1 + \tau_2) \, \lambda^3 + (1 - 2 \tau_1 - \tau_2)  \, \lambda^2 - (3 + 2 \tau_1 + \tau_3)\, \lambda + 1 = 0, \\
p=11 \;: && \lambda^{15} - (15 + 5 \tau + 4 \tau_3 + 3 \tau_5 + 2 \tau_7 + \tau_9) \, \lambda^{14} \\
&& \hspace{5mm} + \; \ldots + (3 - 2 \tau_1 + 5 \tau_2 + 2 \tau_3 + 
 11 \tau_4 + 10 \tau_5) \, \lambda - (1 + \tau_1 + \tau_2 + \tau_3 + \tau_4 + \tau_5) = 0, \nonumber\\
p=12 \;: && \lambda^4 - (6 + 2\tau_2 + \tau_4) \, \lambda^3 - (8 + 8\tau_2 + \tau_4) \, \lambda^2 - (6 + 2\tau_2 + \tau_4) \, \lambda + 1 = 0.
\eea
\label{per}
\end{subequations}

\noindent 
\underline{\it Remark 1.} The conditions for $p=3$ and $p=6$ are related by $\lambda \leftrightarrow \lambda^{-1}$, and the same is true for $p=5$ and $p=10$. This relation between the conditions for $p$ and $2p$ is general for $p$ odd, and is a simple consequence of the fact that the inversion of $\lambda$ keeps the odd terms of the sequence $(r_k)$ invariant but inverts the even terms, $r^{}_{2k+1}(\lambda^{-1},t) = r^{}_{2k+1}(\lambda,t)$ and $r^{}_{2k}(\lambda^{-1},t) = r^{-1}_{2k}(\lambda,t)$. Consequently, when one inverts the even terms in a $p$-periodic sequence, $p$ odd, the new sequence $\tilde r_k$ ceases to be $p$-periodic because it satisfies $\tilde r_{\frac{p+1}2} = \tilde r_{\frac{p-1}2}$, and becomes $2p$-periodic since $\tilde r_p=1$. For $p$ a multiple of 4, a $p$-periodic sequence remains $p$-periodic under $\lambda \leftrightarrow \lambda^{-1}$; the corresponding polynomial equation must therefore be invariant under the inversion of $\lambda$.

\medskip
\noindent 
\underline{\it Remark 2.} The mirror-inversion symmetry  of a periodic sequence implies that for any $p$, the polynomials are invariant under $t \leftrightarrow t^{-1}$ since inverting $t$ is equivalent to run the recurrence backwards, starting from $r_p=1, \, r_{p-1}=t^{-1}$ down to $r_1=t,\,r_0=1$. The middle conditions (\ref{1cond}) are thus also invariant.

\medskip
When the sequence $(r_k)$ is periodic, the explicit calculation of the partition functions simplifies. We will denote the partition function by $T_n^{(p)}(a,b|\lambda)$ in case $r_k$ is $p$-periodic. Let us note that for fixed $p$, this leaves a finite number of possible values for $\lambda$ in terms of $t=\frac ab$, namely the solutions of the polynomial equation for periodicity $p$. These partition functions satisfy the same relation under the exchange of $a$ and $b$,
\be
T^{(p)}_n(b,a|\lambda) = (1 + \lambda)^n \, b_n = r_n \, T^{(p)}_n(a,b|\lambda).
\ee 
Indeed the exchange of $a$ and $b$ inverts the value of $t$ but preserves the polynomial equations, from Remark 2 above, and therefore the value of $\lambda$. The next result provides the explicit expression of $T_n^{(p)}(a,b|\lambda)$ for the three simplest cases, $p=3,6$ and 8, namely those for which $t$ leaves one or two possible values for $\lambda$. In the other cases, the partition functions can be computed but their expressions remain complicated.

\begin{coro}
The partition functions $T_n^{(p)}(a,b|\lambda)$ with periodicity $p=3, \, 6$ and $8$ are given by,
\begin{subequations}
\bea
p=3 \;: && \hspace{-5mm} T_n^{(3)}(a,b|\lambda) = \Big(\frac{a+b}{ab}\Big)^{\lfloor \frac{(2n+1)(n+1)}3 \rfloor} \: \big(a^2 + b^2 \big)^{\lfloor \frac{n^2}3 \rfloor} \times (1,b,a) {\rm \ \ for\ } n=(0,1,2) \bmod 3, \\
\noalign{\medskip}
p=6 \;: && \hspace{-5mm} T_n^{(6)}(a,b|\lambda) = \Big(\frac{a+b}{ab}\Big)^{\lfloor \frac{(2n+1)(n+1)}3 \rfloor} \: \big(a^2 + b^2 \big)^{\lfloor \frac{n^2}3 \rfloor} \: \Big(\frac{ab}{a^2+ab+b^2}\Big)^{\frac{n(n+1)}2} \nonumber\\
\noalign{\smallskip}
&& \hspace{4cm} \times \:  (1,b,b,1,a,a) {\rm \ \ for\ } n=(0,1,2,3,4,5) \bmod 6, \\
\noalign{\medskip}
p=8 \;: && \hspace{-5mm} T_n^{(8)}(a,b|\lambda) = \big( ab \big)^{-\lfloor \frac{n+1}2 \rfloor} \: \Big(\frac{a^2+b^2}{ab}\Big)^{\lfloor \frac{3n^2}8 \rfloor} \:\lambda^{\lfloor \frac{n^2}8 \rfloor} \: (1 + \lambda)^{\lfloor \frac{(n+1)^2}4 \rfloor} \nonumber\\
&& \hspace{2cm} \times \:  (1,b,\textstyle \frac{\lambda a^2 + b^2}{a^2+b^2},b,1,a,\frac{a^2 + \lambda b^2}{a^2+b^2},a) {\rm \ \ for\ } n=(0,1,2,3,4,5,6,7) \bmod 8.
\eea
\end{subequations}
\end{coro}

\medskip
\noindent{\it Proof.} When $p=3$ and $p=6$, $\lambda$ is uniquely fixed to $(1 + t + t^{-1})^{\pm 1}$ and the sequences $(r_k)$ explicitly read $1,\,t,\,t^{-1},\ldots$ and $1,\,t,\,t,\,1,\,t^{-1},\,t^{-1},\ldots$ respectively. For $p=8$, it is given by
\be
1,\,t,\,\frac{\lambda + t^2}{1 + \lambda \, t^2}, \, t, \,1,\, t^{-1}, \, \frac{1 + \lambda \, t^2}{\lambda + t^2}, \, t^{-1}, \, \ldots
\ee
with $\lambda$ one of the two roots of (\ref{per8}). In each case the product in (\ref{tnabl}) is then easily computed. \cqfd

\medskip
Remarkably, the polynomial expressions in (\ref{per}) exactly reproduce\footnote{There are some differences, due to the assumption $0 < \lambda < 1$ made in \cite{BD23}.} 
certain polynomial relations in \cite{BD23}, which the authors obtain when characterizing the finite periodicity of a flow defined in terms of translations in an elliptic curve depending on $\lambda$ and $t$. This is rather surprising because, being related to the iteration of a Wiener-Hopf factorization, the periodicity considered in \cite{BD23} has a totally different origin from ours. It would be interesting to understand and connect the two points of view. We collect in the Appendix some of the facts about the sequence $(r_k)$ which partially clarify the connection; at this stage however, we do not claim a full understanding. 

In fact, it turns out that the two sequences $(a_k)$ and $(b_k)$ exhibit a rather natural relation with the curve studied in \cite{BD23}. As pointed out to us by Michael Somos, $(a_k)$ and $(b_k)$ are Somos-4 sequences, which themselves can be quite generally related to translational flows on elliptic curves \cite{Sw03}. For the specific sequences $(a_k)$ and $(b_k)$ defined in (\ref{akbk}), the relevant elliptic curve is precisely the curve discussed in \cite{BD23}; we refer to the Appendix for more details.


\vskip 0.5truecm
\noindent
{\bf 3.3 Applications to subgraphs of Aztec diamonds}
\addcontentsline{toc}{subsubsection}{3.3 Applications to subgraphs of Aztec diamonds}

\medskip
\noindent
Theorem \ref{thm1} can be used to compute the number of perfect matchings of subgraphs of Aztec graphs, as noticed by Propp and mentioned in the Introduction.

\begin{coro}
\cite{Pr05} The partition function for perfect matchings of a square grid of size $2n$ with vertical bias $\sqrt{\lambda}$ is equal to $\lambda^{-n(n-1)/2} \, \detl P_{\rm sq}$, where $P_{\rm sq}$ is the all-ones matrix of order $2n$ in the corners of which 1's are replaced by 0's.
\end{coro}

\begin{figure}[t]
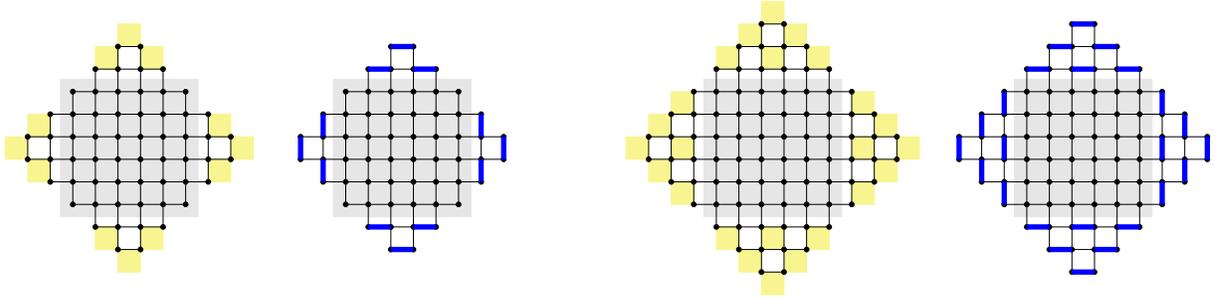

\begin{center}
\psset{unit=.3cm}
\pspicture(0,0)(10,12)
%
\rput(-0.5,4.5){\pspolygon[linewidth=0pt,linecolor=lightyellow,fillstyle=solid,fillcolor=lightyellow](0,0)(0,1)(1,1)(1,0)}
\rput(0.5,5.5){\pspolygon[linewidth=0pt,linecolor=lightyellow,fillstyle=solid,fillcolor=lightyellow](0,0)(0,1)(1,1)(1,0)}
\rput(3.5,8.5){\pspolygon[linewidth=0pt,linecolor=lightyellow,fillstyle=solid,fillcolor=lightyellow](0,0)(0,1)(1,1)(1,0)}
\rput(4.5,9.5){\pspolygon[linewidth=0pt,linecolor=lightyellow,fillstyle=solid,fillcolor=lightyellow](0,0)(0,1)(1,1)(1,0)}
\rput(8.5,3.5){\pspolygon[linewidth=0pt,linecolor=lightyellow,fillstyle=solid,fillcolor=lightyellow](0,0)(0,1)(1,1)(1,0)}
\rput(8.5,5.5){\pspolygon[linewidth=0pt,linecolor=lightyellow,fillstyle=solid,fillcolor=lightyellow](0,0)(0,1)(1,1)(1,0)}
\rput(4.5,-0.5){\pspolygon[linewidth=0pt,linecolor=lightyellow,fillstyle=solid,fillcolor=lightyellow](0,0)(0,1)(1,1)(1,0)}
\rput(3.5,0.5){\pspolygon[linewidth=0pt,linecolor=lightyellow,fillstyle=solid,fillcolor=lightyellow](0,0)(0,1)(1,1)(1,0)}
\rput(0.5,3.5){\pspolygon[linewidth=0pt,linecolor=lightyellow,fillstyle=solid,fillcolor=lightyellow](0,0)(0,1)(1,1)(1,0)}
\rput(5.5,8.5){\pspolygon[linewidth=0pt,linecolor=lightyellow,fillstyle=solid,fillcolor=lightyellow](0,0)(0,1)(1,1)(1,0)}
\rput(5.5,0.5){\pspolygon[linewidth=0pt,linecolor=lightyellow,fillstyle=solid,fillcolor=lightyellow](0,0)(0,1)(1,1)(1,0)}
\rput(9.5,4.5){\pspolygon[linewidth=0pt,linecolor=lightyellow,fillstyle=solid,fillcolor=lightyellow](0,0)(0,1)(1,1)(1,0)}
\rput(2,2){\pspolygon[linewidth=1.2pt,linecolor=mygrey,fillstyle=solid,fillcolor=mygrey](0,0)(0,6)(6,6)(6,0)}
\multido{\nt=4+1}{2}{\rput(0,\nt){\pscircle[linecolor=black,fillstyle=solid,fillcolor=black](0.5,0.5){0.08}}}
\multido{\nt=3+1}{4}{\rput(1,\nt){\pscircle[linecolor=black,fillstyle=solid,fillcolor=black](0.5,0.5){0.08}}}
\multido{\nt=2+1}{6}{\rput(2,\nt){\pscircle[linecolor=black,fillstyle=solid,fillcolor=black](0.5,0.5){0.08}}}
\multido{\nt=1+1}{8}{\rput(3,\nt){\pscircle[linecolor=black,fillstyle=solid,fillcolor=black](0.5,0.5){0.08}}}
\multido{\nt=0+1}{10}{\rput(4,\nt){\pscircle[linecolor=black,fillstyle=solid,fillcolor=black](0.5,0.5){0.08}}}
\multido{\nt=0+1}{10}{\rput(5,\nt){\pscircle[linecolor=black,fillstyle=solid,fillcolor=black](0.5,0.5){0.08}}}
\multido{\nt=1+1}{8}{\rput(6,\nt){\pscircle[linecolor=black,fillstyle=solid,fillcolor=black](0.5,0.5){0.08}}}
\multido{\nt=2+1}{6}{\rput(7,\nt){\pscircle[linecolor=black,fillstyle=solid,fillcolor=black](0.5,0.5){0.08}}}
\multido{\nt=3+1}{4}{\rput(8,\nt){\pscircle[linecolor=black,fillstyle=solid,fillcolor=black](0.5,0.5){0.08}}}
\multido{\nt=4+1}{2}{\rput(9,\nt){\pscircle[linecolor=black,fillstyle=solid,fillcolor=black](0.5,0.5){0.08}}}
\rput(0.5,-0.5){
\psline[linewidth=0.2pt,linecolor=black](4,1)(5,1)
\psline[linewidth=0.2pt,linecolor=black](3,2)(6,2)
\psline[linewidth=0.2pt,linecolor=black](2,3)(7,3)
\psline[linewidth=0.2pt,linecolor=black](1,4)(8,4)
\psline[linewidth=0.2pt,linecolor=black](0,5)(9,5)
\psline[linewidth=0.2pt,linecolor=black](0,6)(9,6)
\psline[linewidth=0.2pt,linecolor=black](1,7)(8,7)
\psline[linewidth=0.2pt,linecolor=black](2,8)(7,8)
\psline[linewidth=0.2pt,linecolor=black](3,9)(6,9)
\psline[linewidth=0.2pt,linecolor=black](4,10)(5,10)
\psline[linewidth=0.2pt,linecolor=black](0,5)(0,6)
\psline[linewidth=0.2pt,linecolor=black](1,4)(1,7)
\psline[linewidth=0.2pt,linecolor=black](2,3)(2,8)
\psline[linewidth=0.2pt,linecolor=black](3,2)(3,9)
\psline[linewidth=0.2pt,linecolor=black](4,1)(4,10)
\psline[linewidth=0.2pt,linecolor=black](5,1)(5,10)
\psline[linewidth=0.2pt,linecolor=black](6,2)(6,9)
\psline[linewidth=0.2pt,linecolor=black](7,3)(7,8)
\psline[linewidth=0.2pt,linecolor=black](8,4)(8,7)
\psline[linewidth=0.2pt,linecolor=black](9,5)(9,6)}
\endpspicture
\hspace{0.5cm}
\pspicture(0,0)(10,12)
\rput(2,2){\pspolygon[linewidth=1.2pt,linecolor=mygrey,fillstyle=solid,fillcolor=mygrey](0,0)(0,6)(6,6)(6,0)}
\multido{\nt=4+1}{2}{\rput(0,\nt){\pscircle[linecolor=black,fillstyle=solid,fillcolor=black](0.5,0.5){0.08}}}
\multido{\nt=3+1}{4}{\rput(1,\nt){\pscircle[linecolor=black,fillstyle=solid,fillcolor=black](0.5,0.5){0.08}}}
\multido{\nt=2+1}{6}{\rput(2,\nt){\pscircle[linecolor=black,fillstyle=solid,fillcolor=black](0.5,0.5){0.08}}}
\multido{\nt=1+1}{8}{\rput(3,\nt){\pscircle[linecolor=black,fillstyle=solid,fillcolor=black](0.5,0.5){0.08}}}
\multido{\nt=0+1}{10}{\rput(4,\nt){\pscircle[linecolor=black,fillstyle=solid,fillcolor=black](0.5,0.5){0.08}}}
\multido{\nt=0+1}{10}{\rput(5,\nt){\pscircle[linecolor=black,fillstyle=solid,fillcolor=black](0.5,0.5){0.08}}}
\multido{\nt=1+1}{8}{\rput(6,\nt){\pscircle[linecolor=black,fillstyle=solid,fillcolor=black](0.5,0.5){0.08}}}
\multido{\nt=2+1}{6}{\rput(7,\nt){\pscircle[linecolor=black,fillstyle=solid,fillcolor=black](0.5,0.5){0.08}}}
\multido{\nt=3+1}{4}{\rput(8,\nt){\pscircle[linecolor=black,fillstyle=solid,fillcolor=black](0.5,0.5){0.08}}}
\multido{\nt=4+1}{2}{\rput(9,\nt){\pscircle[linecolor=black,fillstyle=solid,fillcolor=black](0.5,0.5){0.08}}}
\rput(0.5,-0.5){
\psline[linewidth=0.2pt,linecolor=black](4,1)(5,1)
\psline[linewidth=0.2pt,linecolor=black](3,2)(6,2)
\psline[linewidth=0.2pt,linecolor=black](2,3)(7,3)
\psline[linewidth=0.2pt,linecolor=black](1,4)(8,4)
\psline[linewidth=0.2pt,linecolor=black](0,5)(9,5)
\psline[linewidth=0.2pt,linecolor=black](0,6)(9,6)
\psline[linewidth=0.2pt,linecolor=black](1,7)(8,7)
\psline[linewidth=0.2pt,linecolor=black](2,8)(7,8)
\psline[linewidth=0.2pt,linecolor=black](3,9)(6,9)
\psline[linewidth=0.2pt,linecolor=black](4,10)(5,10)
\psline[linewidth=0.2pt,linecolor=black](0,5)(0,6)
\psline[linewidth=0.2pt,linecolor=black](1,4)(1,7)
\psline[linewidth=0.2pt,linecolor=black](2,3)(2,8)
\psline[linewidth=0.2pt,linecolor=black](3,2)(3,9)
\psline[linewidth=0.2pt,linecolor=black](4,1)(4,10)
\psline[linewidth=0.2pt,linecolor=black](5,1)(5,10)
\psline[linewidth=0.2pt,linecolor=black](6,2)(6,9)
\psline[linewidth=0.2pt,linecolor=black](7,3)(7,8)
\psline[linewidth=0.2pt,linecolor=black](8,4)(8,7)
\psline[linewidth=0.2pt,linecolor=black](9,5)(9,6)}
\psline[linewidth=2.0pt,linecolor=blue](4.5,9.5)(5.5,9.5)
\psline[linewidth=2.0pt,linecolor=blue](3.5,8.5)(4.5,8.5)
\psline[linewidth=2.0pt,linecolor=blue](5.5,8.5)(6.5,8.5)
\psline[linewidth=2.0pt,linecolor=blue](1.5,5.5)(1.5,6.5)
\psline[linewidth=2.0pt,linecolor=blue](0.5,4.5)(0.5,5.5)
\psline[linewidth=2.0pt,linecolor=blue](1.5,3.5)(1.5,4.5)
\psline[linewidth=2.0pt,linecolor=blue](8.5,5.5)(8.5,6.5)
\psline[linewidth=2.0pt,linecolor=blue](9.5,4.5)(9.5,5.5)
\psline[linewidth=2.0pt,linecolor=blue](8.5,3.5)(8.5,4.5)
\psline[linewidth=2.0pt,linecolor=blue](4.5,0.5)(5.5,0.5)
\psline[linewidth=2.0pt,linecolor=blue](3.5,1.5)(4.5,1.5)
\psline[linewidth=2.0pt,linecolor=blue](5.5,1.5)(6.5,1.5)
\endpspicture
\hspace{1.5cm}
\pspicture(0,0)(10,12)
%
\rput(-0.5,4.5){\pspolygon[linewidth=0pt,linecolor=lightyellow,fillstyle=solid,fillcolor=lightyellow](0,0)(0,1)(1,1)(1,0)}
\rput(0.5,5.5){\pspolygon[linewidth=0pt,linecolor=lightyellow,fillstyle=solid,fillcolor=lightyellow](0,0)(0,1)(1,1)(1,0)}
\rput(1.5,6.5){\pspolygon[linewidth=0pt,linecolor=lightyellow,fillstyle=solid,fillcolor=lightyellow](0,0)(0,1)(1,1)(1,0)}
\rput(3.5,8.5){\pspolygon[linewidth=0pt,linecolor=lightyellow,fillstyle=solid,fillcolor=lightyellow](0,0)(0,1)(1,1)(1,0)}
\rput(4.5,9.5){\pspolygon[linewidth=0pt,linecolor=lightyellow,fillstyle=solid,fillcolor=lightyellow](0,0)(0,1)(1,1)(1,0)}
\rput(5.5,10.5){\pspolygon[linewidth=0pt,linecolor=lightyellow,fillstyle=solid,fillcolor=lightyellow](0,0)(0,1)(1,1)(1,0)}
\rput(6.5,9.5){\pspolygon[linewidth=0pt,linecolor=lightyellow,fillstyle=solid,fillcolor=lightyellow](0,0)(0,1)(1,1)(1,0)}
\rput(7.5,8.5){\pspolygon[linewidth=0pt,linecolor=lightyellow,fillstyle=solid,fillcolor=lightyellow](0,0)(0,1)(1,1)(1,0)}
\rput(9.5,6.5){\pspolygon[linewidth=0pt,linecolor=lightyellow,fillstyle=solid,fillcolor=lightyellow](0,0)(0,1)(1,1)(1,0)}
\rput(10.5,5.5){\pspolygon[linewidth=0pt,linecolor=lightyellow,fillstyle=solid,fillcolor=lightyellow](0,0)(0,1)(1,1)(1,0)}
\rput(11.5,4.5){\pspolygon[linewidth=0pt,linecolor=lightyellow,fillstyle=solid,fillcolor=lightyellow](0,0)(0,1)(1,1)(1,0)}
\rput(10.5,3.5){\pspolygon[linewidth=0pt,linecolor=lightyellow,fillstyle=solid,fillcolor=lightyellow](0,0)(0,1)(1,1)(1,0)}
\rput(9.5,2.5){\pspolygon[linewidth=0pt,linecolor=lightyellow,fillstyle=solid,fillcolor=lightyellow](0,0)(0,1)(1,1)(1,0)}
\rput(7.5,0.5){\pspolygon[linewidth=0pt,linecolor=lightyellow,fillstyle=solid,fillcolor=lightyellow](0,0)(0,1)(1,1)(1,0)}
\rput(6.5,-0.5){\pspolygon[linewidth=0pt,linecolor=lightyellow,fillstyle=solid,fillcolor=lightyellow](0,0)(0,1)(1,1)(1,0)}
\rput(5.5,-1.5){\pspolygon[linewidth=0pt,linecolor=lightyellow,fillstyle=solid,fillcolor=lightyellow](0,0)(0,1)(1,1)(1,0)}
\rput(4.5,-0.5){\pspolygon[linewidth=0pt,linecolor=lightyellow,fillstyle=solid,fillcolor=lightyellow](0,0)(0,1)(1,1)(1,0)}
\rput(3.5,0.5){\pspolygon[linewidth=0pt,linecolor=lightyellow,fillstyle=solid,fillcolor=lightyellow](0,0)(0,1)(1,1)(1,0)}
\rput(1.5,2.5){\pspolygon[linewidth=0pt,linecolor=lightyellow,fillstyle=solid,fillcolor=lightyellow](0,0)(0,1)(1,1)(1,0)}
\rput(0.5,3.5){\pspolygon[linewidth=0pt,linecolor=lightyellow,fillstyle=solid,fillcolor=lightyellow](0,0)(0,1)(1,1)(1,0)}
\rput(1.5,4.5){\pspolygon[linewidth=0pt,linecolor=lightyellow,fillstyle=solid,fillcolor=lightyellow](0,0)(0,1)(1,1)(1,0)}
\rput(5.5,8.5){\pspolygon[linewidth=0pt,linecolor=lightyellow,fillstyle=solid,fillcolor=lightyellow](0,0)(0,1)(1,1)(1,0)}
\rput(5.5,0.5){\pspolygon[linewidth=0pt,linecolor=lightyellow,fillstyle=solid,fillcolor=lightyellow](0,0)(0,1)(1,1)(1,0)}
\rput(9.5,4.5){\pspolygon[linewidth=0pt,linecolor=lightyellow,fillstyle=solid,fillcolor=lightyellow](0,0)(0,1)(1,1)(1,0)}
\rput(3,2){\pspolygon[linewidth=1.2pt,linecolor=mygrey,fillstyle=solid,fillcolor=mygrey](0,0)(0,6)(6,6)(6,0)}
\multido{\nt=4+1}{2}{\rput(0,\nt){\pscircle[linecolor=black,fillstyle=solid,fillcolor=black](0.5,0.5){0.08}}}
\multido{\nt=3+1}{4}{\rput(1,\nt){\pscircle[linecolor=black,fillstyle=solid,fillcolor=black](0.5,0.5){0.08}}}
\multido{\nt=2+1}{6}{\rput(2,\nt){\pscircle[linecolor=black,fillstyle=solid,fillcolor=black](0.5,0.5){0.08}}}
\multido{\nt=1+1}{8}{\rput(3,\nt){\pscircle[linecolor=black,fillstyle=solid,fillcolor=black](0.5,0.5){0.08}}}
\multido{\nt=0+1}{10}{\rput(4,\nt){\pscircle[linecolor=black,fillstyle=solid,fillcolor=black](0.5,0.5){0.08}}}
\multido{\nt=-1+1}{12}{\rput(5,\nt){\pscircle[linecolor=black,fillstyle=solid,fillcolor=black](0.5,0.5){0.08}}}
\multido{\nt=-1+1}{12}{\rput(6,\nt){\pscircle[linecolor=black,fillstyle=solid,fillcolor=black](0.5,0.5){0.08}}}
\multido{\nt=0+1}{10}{\rput(7,\nt){\pscircle[linecolor=black,fillstyle=solid,fillcolor=black](0.5,0.5){0.08}}}
\multido{\nt=1+1}{8}{\rput(8,\nt){\pscircle[linecolor=black,fillstyle=solid,fillcolor=black](0.5,0.5){0.08}}}
\multido{\nt=2+1}{6}{\rput(9,\nt){\pscircle[linecolor=black,fillstyle=solid,fillcolor=black](0.5,0.5){0.08}}}
\multido{\nt=3+1}{4}{\rput(10,\nt){\pscircle[linecolor=black,fillstyle=solid,fillcolor=black](0.5,0.5){0.08}}}
\multido{\nt=4+1}{2}{\rput(11,\nt){\pscircle[linecolor=black,fillstyle=solid,fillcolor=black](0.5,0.5){0.08}}}
\rput(0.5,-0.5){
\psline[linewidth=0.2pt,linecolor=black](5,0)(6,0)
\psline[linewidth=0.2pt,linecolor=black](4,1)(7,1)
\psline[linewidth=0.2pt,linecolor=black](3,2)(8,2)
\psline[linewidth=0.2pt,linecolor=black](2,3)(9,3)
\psline[linewidth=0.2pt,linecolor=black](1,4)(10,4)
\psline[linewidth=0.2pt,linecolor=black](0,5)(11,5)
\psline[linewidth=0.2pt,linecolor=black](0,6)(11,6)
\psline[linewidth=0.2pt,linecolor=black](1,7)(10,7)
\psline[linewidth=0.2pt,linecolor=black](2,8)(9,8)
\psline[linewidth=0.2pt,linecolor=black](3,9)(8,9)
\psline[linewidth=0.2pt,linecolor=black](4,10)(7,10)
\psline[linewidth=0.2pt,linecolor=black](5,11)(6,11)
\psline[linewidth=0.2pt,linecolor=black](0,5)(0,6)
\psline[linewidth=0.2pt,linecolor=black](1,4)(1,7)
\psline[linewidth=0.2pt,linecolor=black](2,3)(2,8)
\psline[linewidth=0.2pt,linecolor=black](3,2)(3,9)
\psline[linewidth=0.2pt,linecolor=black](4,1)(4,10)
\psline[linewidth=0.2pt,linecolor=black](5,0)(5,11)
\psline[linewidth=0.2pt,linecolor=black](6,0)(6,11)
\psline[linewidth=0.2pt,linecolor=black](7,1)(7,10)
\psline[linewidth=0.2pt,linecolor=black](8,2)(8,9)
\psline[linewidth=0.2pt,linecolor=black](9,3)(9,8)
\psline[linewidth=0.2pt,linecolor=black](10,4)(10,7)
\psline[linewidth=0.2pt,linecolor=black](11,5)(11,6)}
\endpspicture
\hspace{1cm}
\pspicture(0,0)(10,12)
\rput(3,2){\pspolygon[linewidth=1.2pt,linecolor=mygrey,fillstyle=solid,fillcolor=mygrey](0,0)(0,6)(6,6)(6,0)}
\multido{\nt=4+1}{2}{\rput(0,\nt){\pscircle[linecolor=black,fillstyle=solid,fillcolor=black](0.5,0.5){0.08}}}
\multido{\nt=3+1}{4}{\rput(1,\nt){\pscircle[linecolor=black,fillstyle=solid,fillcolor=black](0.5,0.5){0.08}}}
\multido{\nt=2+1}{6}{\rput(2,\nt){\pscircle[linecolor=black,fillstyle=solid,fillcolor=black](0.5,0.5){0.08}}}
\multido{\nt=1+1}{8}{\rput(3,\nt){\pscircle[linecolor=black,fillstyle=solid,fillcolor=black](0.5,0.5){0.08}}}
\multido{\nt=0+1}{10}{\rput(4,\nt){\pscircle[linecolor=black,fillstyle=solid,fillcolor=black](0.5,0.5){0.08}}}
\multido{\nt=-1+1}{12}{\rput(5,\nt){\pscircle[linecolor=black,fillstyle=solid,fillcolor=black](0.5,0.5){0.08}}}
\multido{\nt=-1+1}{12}{\rput(6,\nt){\pscircle[linecolor=black,fillstyle=solid,fillcolor=black](0.5,0.5){0.08}}}
\multido{\nt=0+1}{10}{\rput(7,\nt){\pscircle[linecolor=black,fillstyle=solid,fillcolor=black](0.5,0.5){0.08}}}
\multido{\nt=1+1}{8}{\rput(8,\nt){\pscircle[linecolor=black,fillstyle=solid,fillcolor=black](0.5,0.5){0.08}}}
\multido{\nt=2+1}{6}{\rput(9,\nt){\pscircle[linecolor=black,fillstyle=solid,fillcolor=black](0.5,0.5){0.08}}}
\multido{\nt=3+1}{4}{\rput(10,\nt){\pscircle[linecolor=black,fillstyle=solid,fillcolor=black](0.5,0.5){0.08}}}
\multido{\nt=4+1}{2}{\rput(11,\nt){\pscircle[linecolor=black,fillstyle=solid,fillcolor=black](0.5,0.5){0.08}}}
\rput(0.5,-0.5){
\psline[linewidth=0.2pt,linecolor=black](5,0)(6,0)
\psline[linewidth=0.2pt,linecolor=black](4,1)(7,1)
\psline[linewidth=0.2pt,linecolor=black](3,2)(8,2)
\psline[linewidth=0.2pt,linecolor=black](2,3)(9,3)
\psline[linewidth=0.2pt,linecolor=black](1,4)(10,4)
\psline[linewidth=0.2pt,linecolor=black](0,5)(11,5)
\psline[linewidth=0.2pt,linecolor=black](0,6)(11,6)
\psline[linewidth=0.2pt,linecolor=black](1,7)(10,7)
\psline[linewidth=0.2pt,linecolor=black](2,8)(9,8)
\psline[linewidth=0.2pt,linecolor=black](3,9)(8,9)
\psline[linewidth=0.2pt,linecolor=black](4,10)(7,10)
\psline[linewidth=0.2pt,linecolor=black](5,11)(6,11)
\psline[linewidth=0.2pt,linecolor=black](0,5)(0,6)
\psline[linewidth=0.2pt,linecolor=black](1,4)(1,7)
\psline[linewidth=0.2pt,linecolor=black](2,3)(2,8)
\psline[linewidth=0.2pt,linecolor=black](3,2)(3,9)
\psline[linewidth=0.2pt,linecolor=black](4,1)(4,10)
\psline[linewidth=0.2pt,linecolor=black](5,0)(5,11)
\psline[linewidth=0.2pt,linecolor=black](6,0)(6,11)
\psline[linewidth=0.2pt,linecolor=black](7,1)(7,10)
\psline[linewidth=0.2pt,linecolor=black](8,2)(8,9)
\psline[linewidth=0.2pt,linecolor=black](9,3)(9,8)
\psline[linewidth=0.2pt,linecolor=black](10,4)(10,7)
\psline[linewidth=0.2pt,linecolor=black](11,5)(11,6)}
\psline[linewidth=2.0pt,linecolor=blue](5.5,10.5)(6.5,10.5)
\psline[linewidth=2.0pt,linecolor=blue](4.5,9.5)(5.5,9.5)
\psline[linewidth=2.0pt,linecolor=blue](6.5,9.5)(7.5,9.5)
\psline[linewidth=2.0pt,linecolor=blue](3.5,8.5)(4.5,8.5)
\psline[linewidth=2.0pt,linecolor=blue](5.5,8.5)(6.5,8.5)
\psline[linewidth=2.0pt,linecolor=blue](7.5,8.5)(8.5,8.5)
\psline[linewidth=2.0pt,linecolor=blue](1.5,5.5)(1.5,6.5)
\psline[linewidth=2.0pt,linecolor=blue](0.5,4.5)(0.5,5.5)
\psline[linewidth=2.0pt,linecolor=blue](1.5,3.5)(1.5,4.5)
\psline[linewidth=2.0pt,linecolor=blue](2.5,2.5)(2.5,3.5)
\psline[linewidth=2.0pt,linecolor=blue](2.5,4.5)(2.5,5.5)
\psline[linewidth=2.0pt,linecolor=blue](2.5,6.5)(2.5,7.5)
\psline[linewidth=2.0pt,linecolor=blue](10.5,5.5)(10.5,6.5)
\psline[linewidth=2.0pt,linecolor=blue](11.5,4.5)(11.5,5.5)
\psline[linewidth=2.0pt,linecolor=blue](10.5,3.5)(10.5,4.5)
\psline[linewidth=2.0pt,linecolor=blue](9.5,2.5)(9.5,3.5)
\psline[linewidth=2.0pt,linecolor=blue](9.5,4.5)(9.5,5.5)
\psline[linewidth=2.0pt,linecolor=blue](9.5,6.5)(9.5,7.5)
\psline[linewidth=2.0pt,linecolor=blue](5.5,-0.5)(6.5,-0.5)
\psline[linewidth=2.0pt,linecolor=blue](4.5,0.5)(5.5,0.5)
\psline[linewidth=2.0pt,linecolor=blue](6.5,0.5)(7.5,0.5)
\psline[linewidth=2.0pt,linecolor=blue](3.5,1.5)(4.5,1.5)
\psline[linewidth=2.0pt,linecolor=blue](5.5,1.5)(6.5,1.5)
\psline[linewidth=2.0pt,linecolor=blue](7.5,1.5)(8.5,1.5)
\endpspicture
\end{center}
\caption{{\it Left:} the figure shows an Aztec diamond of order 5 with, in gray, the square subgraph of size 6. {\it Right:} The same square subgraph is embedded in a diamond of order 6. In both cases, the yellow faces are those for which the face variable is set to zero, forcing the arrangement of blue dimers.}
\label{fig4}
\end{figure}

\noindent
{\it Proof.} Let us consider a square grid of size $2n$-by-$2n$ (the size refers to the number of vertices in each row and each column; it coincides with the size of the square domain tiled by dominos, and materialized in Figure~\ref{fig4} by the shaded square). As one can see in Figure~\ref{fig4} for $n=3$, it can be embedded in an Aztec graph of order $2n-1$ or $2n$. We discuss the first option as the second one is a bit less economical. To reduce the Aztec graph to the square subgraph, the trick is to choose the face weights in such a way that the region which is exterior to the square is totally frozen. A simple way to achieve this is to set to zero the variables of all the faces shown in yellow, because this freezes the dimer arrangement in the four corners, leaving the central square free, as shown in Figure~\ref{fig4}. To see this, one notes that the yellow boundary faces must be adjacent to a single dimer (they cannot be adjacent to two dimers, and if one was adjacent to no dimer, its zero face weight would bring a vanishing contribution). This fixes the dimers along the boundaries of the four corners, and in turn forces a similar arrangement in the corners. Thus each yellow face is adjacent to a single dimer and contributes a face factor equal to $1$. In terms of the $(P,Q)$ weighting, this choice of face weights corresponds to take $Q=1$ and $P=P_{\rm sq}$ obtained from the all-ones matrix of size $2n$ by inserting in the four corners triangular arrays full of 0's. The number of 0's in each corner is $\frac{n(n-1)}2$.

Finally the \ldet of $P_{\rm sq}$ includes the vertical bias $\sqrt{\lambda}$ for each of the $n(n-1)$ vertical dimers in the W and E corners. Dividing by $\lambda^{n(n-1)/2}$ yields the required partition function. 

We note that because $P_{\rm sq}$ contains many zeros, the condensation algorithm will output undeterminate ratios at intermediate steps. It may be regularized by replacing all zeros in $P_{\rm sq}$ by a formal variable $t$. The above argument (no yellow face can be adjacent to two dimers) ensures that the \ldet of the new $P_{\rm sq}$ is a polynomial in $t$. Its limit for $t$ going to zero yields the correct result. \cqfd

\bigskip
More examples of this kind may be given, including rectangular subgraphs. For instance, taking $Q=1$ and for $P$ the tridiagonal matrix of size $n+1$ with ones on the three main diagonals, we obtain the Fibonacci polynomials\footnote{The Fibonacci polynomials satisfy $F_{n+1}(x) = x F_n(x) + F_{n-1}(x)$ with $F_1(x)=1$ and $F_2=x$.}, $\detl P = F_{2n+1}(\sqrt{\lambda})$, namely the partition function for biased perfect matchings of a $2 \times 2n$ rectangular graph. 
The following example concerns the holey Aztec diamond.

\smallskip
Let us denote by $p_i(n)$, for $i=0,1,2$, the fractions  of perfect matchings of the Aztec graph of order $n$ which have $i$ dimers adjacent to the central face. In order to compute these fractions, we attach a weight 1 to all faces of the Aztec graph, except the central face which is assigned a weight $t$. The central face belongs to a size $n+1$ matrix $P_t$ if $n$ is even or to a size $n$ matrix $Q_t$ if $n$ is odd, where both $P_t$ and $Q_t$ are all-ones matrices with $t$ as central entry. The corresponding partition function $T_n[t]$ takes the form (recall that $T_n=2^{n(n+1)/2}$ is the partition function for $t=1$)
\be
T_n[t] = \Big(p_0(n) \, t + p_1(n) + \frac{p_2(n)}{t}\Big) \, T_n,
\ee
and, from Theorem \ref{thm1}, can be computed as $T_n[t] = \det_1 P_t$ if $n$ is even, or $T_n[t] = 2^n  \det_1 Q^{-1}_t = 2^n \det_1 Q_{t^{-1}}$ if $n$ is odd.

In addition to the fact that the three numbers $p_i(n)$ sum up to 1, they satisfy an additional non-trivial identity, derived by Propp \cite{Pr03} and generalized by Kuo \cite{Ku04}. For any face of a planar graph bordered by four edges, it relates the fraction of perfect matchings for which the face is adjacent to two dimers and the fractions of those for which a dimer is on one of the four edges. In the present case, the identity reads
\be
4 p_2(n) = \big[p_1(n) + 2 p_2(n)\big]^2.
\label{proppid}
\ee
Expressing $p_0 = (1 - \sqrt{p_2})^2$ and $p_1 = 2\sqrt{p_2} ( 1- \sqrt{p_2})$ in terms of $p_2$, we obtain 
\be
L_n(t) \equiv \frac{T_n[t]}{T_n} = \frac 1t \, \Big[\sqrt{p_2(n)} + \big(1-\sqrt{p_2(n)}\big)\,t\Big]^2.
\label{square}
\ee

Let us now fix $n$ to be even. Since the matrix $P_t$ used for $n$ even is identical to the matrix $Q_t$ used for $n+1$, we obtain, using $T_{n+1}=2^{n+1} \, T_n$,
\be
L_n(t) = \frac{\textstyle \det_1 P_t}{T_n} = \frac{\textstyle \det_1 Q_t}{T_n} = 2^{-(n+1)} \, \frac{T_{n+1}[t^{-1}]}{T_n} = \frac{T_{n+1}[t^{-1}]}{T_{n+1}} = L_{n+1}(t^{-1}).
\ee
This provides a very simple proof that\footnote{This relation also directly follows by using the shuffling algorithm to go from order $n$ to order $n+1$.}
\be
p_i(n+1) = p_{2-i}(n), \qquad \text{for all $n$ even}.
\label{even}
\ee
It is therefore sufficient to focus on the $n$ even. Let us write
\be
L_n(t) = \frac 1{4t} \Big[(1-\alpha_n) + (1 + \alpha_n)t\Big]^2, \qquad 1-\alpha_n = 2 \sqrt{p_2(n)}.
\label{Ln}
\ee 
Explicit numerical calculations of 1-determinants led us to formulate a conjecture for the numbers $\alpha_n$, which we subsequently proved.

\begin{coro}
The numbers $\alpha_n$ in the sequence given in (\ref{Ln}) are given by $2^{-n} \, {n/2 \choose n/4}^2$ for $n=0 \bmod 4$, and by $0$ for $n=2 \bmod 4$; for $n$ odd, they are equal to $\alpha_n = -\alpha_{n-1}$.\\
It follows that the three fractions $p_i(n)$, for $n$ even, are given by
\be
(p_0,p_1,p_2) = \begin{cases} \displaystyle
\frac 14 \left[1 + 2^{-n} \, {\frac n2 \choose \frac n4}^{\!2}\right]^2, \, \frac 12 \left[1 - 2^{-2n} \, {\frac n2 \choose \frac n4}^{\!4}\right], \, \frac 14 \left[1 - 2^{-n} \, {\frac n2 \choose \frac n4}^{\!2}\right]^2, & \text{for } n=0 \bmod 4, \\
\noalign{\medskip}
\frac 14, \, \frac 12, \, \frac 14, & \text{for } n=2 \bmod 4.
\end{cases}
\label{p012}
\ee
For $n$ odd, the fractions are related to the previous ones by $p_i(n) = p_{2-i}(n-1)$.\\
From the Robbins-Rumsey formula (\ref{RR}), the numbers $2^{n(n+1)/2}p_i(n)$ for $n$ even are the 2-enumeration of the alternating sign matrices of size $n+1$ having their central entry equal to $1-i$, and similarly, $2^{n(n-1)/2}p_i(n)$ for $n$ odd yields the 2-enumeration of the alternating sign matrices of size $n$ having their central entry equal to $i-1$. 
\end{coro}

\noindent 
{\it Proof.} Ciucu has provided a proof that $p_2(n) = \frac 14$ when $n=2,3 \bmod 4$ \cite{Ci97}, and also mentions a proof by Propp for all values of $n$, which, as far as we know, has remained unpublished. Here we present a unified proof based on the trivariate generating function for one minus the average number of dimers adjacent to the face $(i,j)$ in the Aztec graph of order $n$, that is, for the numbers $p_0(i,j;n) - p_2(i,j;n)$. 

By using the octahedral recurrence, they satisfy a Laplacian-like linear recurrence which allows one to determine the generating function. This function has been computed in \cite{DFSG14} to which we refer for the details. In our notations, it reads
\be
G(x,y,z) = \sum_{n = 0}^\infty \sum_{i,j = -n}^n \big[p_0(i,j;n) - p_2(i,j;n)\big] x^i \, y^j \, z^n = \frac{1-z}{1 + z^2 - \frac z2(x + x^{-1} + y + y^{-1})}.
\ee
For $i=j=0$, the coefficients reduce to $p_0(n) - p_2(n)$. With the identity (\ref{proppid}) and the relation $p_0 + p_1 + p_2 = 1$, a proof that $p_0 - p_2 = 0$ or $ \pm 2^{-n} \, {n/2 \choose n/4}^2$ is enough to prove (\ref{p012}).

Let us first omit the term $-z$ in the numerator of $G$. The $n$th $z$-derivative at $z=0$ yields (using e.g. Fa\`a di Bruno's formula for the multiple derivative of the composition of two functions),
\bea
p_0(n) - p_2(n) \!\egal \!\frac 1{1 + z^2 - \frac z2 (\ldots)}\Big|_{x^0,y^0,z^n} = 2^{-n} \sum_{r=0}^{\lfloor \frac n2 \rfloor} (-4)^r \, {n-r \choose r} \big(x + x^{-1} + y + y^{-1}\big)^{n-2r}\Big|_{x^0,y^0} \nonumber\\
&& \hspace{-25mm} = \: 2^{-n} \sum_{r=0}^{\frac n2} \: (-4)^r \, {n-r \choose r} \, {n-2r \choose \frac n2 -r}^2 \, \delta_{n,{\rm even}} = (-1)^{\frac n2} \: \sum_{k=0}^{\frac n2} \: \big({\textstyle -\frac 14}\big)^k \, {\frac n2 + k \choose 2k} \, {2k \choose k}^2 \, \delta_{n,{\rm even}}.
\label{neven}
\eea
This proves that the function $\big[1 + z^2 - \frac z2(x + x^{-1} + y + y^{-1})\big]^{-1}$ has only even powers of $z$ upon taking the constant terms in $x,y$. The term $-z$ in the numerator of $G(x,y,z)$ then produces the odd terms and readily implies $p_0(n) - p_2(n) = -\big[p_0(n-1) - p_2(n-1)\big]$ for $n$ odd. The remaining sum in (\ref{neven}) for $n$ even is, up to a sign, a specialization of the following family of polynomials,
\be
f_m(x) = \sum_{k=0}^m \; {m + k \choose 2k} \, {2k \choose k}^2 \, x^k.
\ee
These have been shown \cite{Su12} to satisfy the identity $f_m\big(y(1+y)\big) = \big[P_m(2y+1)\big]^2$ in terms of the Legendre polynomials $P_m$. Substituting $m$ for $\frac n2$ and taking $x=-\frac 14, \, y=-\frac12$, we obtain, for $n$ even,
\be
p_0(n) - p_2(n) = (-1)^{\frac n2} \: f_{\frac n2}\big(\textstyle \!\!-\!\frac 14 \big) = (-1)^{\frac n2} \: \big[P_{\frac n2}(0)\big]^2 = \begin{cases} 
2^{-n} \, \displaystyle {\frac n2 \choose \frac n4}^2 & \text{if $n = 0 \bmod 4$,} \\
0 & \text{if $n = 2 \bmod 4$}.
\end{cases}  
\ee
This concludes the proof. \cqfd


\vskip 0.5truecm
\noindent
{\bf 3.4 General weights: the generalized Robbins-Rumsey formula}
\addcontentsline{toc}{subsubsection}{3.4 General weights: the generalized Robbins-Rumsey formula}

\medskip
\noindent
Theorem \ref{thm1} readily implies that to any perfect matching of the Aztec diamond of order $n$, one can associate two alternating sign matrices, of size $n$ and $n+1$ \cite{EKLP92}. 

\begin{figure}[t]
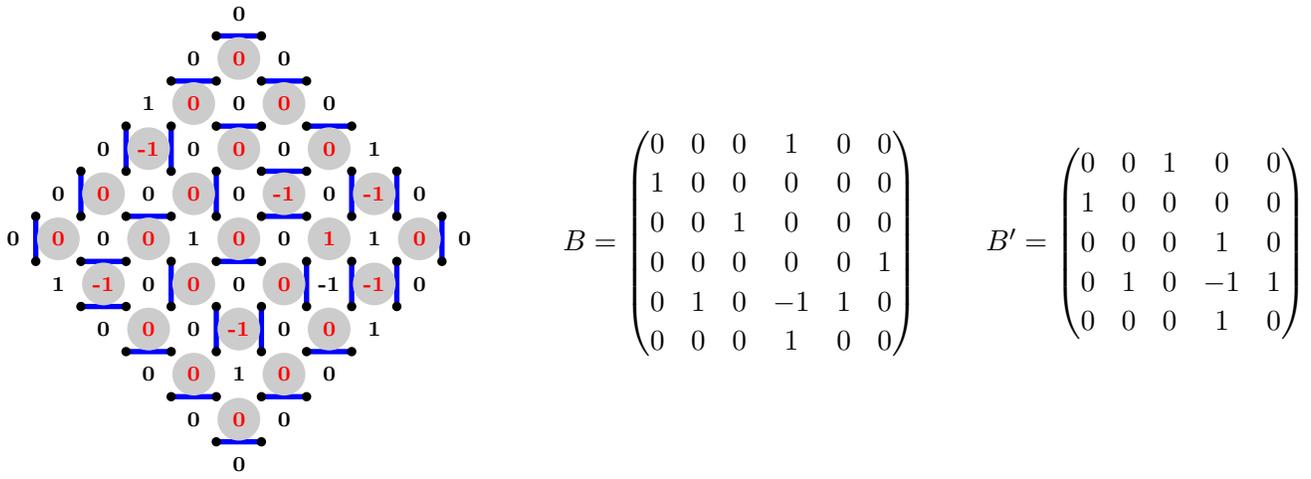

\begin{equation*}
\pspicture(0,-3.5)(6,1.5)
\psset{unit=.6cm}
\rput(-2.5,-7){
\rput(5,10){\psline[linewidth=2.0pt,linecolor=blue](0.5,0.5)(1.5,0.5)}
\rput(4,9){\psline[linewidth=2.0pt,linecolor=blue](0.5,0.5)(1.5,0.5)}
\rput(6,9){\psline[linewidth=2.0pt,linecolor=blue](0.5,0.5)(1.5,0.5)}
\rput(4,7){\psline[linewidth=2.0pt,linecolor=blue](0.5,0.5)(0.5,1.5)}
\rput(5,8){\psline[linewidth=2.0pt,linecolor=blue](0.5,0.5)(1.5,0.5)}
\rput(7,8){\psline[linewidth=2.0pt,linecolor=blue](0.5,0.5)(1.5,0.5)}
\rput(2,6){\psline[linewidth=2.0pt,linecolor=blue](0.5,0.5)(0.5,1.5)}
\rput(1,5){\psline[linewidth=2.0pt,linecolor=blue](0.5,0.5)(0.5,1.5)}
\rput(10,5){\psline[linewidth=2.0pt,linecolor=blue](0.5,0.5)(0.5,1.5)}
\rput(9,6){\psline[linewidth=2.0pt,linecolor=blue](0.5,0.5)(0.5,1.5)}
\rput(8,6){\psline[linewidth=2.0pt,linecolor=blue](0.5,0.5)(0.5,1.5)}
\rput(6,7){\psline[linewidth=2.0pt,linecolor=blue](0.5,0.5)(1.5,0.5)}
\rput(5,6){\psline[linewidth=2.0pt,linecolor=blue](0.5,0.5)(0.5,1.5)}
\rput(3,7){\psline[linewidth=2.0pt,linecolor=blue](0.5,0.5)(0.5,1.5)}
\rput(9,4){\psline[linewidth=2.0pt,linecolor=blue](0.5,0.5)(0.5,1.5)}
\rput(7,4){\psline[linewidth=2.0pt,linecolor=blue](0.5,0.5)(0.5,1.5)}
\rput(8,4){\psline[linewidth=2.0pt,linecolor=blue](0.5,0.5)(0.5,1.5)}
\rput(6,6){\psline[linewidth=2.0pt,linecolor=blue](0.5,0.5)(1.5,0.5)}
\rput(5,1){\psline[linewidth=2.0pt,linecolor=blue](0.5,0.5)(1.5,0.5)}
\rput(6,2){\psline[linewidth=2.0pt,linecolor=blue](0.5,0.5)(1.5,0.5)}
\rput(7,3){\psline[linewidth=2.0pt,linecolor=blue](0.5,0.5)(1.5,0.5)}
\rput(6,3){\psline[linewidth=2.0pt,linecolor=blue](0.5,0.5)(0.5,1.5)}
\rput(5,5){\psline[linewidth=2.0pt,linecolor=blue](0.5,0.5)(1.5,0.5)}
\rput(2,5){\psline[linewidth=2.0pt,linecolor=blue](0.5,0.5)(1.5,0.5)}
\rput(2,4){\psline[linewidth=2.0pt,linecolor=blue](0.5,0.5)(1.5,0.5)}
\rput(3,6){\psline[linewidth=2.0pt,linecolor=blue](0.5,0.5)(1.5,0.5)}
\rput(4,4){\psline[linewidth=2.0pt,linecolor=blue](0.5,0.5)(0.5,1.5)}
\rput(3,3){\psline[linewidth=2.0pt,linecolor=blue](0.5,0.5)(1.5,0.5)}
\rput(4,2){\psline[linewidth=2.0pt,linecolor=blue](0.5,0.5)(1.5,0.5)}
\rput(5,3){\psline[linewidth=2.0pt,linecolor=blue](0.5,0.5)(0.5,1.5)}
\multido{\nt=5+1}{2}{\rput(1,\nt){\pscircle[linecolor=black,fillstyle=solid,fillcolor=black](0.5,0.5){0.08}}}
\multido{\nt=4+1}{4}{\rput(2,\nt){\pscircle[linecolor=black,fillstyle=solid,fillcolor=black](0.5,0.5){0.08}}}
\multido{\nt=3+1}{6}{\rput(3,\nt){\pscircle[linecolor=black,fillstyle=solid,fillcolor=black](0.5,0.5){0.08}}}
\multido{\nt=2+1}{8}{\rput(4,\nt){\pscircle[linecolor=black,fillstyle=solid,fillcolor=black](0.5,0.5){0.08}}}
\multido{\nt=1+1}{10}{\rput(5,\nt){\pscircle[linecolor=black,fillstyle=solid,fillcolor=black](0.5,0.5){0.08}}}
\multido{\nt=1+1}{10}{\rput(6,\nt){\pscircle[linecolor=black,fillstyle=solid,fillcolor=black](0.5,0.5){0.08}}}
\multido{\nt=2+1}{8}{\rput(7,\nt){\pscircle[linecolor=black,fillstyle=solid,fillcolor=black](0.5,0.5){0.08}}}
\multido{\nt=3+1}{6}{\rput(8,\nt){\pscircle[linecolor=black,fillstyle=solid,fillcolor=black](0.5,0.5){0.08}}}
\multido{\nt=4+1}{4}{\rput(9,\nt){\pscircle[linecolor=black,fillstyle=solid,fillcolor=black](0.5,0.5){0.08}}}
\multido{\nt=5+1}{2}{\rput(10,\nt){\pscircle[linecolor=black,fillstyle=solid,fillcolor=black](0.5,0.5){0.08}}}
\rput(1,6){\scriptsize \bf 0}
\rput(2,7){\scriptsize \bf 0}
\rput(3,8){\scriptsize \bf 0}
\rput(4,9){\scriptsize \bf 1}
\rput(5,10){\scriptsize \bf 0}
\rput(6,11){\scriptsize \bf 0}
\rput(2,6){\pscircle[linecolor=mygray,fillstyle=solid,fillcolor=mygray](0,0){.45}}
\rput(3,7){\pscircle[linecolor=mygray,fillstyle=solid,fillcolor=mygray](0,0){.45}}
\rput(4,8){\pscircle[linecolor=mygray,fillstyle=solid,fillcolor=mygray](0,0){.45}}
\rput(5,9){\pscircle[linecolor=mygray,fillstyle=solid,fillcolor=mygray](0,0){.45}}
\rput(6,10){\pscircle[linecolor=mygray,fillstyle=solid,fillcolor=mygray](0,0){.45}}
\rput(2,6){\red \scriptsize \bf 0}
\rput(3,7){\red \scriptsize \bf 0}
\rput(4,8){\red \scriptsize \bf -1}
\rput(5,9){\red \scriptsize \bf 0}
\rput(6,10){\red \scriptsize \bf 0}
\rput(2,5){\scriptsize \bf 1}
\rput(3,6){\scriptsize \bf 0}
\rput(4,7){\scriptsize \bf 0}
\rput(5,8){\scriptsize \bf 0}
\rput(6,9){\scriptsize \bf 0}
\rput(7,10){\scriptsize \bf 0}
\rput(3,5){\pscircle[linecolor=mygray,fillstyle=solid,fillcolor=mygray](0,0){.45}}
\rput(4,6){\pscircle[linecolor=mygray,fillstyle=solid,fillcolor=mygray](0,0){.45}}
\rput(5,7){\pscircle[linecolor=mygray,fillstyle=solid,fillcolor=mygray](0,0){.45}}
\rput(6,8){\pscircle[linecolor=mygray,fillstyle=solid,fillcolor=mygray](0,0){.45}}
\rput(7,9){\pscircle[linecolor=mygray,fillstyle=solid,fillcolor=mygray](0,0){.45}}
\rput(3,5){\red \scriptsize \bf -1}
\rput(4,6){\red \scriptsize \bf 0}
\rput(5,7){\red \scriptsize \bf 0}
\rput(6,8){\red \scriptsize \bf 0}
\rput(7,9){\red \scriptsize \bf 0}
\rput(3,4){\scriptsize \bf 0}
\rput(4,5){\scriptsize \bf 0}
\rput(5,6){\scriptsize \bf 1}
\rput(6,7){\scriptsize \bf 0}
\rput(7,8){\scriptsize \bf 0}
\rput(8,9){\scriptsize \bf 0}
\rput(4,4){\pscircle[linecolor=mygray,fillstyle=solid,fillcolor=mygray](0,0){.45}}
\rput(5,5){\pscircle[linecolor=mygray,fillstyle=solid,fillcolor=mygray](0,0){.45}}
\rput(6,6){\pscircle[linecolor=mygray,fillstyle=solid,fillcolor=mygray](0,0){.45}}
\rput(7,7){\pscircle[linecolor=mygray,fillstyle=solid,fillcolor=mygray](0,0){.45}}
\rput(8,8){\pscircle[linecolor=mygray,fillstyle=solid,fillcolor=mygray](0,0){.45}}
\rput(4,4){\red \scriptsize \bf 0}
\rput(5,5){\red \scriptsize \bf 0}
\rput(6,6){\red \scriptsize \bf 0}
\rput(7,7){\red \scriptsize \bf -1}
\rput(8,8){\red \scriptsize \bf 0}
\rput(4,3){\scriptsize \bf 0}
\rput(5,4){\scriptsize \bf 0}
\rput(6,5){\scriptsize \bf 0}
\rput(7,6){\scriptsize \bf 0}
\rput(8,7){\scriptsize \bf 0}
\rput(9,8){\scriptsize \bf 1}
\rput(5,3){\pscircle[linecolor=mygray,fillstyle=solid,fillcolor=mygray](0,0){.45}}
\rput(6,4){\pscircle[linecolor=mygray,fillstyle=solid,fillcolor=mygray](0,0){.45}}
\rput(7,5){\pscircle[linecolor=mygray,fillstyle=solid,fillcolor=mygray](0,0){.45}}
\rput(8,6){\pscircle[linecolor=mygray,fillstyle=solid,fillcolor=mygray](0,0){.45}}
\rput(9,7){\pscircle[linecolor=mygray,fillstyle=solid,fillcolor=mygray](0,0){.45}}
\rput(5,3){\red \scriptsize \bf 0}
\rput(6,4){\red \scriptsize \bf -1}
\rput(7,5){\red \scriptsize \bf 0}
\rput(8,6){\red \scriptsize \bf 1}
\rput(9,7){\red \scriptsize \bf -1}
\rput(5,2){\scriptsize \bf 0}
\rput(6,3){\scriptsize \bf 1}
\rput(7,4){\scriptsize \bf 0}
\rput(8,5){\scriptsize \bf -1}
\rput(9,6){\scriptsize \bf 1}
\rput(10,7){\scriptsize \bf 0}
\rput(6,2){\pscircle[linecolor=mygray,fillstyle=solid,fillcolor=mygray](0,0){.45}}
\rput(7,3){\pscircle[linecolor=mygray,fillstyle=solid,fillcolor=mygray](0,0){.45}}
\rput(8,4){\pscircle[linecolor=mygray,fillstyle=solid,fillcolor=mygray](0,0){.45}}
\rput(9,5){\pscircle[linecolor=mygray,fillstyle=solid,fillcolor=mygray](0,0){.45}}
\rput(10,6){\pscircle[linecolor=mygray,fillstyle=solid,fillcolor=mygray](0,0){.45}}
\rput(6,2){\red \scriptsize \bf 0}
\rput(7,3){\red \scriptsize \bf 0}
\rput(8,4){\red \scriptsize \bf 0}
\rput(9,5){\red \scriptsize \bf -1}
\rput(10,6){\red \scriptsize \bf 0}
\rput(6,1){\scriptsize \bf 0}
\rput(7,2){\scriptsize \bf 0}
\rput(8,3){\scriptsize \bf 0}
\rput(9,4){\scriptsize \bf 1}
\rput(10,5){\scriptsize \bf 0}
\rput(11,6){\scriptsize \bf 0}
}
\rput(14.5,-1.1){
$B = \begin{pmatrix}
0 & 0 & 0 & 1 & 0 & 0 \\
1 & 0 & 0 & 0 & 0 & 0 \\
0 & 0 & 1 & 0 & 0 & 0 \\
0 & 0 & 0 & 0 & 0 & 1 \\
0 & 1 & 0 & -1 & 1 & 0 \\
0 & 0 & 0 & 1 & 0 & 0
\end{pmatrix}$}
\rput(23.5,-1.1){
$B' = \begin{pmatrix}
0 & 0 & 1 & 0 & 0 \\
1 & 0 & 0 & 0 & 0 \\
0 & 0 & 0 & 1 & 0 \\
0 & 1 & 0 & -1 & 1 \\
0 & 0 & 0 & 1 & 0
\end{pmatrix}$}
\endpspicture
\end{equation*}
\caption{On the left are shown a perfect matching of the Aztec diamond of order 5 and, on each face, the value of 1 minus the number of dimers adjacent to that face. The associated alternating sign matrices $B$ and $B'$ contain respectively the numbers in black and the opposite of those in red in shaded circles.}
\label{fig5}
\end{figure}

In the case of a $(P,1)$ weighting, the \ldet formula implies that every perfect matching of an Aztec diamond of order $n$ can be associated with an alternating sign matrix $B$ of order $n+1$. Indeed, by comparing the weight $\prod_{i,j} \, p_{ij}^{1-N_{i,j}}$ of a perfect matching with the factor $P^{B} = \prod_{i,j} p_{ij}^{b_{ij}}$ in the Robbins-Rumsey formula, we obtain $b_{ij} = 1-N_{i,j}$ for all $p_{ij}$ faces (yellow faces in Figure~\ref{fig3}). An example is shown in Figure~\ref{fig5} in which the matrix $B$ is written in black. While it is manifest from this correspondence that the first and last rows and columns contain no $-1$ (and therefore a single $1$), it is not so clear that the matrix so obtained is actually sign alternating; by thinking about the ways the dimers must be arranged around a face with $N_{i,j}=0$ or 2, it is not too difficult to convince oneself that it is.  The correspondence is not bijective: a matrix $B$ is associated with $2^{N_-(B)}$ distinct matchings. Indeed an entry $b_{ij}=-1$ is attached to a face which is adjacent to two dimers; such a face can be flipped to produce a matching that is distinct but of equal weight (a flip is a rotation of two parallel dimers located around a face). With a vertical bias $\sqrt{\lambda}$, the two matchings related by a flip contribute a relative factor equal to 1 and $\lambda$, so that every entry $-1$ in $B$ contributes an overall factor $1+\lambda$; this gives a simple combinatorial explanation of the factor $(1+\lambda)^{N_-(B)}$ in the Robbins-Rumsey formula (\ref{RR}). As for the factor $\lambda^{P(B)}=\lambda^{{\rm Inv}(B)-N_-(B)}$ we note that ${\rm Inv}(B)$ is half the maximal number of vertical dimers among the $2^{N_-(B)}$ distinct matchings obtained from each other by flips at the faces where $b_{ij}=-1$. 

The \ldet formula for the $(1,Q)$ weighting leads to a similar correspondence. However because $T_n(1,Q|\lambda)$ is proportional to the \ldet of $Q^{-1}$, identifying the weight $\prod_{i,j} \, q_{ij}^{1-N_{i,j}}$ of a perfect matching with the factor $Q^{-B'}$ in the Robbins-Rumsey formula, leads to an alternating sign matrix $B'$ of order $n$ given by $b'_{ij} = N_{i,j}-1$ for all the $q_{ij}$ faces, see Figure~\ref{fig5}. The prefactor $(1+\lambda)^n$ and the term $(1+\lambda)^{N_-(B')}$ nicely combine to give
\be
T_n(1,Q|\lambda) = (1+\lambda)^n \displaystyle \sum_{B' \,\in\, {\rm ASM}_n}  \lambda^{P(B')} \: (1 + \lambda)^{N_-(B')} \; Q^{-B'} = \displaystyle \sum_{B' \,\in\, {\rm ASM}_n}  \lambda^{P(B')} \: (1 + \lambda)^{N_+(B')} \; Q^{(-B')}.
\ee
This preserves the above combinatorial interpretation of the power of $1+\lambda$, since the flippable faces are now associated to the entries $+1$ of $B'$, in number equal to $N_+(B')$.

Except if $B$ is a permutation matrix, neither $B$ nor $B'$ completely determines the perfect matching it is associated with, but the pair $(B,B')$ does; one of the proofs in \cite{EKLP92} for the number of domino tilings of Aztec diamonds is precisely based on this bijection (from what follows, the smaller matrix $B'$ alone never determines completely the perfect matching). Because the two matrices $B$ and $B'$ are defined from a single matching, they ought to be related in some way; indeed such a pair of alternating sign matrices has been called compatible in \cite{RR86}. An algebraic criterion for compatibility has been given in \cite{RR86}, in the form of inequalities (see below), but the notion of compatibility is easier to understand in terms of perfect matchings, since it merely reduces to the fact that the pair $(B,B')$ is unambiguously associated with a perfect matching in the way described above. The number of pairs of compatible matrices, indicated by $B \approx B'$, and their explicit construction when one of the two is given, is as follows.

As explained above, an arbitrary alternating sign matrix $B$ of size $n+1$ fixes the arrangement of dimers around the $p$-faces, except at the faces $p_{ij}$ adjacent to two dimers, where $b_{ij}=-1$. Any choice at each of these faces, two horizontal or two vertical dimers, fixes one of the $2^{N_-(B)}$ perfect matchings compatible with $B$, and produces that many different alternating sign matrices $B'$ by reading off the values $N_{ij}-1$ at the $q_{ij}$ faces. If $B$ is a permutation matrix, $N_-(B)=0$, there is a unique matrix $B'$ compatible with $B$ because $B$ alone determines a unique perfect matching.

In a similar way, an alternating sign matrix $B'$ of size $n$ completely determines the dimer arrangement around the $q$-faces, except when a face $q_{ij}$ is adjacent to two dimers, where $b'_{ij}=1$. This time, all possible choices at those faces yield a total of $2^{N_+(B')}$ different perfect matchings and so many matrices $B$. In conclusion, we have
\be
\#\,\text{compatible pairs }(B,B') = \begin{cases}
2^{N_-(B)} & {\rm for\ fixed\ }B,\\
2^{N_+(B')} & {\rm for\ fixed\ }B'.
\end{cases}
\ee

At this stage, the picture we have is the following. We have two (partial) face weightings for perfect matchings of the Aztec diamond of order $n$, defined by $(P,1)$ and $(1,Q)$ respectively. Upon the addition of a vertical bias $\sqrt{\lambda}$, the corresponding partition functions are given by $\lambda$-determinants, which, together, give rise to a bijection between the set of perfect matchings and pairs of compatible alternating sign matrices $(B,B')$, of size $n+1$ and $n$. Can one then write the partition function for a general weighting $(P,Q)$ in terms of pairs of compatible alternating sign matrices ?

To see this, let us come back to the expressions, for $n=0,1,2$, of the partition functions for general weightings, given earlier in (\ref{initPQ}) and (\ref{T2}),
\bea
&& \hspace{-1cm} T_0(p_{11},-|\lambda) = p_{11}, \qquad T_1\left(\SmallMatrix{p_{11} & p_{12} \\ p_{21} & p_{22}},q_{11}|\:\lambda\right) = \frac{p_{11}\,p_{22} + \lambda \, p_{12}\,p_{21}}{q_{11}}, \\
\noalign{\medskip}
&& \hspace{-1cm}T_2\left(\SmallMatrix{p_{11} & p_{12} & p_{13} \\ p_{21} & p_{22} & p_{23} \\ p_{31} & p_{32} & p_{33}},
\SmallMatrix{q_{11} & q_{12} \\ q_{21} & q_{22}}\Big|\: \lambda\right) \nonumber\\ 
\noalign{\medskip}
&& \hspace{-5mm} = \frac 1{p_{22}} \Big[\frac{p_{11}\,p_{22} + \lambda \, p_{12}\,p_{21}}{q_{11}} \cdot \frac{p_{22}\,p_{33} + \lambda \, p_{23}\,p_{32}}{q_{22}} + \lambda \, \frac{p_{21}\,p_{32} + \lambda \, p_{22}\,p_{31}}{q_{21}} \cdot \frac{p_{12}\,p_{23} + \lambda \, p_{13}\,p_{22}}{q_{12}}\Big].
\label{T2bis}
\eea

One may observe that $T_1$ and $T_2$ are obtained from $T_0$ by successive substitutions. Indeed $T_1$ is obtained from $T_0$ by the substitution
\be
p_{11} \longrightarrow \frac{p_{11}\,p_{22} + \lambda \, p_{12}\,p_{21}}{q_{11}}.
\ee
Then $T_2$ is obtained from $T_1$ by the further substitution
\be
q_{11} \longrightarrow p_{22} \qquad \text{and} \qquad p_{ij} \longrightarrow \frac{p_{ij}\,p_{i+1,j+1} + \lambda \, p_{i,j+1}\,p_{i+1,j}}{q_{ij}} \qquad \text{for }1 \le i,j \le 2.
\label{subs}
\ee
In the next step, $T_3$, given by

\bea
&& \hspace{-1cm} T_3\left(\SmallMatrix{p_{11} & p_{12} & p_{13} & p_{14} \\ p_{21} & p_{22} & p_{23} & p_{24} \\ p_{31} & p_{32} & p_{33} & p_{34} \\ p_{41} & p_{42} & p_{43} & p_{44}},
\SmallMatrix{q_{11} & q_{12} & q_{13} \\ q_{21} & q_{22} & q_{23} \\ q_{31} & q_{32} & q_{33}}\Big|\: \lambda\right) = \nonumber\\
&& \hspace{-5mm}\frac 1{T_1\left(\SmallMatrix{p_{22} & p_{23} \\ p_{32} & p_{33}},q_{22}|\:\lambda\right)} \left\{T_2\left(\SmallMatrix{p_{11} & p_{12} & p_{13} \\ p_{21} & p_{22} & p_{23} \\ p_{31} & p_{32} & p_{33}},
\SmallMatrix{q_{11} & q_{12} \\ q_{21} & q_{22}}\Big|\: \lambda\right) \cdot T_2\left(\SmallMatrix{p_{22} & p_{23} & p_{24} \\ p_{32} & p_{33} & p_{34} \\ p_{42} & p_{43} & p_{44}},
\SmallMatrix{q_{22} & q_{23} \\ q_{32} & q_{33}}\Big|\: \lambda\right) \right. \nonumber\\
\noalign{\medskip}
&& \left. \hspace{2cm} + \: \lambda \: T_2\left(\SmallMatrix{p_{21} & p_{22} & p_{23} \\ p_{31} & p_{32} & p_{33} \\ p_{41} & p_{42} & p_{43}},
\SmallMatrix{q_{21} & q_{22} \\ q_{31} & q_{32}}\Big|\: \lambda\right) \cdot T_2\left(\SmallMatrix{p_{12} & p_{13} & p_{14} \\ p_{22} & p_{23} & p_{24} \\ p_{32} & p_{33} & p_{34}},
\SmallMatrix{q_{12} & q_{13} \\ q_{22} & q_{23}}\Big|\: \lambda\right) \right\},
\eea
can itself be obtained from $T_2$ by similar substitutions, namely
\be
p_{22} \longrightarrow T_1\left(\SmallMatrix{p_{22} & p_{23} \\ p_{32} & p_{33}},q_{22}|\:\lambda\right) = \frac{p_{22}\,p_{33} + \lambda \, p_{23}\,p_{32}}{q_{22}},
\ee
and the replacement of each of the four ratios in (\ref{T2bis}) by $T_2$ functions, for instance,
\be
\frac{p_{11}\,p_{22} + \lambda \, p_{12}\,p_{21}}{q_{11}} \longrightarrow T_2\left(\SmallMatrix{p_{11} & p_{12} & p_{13} \\ p_{21} & p_{22} & p_{23} \\ p_{31} & p_{32} & p_{33}},
\SmallMatrix{q_{11} & q_{12} \\ q_{21} & q_{22}}\Big|\: \lambda\right).
\ee
This replacement of a $T_1$ by a $T_2$ function is precisely obtained by the substitutions mentioned in (\ref{subs}), and the same holds for the other three ratios.

This is the general pattern: the partition function $T_n(P,Q|\lambda)$ for a general weighting $(P,Q)$ and a vertical bias $\sqrt{\lambda}$ is the result of the repeated application to $T_0 = p_{11}$ of the substitution $S$ given by,
\be
S \; : \quad q_{ij} \longrightarrow p_{i+1,j+1} \qquad \text{and} \qquad p_{ij} \longrightarrow \frac{p_{ij}\,p_{i+1,j+1} + \lambda \, p_{i,j+1}\,p_{i+1,j}}{q_{ij}} \qquad \text{for } i,j \ge 1.
\label{subsgen}
\ee

The substitution law given by $S$ was precisely at the heart of \cite{RR86}, whose main purpose was to give a non-recursive expression of the $n$-th iterate of $S$. In the present notations, their Theorem 1 yields the following combinatorial result.

We define, following \cite{RR86}, the corner sum matrix $\bar A$ of a matrix $A=(a_{ij})_{1 \le i,j \le n}$ of size $n$, by 
\be
\bar a_{ij} = \sum_{k \le i}\sum_{\ell \le j} \: a_{k\ell}.
\label{abar}
\ee 
and note that $A$ can be fully recovered from $\bar A$ by
\be
a_{ij} = \bar a_{ij} - \bar a_{i,j-1} - \bar a_{i-1,j} + \bar a_{i-1,j-1}, \qquad \text{with }a_{0j} = a_{i0}=0.
\ee

\begin{theo} \label{thm2}
The partition function for perfect matchings of the order $n$ Aztec graph with respect to the general face weighting $(P,Q)$ and vertical bias $\sqrt{\lambda}$, is given recursively by $T_n(P,Q|\lambda) = S^n(p_{11})$ for the substitution $S$ defined in (\ref{subsgen}), and non-recursively by the two alternative expressions,
\begin{subequations}
\bea
T_n(P,Q|\lambda) \egal \sum_{B  \, \in  \,{\rm ASM}_{n+1}} \sum_{B' \, \in  \,{\rm ASM}_{n} \atop B' \approx B}\; \lambda^{P(B) + |\bar B'| - |\bar B'|_{\min}} \, P^B \, Q^{-B'}, \\
\egal \sum_{B'  \, \in  \,{\rm ASM}_{n}} \sum_{B \, \in  \,{\rm ASM}_{n+1} \atop B \approx B'}\; \lambda^{P(B') + |\bar B|_{\max} - |\bar B|} \, P^B \, Q^{-B'}.
\eea
\label{eqthm2}
\end{subequations}
$\!\!P(B)$ has been defined earlier, right after (\ref{RR}), and for any matrix $A$, $|\bar A|$ is the sum of all elements of $\bar A$ defined in (\ref{abar}). Finally $|\bar B'|_{\min} = \min_{B' \approx B} |\bar B'|$ and $|\bar B|_{\max} = \max_{B \approx B'} |\bar B|$.
\end{theo}

\noindent {\it Proof.} We refer the reader to the proof given in \cite{RR86}. Their Theorem 1 is formulated differently, with, inside the summation, the exponent of $\lambda$ given by $F(B) - F(B') = (n+1)^2 + |\bar B'| - |\bar B|$. Using the various relations derived in \cite{RR86}, their formula can however be recast in the above two forms, more suitable to recover the special cases $P=1$ and $Q=1$. \cqfd

\medskip
As the expressions (\ref{eqthm2}) involve not only the pair $(B,B')$ but also their corner sum matrices, we recall here some of the results of \cite{RR86}. 

For a given $B \in {\rm ASM}_{n+1}$, an alternating sign matrix $B'$ in ASM$_n$ is compatible with $B$ if and only if the following conditions hold,
\be
\max\,(\bar b_{ij},\bar b_{i+1,j+1}-1) \le \bar b'_{ij} \le \min\,(\bar b_{i,j+1},\bar b_{i+1,j}), \qquad \text{for all } 1 \le i,j \le n.
\label{ineq}
\ee
Moreover Lemma 2 of \cite{RR86} states that for every $i,j$ between 1 and $n$, the lower and upper bounds in the previous inequality are equal unless $b_{i+1,j+1}=-1$, in which case they differ by 1. It follows that for those $i,j$ such that $b_{i+1,j+1} \neq -1$, the entries $\bar b'_{i,j} = \max\,(\bar b_{ij},\bar b_{i+1,j+1}-1)$ are completely fixed by $\bar B$; for those $i,j$ such that $b_{i+1,j+1}=-1$, the entries $\bar b'_{ij}$ can be either $\max\,(\bar b_{ij},\bar b_{i+1,j+1}-1)$ or $\max\,(\bar b_{ij},\bar b_{i+1,j+1}-1)+1$. This yields, as anticipated, a total of $2^{N_-(B)}$ distinct matrices $\bar B'$ from which the $B'$ themselves can be computed. 

When $B'$ ranges over the set of matrices compatible with $B$, the difference $\Delta(B') \equiv|\bar B'| - |\bar B'|_{\min}$ takes the integer values from 0 to $N_-(B)$. There is a unique $B'$ with $\Delta(B')=0$, there are $N_-(B)$ matrices $B'$ with $\Delta(B')=1$, and in general exactly $N_-(B) \choose k$ matrices $B'$ with $\Delta(B')=k$. We therefore obtain
\be
\sum_{B' \, \in  \,{\rm ASM}_n, \, B' \approx B } \lambda^{\Delta(B')} = (1 + \lambda)^{N_-(B)},
\ee
and recover the expression $T_n(P,1) = \detl P$.

Similarly for a fixed $B' \in {\rm ASM}_n$, a matrix $B$ in ASM$_{n+1}$ is compatible with $B'$ if and only if its corner sum matrix $\bar B$ satisfies
\be
\max\,(\bar b'_{i,j-1},\bar b'_{i-1,j}) \le \bar b_{ij} \le \min\,(\bar b'_{i,j},\bar b'_{i-1,j-1}+1), \qquad \text{for all } 1 \le i,j \le n,
\label{ineq2}
\ee
with the convention $\bar b'_{i,0}=\bar b'_{0,j}=0$. From Lemma 4 of \cite{RR86}, the lower and upper bounds are equal if $b'_{i,j} \neq 1$, and differ by 1 if $b'_{ij}=1$, so that following the same arguments as above, one finds that there are $2^{N_+(B')}$ matrices $B$ which are compatible with $B'$. One also finds that $\Delta(B) \equiv |\bar B|_{\max} - |\bar B|$ ranges over the integer values form $0$ to $N_+(B')$, to obtain the following identity,
\be
\sum_{B \, \in  \,{\rm ASM}_{n+1}, \, B \approx B'} \lambda^{\Delta(B)} = (1 + \lambda)^{N_+(B')} = (1 + \lambda)^{n} \, (1 + \lambda)^{N_-(B')}.
\ee
It implies the expression found earlier for $T_n(1,Q|\lambda) = (1+\lambda)^{n} \, \detl Q^{-1}$.

As a final remark to close this section, we observe that with a general weighting $(P,Q)$, the vertical bias $\sqrt\lambda$ is redundant since the weighting $(P,Q)$ is complete in itself. This implies that the dependence in $\lambda$ may be absorbed into a redefinition of $P$ and $Q$, yielding an identity of the form $T_n(P,Q|\lambda) \sim T_n(P_\lambda,Q_\lambda|1)$ where the entries of $P_\lambda$ and $Q_\lambda$ are those of $P$ and $Q$ multiplied by appropriate powers of $\lambda$. Interestingly, this point of view allows us to re-derive the dependence in $\lambda$ in the generalized Robbins-Rumsey formula. Let us see how this works.

Because we want to absorb edge weights into face weights, it is more convenient to consider the following modified face weighting given by
\be
\widetilde w_F(M) = \prod_{(k,\ell) \in \widehat{\cal A}_n} x_{k,\ell}^{-N_{k,\ell}}.
\ee
It differs from the weighting used so far by the global factor $\prod_{k,\ell} x_{k,\ell} = \big(\prod_{i,j} \, p_{ij}\big) \big(\prod_{i,j} \, q_{ij}\big)$, which can be reinstated at the end. With respect to $\widetilde w_F(M)$, the edge which is adjacent to the $p_{ij}$ and $q_{i'j'}$ faces contributes a face weight equal to $(p_{ij} q_{i'j'})^{-1}$ times 1 if the edge is horizontal, or $\sqrt{\lambda}$ if it is vertical. 

\begin{figure}[t]
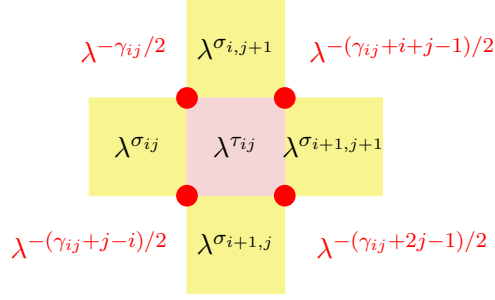

\begin{center}
\psset{unit=1.3cm}
\pspicture(0,-.7)(3,2)
\rput(0,0){\pspolygon[linewidth=0pt,linecolor=lightyellow,fillstyle=solid,fillcolor=lightyellow](0,0)(0,1)(1,1)(1,0)}
\rput(2,0){\pspolygon[linewidth=0pt,linecolor=lightyellow,fillstyle=solid,fillcolor=lightyellow](0,0)(0,1)(1,1)(1,0)}
\rput(1,-1){\pspolygon[linewidth=0pt,linecolor=lightyellow,fillstyle=solid,fillcolor=lightyellow](0,0)(0,1)(1,1)(1,0)}
\rput(1,1){\pspolygon[linewidth=0pt,linecolor=lightyellow,fillstyle=solid,fillcolor=lightyellow](0,0)(0,1)(1,1)(1,0)}
\rput(1,0){\pspolygon[linewidth=0pt,linecolor=lightred,fillstyle=solid,fillcolor=lightred](0,0)(0,1)(1,1)(1,0)}
\rput(1.5,0.5){$\lambda^{\tau_{ij}}$}
\rput(0.5,0.5){$\lambda^{\sigma_{ij}}$}
\rput(1.5,1.5){$\lambda^{\sigma_{i,j+1}}$}
\rput(1.5,-0.5){$\lambda^{\sigma_{i+1,j}}$}
\rput(2.5,0.5){$\lambda^{\sigma_{i+1,j+1}}$}
\rput(1,1){\pscircle[linecolor=red,fillstyle=solid,fillcolor=red](0,0){0.1}}
\rput(2,1){\pscircle[linecolor=red,fillstyle=solid,fillcolor=red](0,0){0.1}}
\rput(1,0){\pscircle[linecolor=red,fillstyle=solid,fillcolor=red](0,0){0.1}}
\rput(2,0){\pscircle[linecolor=red,fillstyle=solid,fillcolor=red](0,0){0.1}}
\rput(0.35,1.5){\red $\lambda^{-\gamma_{ij}/2}$}
\rput(3.2,1.5){\red $\lambda^{-(\gamma_{ij}+i+j-1)/2}$}
\rput(0,-0.5){\red $\lambda^{-(\gamma_{ij}+j-i)/2}$}
\rput(3.2,-0.5){\red $\lambda^{-(\gamma_{ij}+2j-1)/2}$}
\endpspicture
\end{center}
\caption{Illustration of the redefinition of the vertical bias into face weights and gauge transformations: the central red face is the face $q_{ij}$ and the four yellow surrounding faces are the faces $p_{ij}$ (left), $p_{i,j+1}$ (top), $p_{i+1,j}$ (bottom) and $p_{i+1,j+1}$ (right). The powers of $\lambda$ marked in the faces are those appearing  in the redefined matrices $P_\lambda$ and $Q_\lambda$. The gauge transformations at the four vertices around the face $q_{ij}$ have parameters written in red.}
\label{fig6}
\end{figure}

The goal is thus to transfer the edge weights $\sqrt{\lambda}$ to the face weights. We introduce the following matrices $P_\lambda$ and $Q_\lambda$, containing the new face weights,
\be
(P_\lambda)_{ij} = \lambda^{\sigma_{ij}}\,p_{ij}, \quad \textstyle \sigma_{ij}= i+j-1-ij, \qquad (Q_\lambda)_{ij} = \lambda^{\tau_{ij}}\,q_{ij}, \quad \textstyle \tau_{ij}=\frac{i+j-1}2 -ij. 
\ee
This must be accompanied by gauge transformations \cite{Be21}. A gauge transformation consists in assigning all the edges around a vertex $v$ a weight $w_v$. Because exactly one of these edges will belong to an arbitrary matching, this will simply multiply the partition function by an overall factor $w_v$. As gauge transformations can be performed independently on all vertices, the weights $w_v$ can be chosen so that, when combined with the powers of $\lambda$ in $P_\lambda$ and $Q_\lambda$, they yield the desired result. For that, we use the following factors for the gauge transformations on the four vertices around the face $q_{ij}$,
\be
\lambda^{-\gamma_{ij}/2}, \quad \lambda^{-(\gamma_{ij}+i+j-1)/2}, \quad \lambda^{-(\gamma_{ij}+j-i)/2}, \quad \lambda^{-(\gamma_{ij}+2j-1)/2}, \qquad \gamma_{ij}=(i-1)(2j-1),
\label{gauge}
\ee
in the order top-left, top-right, bottom-left and bottom-right, see Figure~\ref{fig6}. 

One can check that the powers of $\lambda$ attached to the faces together with the extra edge weights coming from the gauge transformations guarantee that all horizontal edges have weight 1 and all vertical edges have weight $\sqrt{\lambda}$ (in addition to the weight $(p_{ij} q_{i'j'})^{-1}$ coming from adjacent faces). For instance, for the left (vertical) edge of the face $q_{ij}$, we obtain the factor
\be
\lambda^{-\sigma_{ij}} \cdot \lambda^{-\tau_{ij}} \cdot \lambda^{-\gamma_{ij}/2} \cdot \lambda^{-(\gamma_{ij}+j-i)/2} = \sqrt{\lambda},
\ee
while for the upper (horizontal) edge of the same face, we have
\be
\lambda^{-\sigma_{i,j+1}} \cdot \lambda^{-\tau_{ij}} \cdot \lambda^{-\gamma_{ij}/2} \cdot \lambda^{-(\gamma_{ij}+i+j-1)/2} = 1.
\ee

The identity relating the two partition functions $T_n(P,Q|\lambda)$ and $T_n(P_\lambda,Q_\lambda|1)$ requires to include two factors. One is related to the gauge transformations and is the product of the factors at all sites; one finds that it is equal to $\lambda^{-n^3(n+1)/2}$. The other comes from the difference between the two measures $\widetilde w_F(M)$ and $w_F(M)$. The $\lambda$-dependence of the global factor $\prod_{k,\ell} x_{k,\ell}$ is given by $\prod_{i,j=1}^{n+1} \lambda^{\sigma_{ij}} \cdot \prod_{i,j=1}^{n} \lambda^{\tau_{ij}} = \lambda^{-n^2(n^2+n+1)/2}$; we divide by this factor to eliminate it. Altogether we obtain
\bea
T_n(P,Q|\lambda) \egal \lambda^{-n^3(n+1)/2} \, \lambda^{n^2(n^2+n+1)/2} \: T_n(P_\lambda,Q_\lambda|1) \nonumber\\
\noalign{\medskip}
&& \hspace{-1.2cm} = \: \lambda^{n^2/2} \: \sum_{B \approx B'} \lambda^{\sum_{i,j=1}^{n+1} \sigma_{ij} b_{ij} - \sum_{i,j=1}^n \tau_{ij} b'_{ij}} \; P^B \, Q^{-B'} = \sum_{B \approx B'} \lambda^{(n+1)^2 + |\bar B'| - |\bar B|} \; P^B \, Q^{-B'},
\label{genRR}
\eea
where we have used $\sum_{ij} ij \, b_{ij} = |\bar B|$, valid for any alternating sign matrix. The power of $\lambda$ in the sum is the one given in \cite{RR86}.

%


\vskip 0.5truecm
\noindent
{\bf 3.5 General weights: shuffling algorithm and condensation}
\addcontentsline{toc}{subsubsection}{3.5 General weights: shuffling algorithm and condensation}

\medskip
\noindent
Basic and local modifications of graphs can lead to efficient recursive algorithms which allow one to compute partition functions of weighted perfect matchings. Such algorithms have been described and used by many authors \cite{EKLP92,Ci98,Pr05,JRV06,Sp07}, under different names; in the context of Aztec graphs, we will refer to the shuffling algorithm. 

\begin{figure}[h]
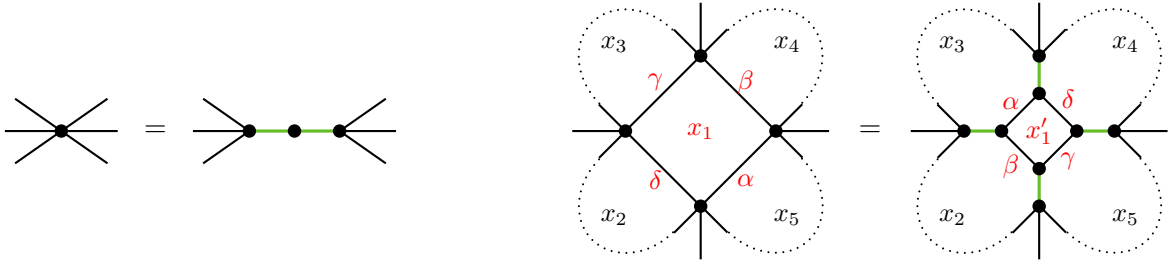

\begin{center}
\pspicture(4,-1)(10,1.7)
\psset{unit=.5cm}
\def\face{\psline[linewidth=0.8pt,linecolor=black](0,0)(-1,1) 
\psline[linewidth=0.8pt,linecolor=black](0,0)(0,1) 
\psline[linewidth=0.8pt,linecolor=black](2.82,0)(2.82,1) 
\psellipticarc[linewidth=0.8pt,linecolor=black,linestyle=dotted,dotsep=2pt](1.41,1)(1.41,1.7){0}{180}}
\psline[linewidth=0.8pt,linecolor=black](0,0)(1.5,0) 
\rput{35}(0,0){\psline[linewidth=0.8pt,linecolor=black](0,0)(1.5,0)}
\rput{-35}(0,0){\psline[linewidth=0.8pt,linecolor=black](0,0)(1.5,0)}
\rput{145}(0,0){\psline[linewidth=0.8pt,linecolor=black](0,0)(1.5,0)}
\rput{180}(0,0){\psline[linewidth=0.8pt,linecolor=black](0,0)(1.5,0)}
\rput{215}(0,0){\psline[linewidth=0.8pt,linecolor=black](0,0)(1.5,0)}
\pscircle[linecolor=black,fillstyle=solid,fillcolor=black](0,0){0.15}
\rput(2.5,0){$=$}
\rput(5,0){
\psline[linewidth=0.8pt,linecolor=black](2.4,0)(3.9,0)
\psline[linewidth=1.2pt,linecolor=mygreen](0,0)(1.2,0) 
\psline[linewidth=1.2pt,linecolor=mygreen](1.2,0)(2.4,0)  
\rput{35}(2.4,0){\psline[linewidth=0.8pt,linecolor=black](0,0)(1.5,0)}
\rput{-35}(2.4,0){\psline[linewidth=0.8pt,linecolor=black](0,0)(1.5,0)}
\rput{145}(0,0){\psline[linewidth=0.8pt,linecolor=black](0,0)(1.5,0)}
\rput{180}(0,0){\psline[linewidth=0.8pt,linecolor=black](0,0)(1.5,0)}
\rput{215}(0,0){\psline[linewidth=0.8pt,linecolor=black](0,0)(1.5,0)}
\pscircle[linecolor=black,fillstyle=solid,fillcolor=black](0,0){0.15}
\pscircle[linecolor=black,fillstyle=solid,fillcolor=black](1.2,0){0.15}
\pscircle[linecolor=black,fillstyle=solid,fillcolor=black](2.4,0){0.15}
}
\rput(15,0){
\pspolygon[linewidth=0.8pt,linecolor=black](0,0)(2,2)(4,0)(2,-2)(0,0)
\rput{-45}(2,2){\face}
\rput{45}(0,0){\face}
\rput{135}(2,-2){\face}
\rput{-135}(4,0){\face}
\pscircle[linecolor=black,fillstyle=solid,fillcolor=black](0,0){0.15}
\pscircle[linecolor=black,fillstyle=solid,fillcolor=black](2,2){0.15}
\pscircle[linecolor=black,fillstyle=solid,fillcolor=black](4,0){0.15}
\pscircle[linecolor=black,fillstyle=solid,fillcolor=black](2,-2){0.15}
\rput(.8,1.3){\red \small $\gamma$}
\rput(3.2,1.3){\red \small $\beta$}
\rput(3.2,-1.3){\red \small $\alpha$}
\rput(.8,-1.3){\red \small $\delta$}
\rput(2,0){\red \small $x_1$}
\rput(-0.3,-2.3){\small $x_2$}
\rput(-0.3,2.3){\small $x_3$}
\rput(4.3,2.3){\small $x_4$}
\rput(4.3,-2.3){\small $x_5$}
\rput(6.5,0){$=$}
\rput(9,0){
\rput{-45}(2,2){\face}
\rput{45}(0,0){\face}
\rput{135}(2,-2){\face}
\rput{-135}(4,0){\face}
\rput(0,0){\urban}
\rput(1.25,0.75){\red \small $\alpha$}
\rput(2.77,0.85){\red \small $\delta$}
\rput(2.77,-0.8){\red \small $\gamma$}
\rput(1.25,-0.95){\red \small $\beta$}
\rput(2,0){\red \small $x'_1$}
\rput(-0.3,-2.3){\small $x_2$}
\rput(-0.3,2.3){\small $x_3$}
\rput(4.3,2.3){\small $x_4$}
\rput(4.3,-2.3){\small $x_5$}
}}
\endpspicture
\end{center}
\caption{Illustration of the two local graph modifications, the vertex splitting on the left, the urban renewal on the right. The dotted lines indicate that some of the faces adjacent to the 4-cycle may be boundary (open) faces.}
\label{fig7}
\end{figure}

The two basic results underlying the shuffling algorithm are related to two local graph modifications called the vertex splitting and the urban renewal; they are illustrated in Figure~\ref{fig7}. Proofs can be found in \cite{Pr05,Sp07}.

In the vertex splitting, any multivalent vertex can be expanded into a linear chain with three vertices, by inserting two new vertices and two new edges, in green. Note that the inverse transformation, the vertex merging, may also be applied. The urban renewal replaces each vertex of a 4-cycle by two vertices and one new edge, also in green. In both cases, the partition functions for the perfect matchings of the original graph $G$ and of the modified graph $G'$ are equal provided (1) the dimers occupying the green edges of $G'$ have weight 1, whatever the edge and face weights in $G$ are, and (2) the dimers occupying the black edges of $G'$ are weighted in the same way as in $G$, except, in the case of the urban renewal, that the edge and face weights of the 4-cell must be modified (they are marked in red, before and after the change). The opposite edge weights are exchanged by pairs, $\alpha \leftrightarrow \gamma$ and $\beta \leftrightarrow \delta$, and the face weight of the 4-cycle is changed from $x_1$ to $x_1'$, given in terms of the weights $x_2, \ldots,x_5$ of the four adjacent faces, by,
\be
x_1' = \frac{\alpha\gamma \, x_2 x_4 + \beta\delta \, x_3 x_5}{x_1}.
\label{urb}
\ee

\begin{figure}[t]
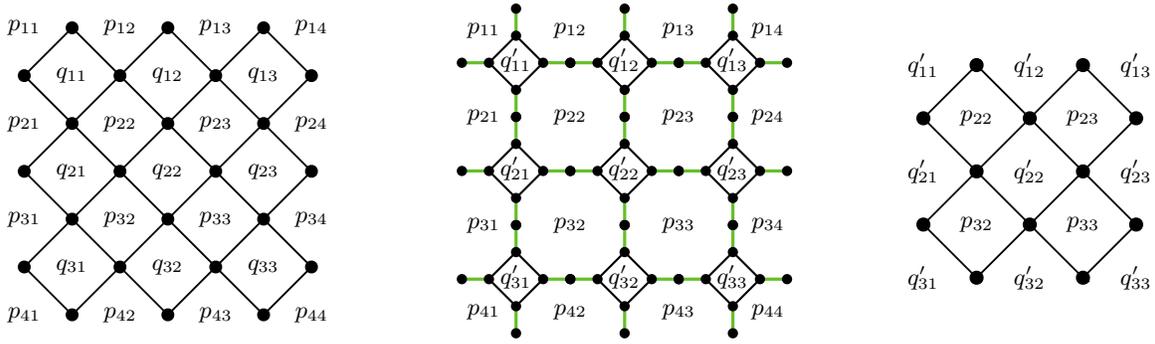

\pspicture(0,2.2)(8,6.5)
\psset{unit=.9cm}
\rput{-45}(-2,3.5){
\multido{\nt=4+1}{2}{\rput(0,\nt){\pscircle[linecolor=black,fillstyle=solid,fillcolor=black](0.5,0.5){0.08}}}
\multido{\nt=3+1}{4}{\rput(1,\nt){\pscircle[linecolor=black,fillstyle=solid,fillcolor=black](0.5,0.5){0.08}}}
\multido{\nt=2+1}{6}{\rput(2,\nt){\pscircle[linecolor=black,fillstyle=solid,fillcolor=black](0.5,0.5){0.08}}}
\multido{\nt=2+1}{6}{\rput(3,\nt){\pscircle[linecolor=black,fillstyle=solid,fillcolor=black](0.5,0.5){0.08}}}
\multido{\nt=3+1}{4}{\rput(4,\nt){\pscircle[linecolor=black,fillstyle=solid,fillcolor=black](0.5,0.5){0.08}}}
\multido{\nt=4+1}{2}{\rput(5,\nt){\pscircle[linecolor=black,fillstyle=solid,fillcolor=black](0.5,0.5){0.08}}}
\psline[linewidth=.7pt,linecolor=black](2.5,7.5)(3.5,7.5)
\psline[linewidth=.7pt,linecolor=black](1.5,6.5)(4.5,6.5)
\psline[linewidth=.7pt,linecolor=black](0.5,5.5)(5.5,5.5)
\psline[linewidth=.7pt,linecolor=black](0.5,4.5)(5.5,4.5)
\psline[linewidth=.7pt,linecolor=black](1.5,3.5)(4.5,3.5)
\psline[linewidth=.7pt,linecolor=black](2.5,2.5)(3.5,2.5)
\psline[linewidth=.7pt,linecolor=black](0.5,5.5)(0.5,4.5)
\psline[linewidth=.7pt,linecolor=black](1.5,6.5)(1.5,3.5)
\psline[linewidth=.7pt,linecolor=black](2.5,7.5)(2.5,2.5)
\psline[linewidth=.7pt,linecolor=black](3.5,7.5)(3.5,2.5)
\psline[linewidth=.7pt,linecolor=black](4.5,6.5)(4.5,3.5)
\psline[linewidth=.7pt,linecolor=black](5.5,5.5)(5.5,4.5)
\rput{45}(3,5){\footnotesize $q_{22}$}
\rput{45}(4,5){\footnotesize $p_{33}$}
\rput{45}(5,5){\footnotesize $q_{33}$}
\rput{45}(6,5){\footnotesize $p_{44}$}
\rput{45}(2,5){\footnotesize $p_{22}$}
\rput{45}(1,5){\footnotesize $q_{11}$}
\rput{45}(0,5){\footnotesize $p_{11}$}
\rput{45}(1,6){\footnotesize $p_{12}$}
\rput{45}(2,6){\footnotesize $q_{12}$}
\rput{45}(3,6){\footnotesize $p_{23}$}
\rput{45}(4,6){\footnotesize $q_{23}$}
\rput{45}(5,6){\footnotesize $p_{34}$}
\rput{45}(2,7){\footnotesize $p_{13}$}
\rput{45}(3,7){\footnotesize $q_{13}$}
\rput{45}(4,7){\footnotesize $p_{24}$}
\rput{45}(3,8){\footnotesize $p_{14}$}
\rput{45}(1,4){\footnotesize $p_{21}$}
\rput{45}(2,4){\footnotesize $q_{21}$}
\rput{45}(3,4){\footnotesize $p_{32}$}
\rput{45}(4,4){\footnotesize $q_{32}$}
\rput{45}(5,4){\footnotesize $p_{43}$}
\rput{45}(2,3){\footnotesize $p_{31}$}
\rput{45}(3,3){\footnotesize $q_{31}$}
\rput{45}(4,3){\footnotesize $p_{42}$}
\rput{45}(3,2){\footnotesize $p_{41}$}
}
\psset{unit=0.36cm}
\rput(20,8.3){
\multirput(0,0)(0,4){3}{
\multirput(0,0)(4,0){3}{\urban}}
\rput(2,8.1){\footnotesize $q'_{11}$}
\rput(6,8.1){\footnotesize $q'_{12}$}
\rput(10,8.1){\footnotesize $q'_{13}$}
\rput(2,4.1){\footnotesize $q'_{21}$}
\rput(6,4.1){\footnotesize $q'_{22}$}
\rput(10,4.1){\footnotesize $q'_{23}$}
\rput(2,0.1){\footnotesize $q'_{31}$}
\rput(6,0.1){\footnotesize $q'_{32}$}
\rput(10,0.1){\footnotesize $q'_{33}$}
\rput(0.8,9.2){\footnotesize $p_{11}$}
\rput(4,9.2){\footnotesize $p_{12}$}
\rput(8,9.2){\footnotesize $p_{13}$}
\rput(11.3,9.2){\footnotesize $p_{14}$}
\rput(0.8,6){\footnotesize $p_{21}$}
\rput(4,6){\footnotesize $p_{22}$}
\rput(8,6){\footnotesize $p_{23}$}
\rput(11.3,6){\footnotesize $p_{24}$}
\rput(0.8,2){\footnotesize $p_{31}$}
\rput(4,2){\footnotesize $p_{32}$}
\rput(8,2){\footnotesize $p_{33}$}
\rput(11.3,2){\footnotesize $p_{34}$}
\rput(0.8,-1.2){\footnotesize $p_{41}$}
\rput(4,-1.2){\footnotesize $p_{42}$}
\rput(8,-1.2){\footnotesize $p_{43}$}
\rput(11.3,-1.2){\footnotesize $p_{44}$}
}
\psset{unit=1cm}
\rput{-45}(9.8,2.3){
\multido{\nt=4+1}{2}{\rput(0,\nt){\pscircle[linecolor=black,fillstyle=solid,fillcolor=black](0.5,0.5){0.08}}}
\multido{\nt=3+1}{4}{\rput(1,\nt){\pscircle[linecolor=black,fillstyle=solid,fillcolor=black](0.5,0.5){0.08}}}
\multido{\nt=3+1}{4}{\rput(2,\nt){\pscircle[linecolor=black,fillstyle=solid,fillcolor=black](0.5,0.5){0.08}}}
\multido{\nt=4+1}{2}{\rput(3,\nt){\pscircle[linecolor=black,fillstyle=solid,fillcolor=black](0.5,0.5){0.08}}}
\psline[linewidth=.7pt,linecolor=black](1.5,6.5)(2.5,6.5)
\psline[linewidth=.7pt,linecolor=black](0.5,5.5)(3.5,5.5)
\psline[linewidth=.7pt,linecolor=black](0.5,4.5)(3.5,4.5)
\psline[linewidth=.7pt,linecolor=black](1.5,3.5)(2.5,3.5)
\psline[linewidth=.7pt,linecolor=black](0.5,5.5)(0.5,4.5)
\psline[linewidth=.7pt,linecolor=black](1.5,6.5)(1.5,3.5)
\psline[linewidth=.7pt,linecolor=black](2.5,6.5)(2.5,3.5)
\psline[linewidth=.7pt,linecolor=black](3.5,5.5)(3.5,4.5)
\rput{45}(3,5){\footnotesize $p_{33}$}
\rput{45}(4,5){\footnotesize $q'_{33}$}
\rput{45}(2,5){\footnotesize $q'_{22}$}
\rput{45}(1,5){\footnotesize $p_{22}$}
\rput{45}(0,5){\footnotesize $q'_{11}$}
\rput{45}(1,6){\footnotesize $q'_{12}$}
\rput{45}(2,6){\footnotesize $p_{23}$}
\rput{45}(3,6){\footnotesize $q'_{23}$}
\rput{45}(2,7){\footnotesize $q'_{13}$}
\rput{45}(1,4){\footnotesize $q'_{21}$}
\rput{45}(2,4){\footnotesize $p_{32}$}
\rput{45}(3,4){\footnotesize $q'_{32}$}
\rput{45}(2,3){\footnotesize $q'_{31}$}
}
\endpspicture
\caption{Illustration, on an order 3 Aztec graph, of the two steps involved in the shuffling algorithm.}
\label{fig8}
\end{figure}

The shuffling algorithm applied to Aztec graphs uses recursively the vertex splitting and the urban renewal, and establishes a recurrence relation between the partition functions at order $n$ and $n-1$. Starting from an Aztec graph order $n$, the algorithm outputs an Aztec graph of order $n-1$, with modified face weights, such that the partition functions of both graphs are equal. Let us recall how it works \cite{Pr05}; the full procedure is illustrated in Figure~\ref{fig8} for an Aztec graph of order 3, which we have rotated clockwise by 45 degrees for graphical convenience. 

Thus we start with an Aztec graph $\hat{\cal A}_n$ of order $n$, equipped with a general face weighting $(P,Q)$ and an additional weight $\sqrt{\lambda}$ attached to each vertical edge. 

The first step applies the urban renewal to all $q$-faces and outputs a modified graph $\hat{\cal A}_n^{\rm mod}$, as shown in the middle panel of Figure~\ref{fig8}. Referring to the right panel of Figure~\ref{fig7}, the edge weights around the $q_{ij}$-face are equal to $\alpha_{ij} = \gamma_{ij} = \sqrt\lambda$ and $\beta_{ij} = \delta_{ij} = 1$, so that they get simply transferred to the new $q'_{ij}$-face. The weights $p_{ij}$ attached to the $p$-faces do not change, whereas the new weights $q'_{ij}$, from (\ref{urb}), are given by
\be
q'_{ij} = \frac{p_{ij} \, p_{i+1,j+1} + \lambda \, p_{i,j+1} \, p_{i+1,j}}{q_{ij}}.
\ee
The partition function $T_n(P,Q|\lambda)$ on $\hat{\cal A}_n$ is equal to that for the perfect matchings on $\hat{\cal A}^{\rm mod}_n$ with the weights just described. Remember that the green edges have all weight 1 and are not affected by the weights of adjacent faces.

In the second step, the graph $\hat{\cal A}^{\rm mod}_n$ can be simplified. All pendant, peripheral green edges must necessarily be covered by dimers. Since all of them contribute a weight 1, these edges and their end nodes may simply be removed without changing the partition function. Then the linear chains with three vertices and two green edges, which separate the 4-cells with weights $q'_{ij}$, can be merged into a single vertex. After these two operations, one is left with an Aztec graph of order $n-1$, in which the $p$-faces have weights $(q'_{ij})_{1 \le i,j \le n}$ and the $q$-faces have weights $(p_{ij})_{2 \le i,j \le n}$; in addition the same vertical bias $\sqrt\lambda$ is obtained.

The shuffling algorithm as described above therefore implies the following identity on the partition functions,
\bea
T_n\big((p_{ij})_{1 \le i,j \le n+1},(q_{ij})_{1 \le i,j \le n}|\lambda\big) \egal T_{n-1}\big((q'_{ij})_{1 \le i,j \le n},(p_{i+1,j+1})_{1 \le i,j \le n-1}|\lambda\big) \nonumber\\
\noalign{\smallskip}
\egal S \, T_{n-1}\big((p_{ij})_{1 \le i,j \le n},(q_{ij})_{1 \le i,j \le n-1}|\lambda\big),
\eea
where $S$ is the substitution discussed in Section 3.4. Thus the shuffling algorithm gives the general Robbins-Rumsey formula a combinatorial content.

\smallskip
From the general discussion of Section 3.4, the partition function for a general face weighting $(P,Q)$ and vertical bias $\sqrt\lambda$ does not reduce to the computation of a $\lambda$-determinant. Let us however observe that it can be calculated by using the condensation algorithm (\ref{cond}). Instead of starting from $A_0=(1)_{1 \le i,j \le n+2}$ and $A_1=P=(p_{ij})_{1 \le i,j \le n+1}$ and running the algorithm to compute $\detl P$, one simply replaces the $n \times n$ central submatrix of $A_0$ by $Q$, and then runs the algorithm as before to produce the sequence $(A_k)$ and eventually obtain $A_{n+1} = T_n(P,Q|\lambda)$. 

Let us denote by Cond$_\lambda(A_0,A_1)$ the output of the condensation algorithm applied to initial matrices $A_0$ and $A_1$. If $A_0$ and $A_1$ have size $m+1$ and $m$ respectively, the algorithm produces a sequence of matrices $(A_k)$, recursively computed from (\ref{cond}). We set Cond$_\lambda(A_0,A_1) \equiv A_m$, so that Cond$_\lambda(1,P) = \detl P$. It is not difficult to see that the above partition functions are given by
\be
T_n(P,Q|\lambda) = {\rm Cond}_\lambda(\xbar Q,P),
\ee
where $\xbar Q$ is any $(n+2) \times (n+2)$ completion of $Q$ whose central $n \times n$ restriction is $Q$. 

The proof of the previous formula is straightforward. Initiating the condensation algorithm by setting $A_0=\xbar Q$ and $A_1=P$, we obtain, after one step, the matrix $A_2$ with entries $(A_2)_{ij} = q'_{ij}$. From the way the sequence $(A_k)$ is constructed, we have Cond$_\lambda(A_{k-1},A_k) = $ Cond$_\lambda(A_k,A_{k+1})$, and therefore we conclude 
\be
T_n(P,Q|\lambda) = {\rm Cond}_\lambda(A_0,A_1) = {\rm Cond}_\lambda(A_1,A_2) = T_{n-1}((q'_{ij}),(p_{i+1,j+1})|\lambda).
\ee

From this, one readily recovers the result quoted earlier in Theorem \ref{thm1} when $P=1$, namely
\be
T_n(1,Q|\lambda) = T_{n-1}\Big(\big(\frac{1+\lambda}{q_{ij}}\big),1 \Big|\lambda\Big) = \detl \displaystyle \Big(\big(\frac{1+\lambda}{q_{ij}}\big)\Big) = (1+\lambda)^n \detl (Q^{-1}).
\ee


\vskip .5truecm
\noindent
{\bf \large 4. Inhomogeneous $\lambda$-determinants}
\addcontentsline{toc}{subsection}{4. Inhomogeneous $\lambda$-determinants}
\setcounter{section}{4}
\setcounter{equation}{0}
\setcounter{theo}{0}

\medskip
\noindent
Apart for the vertical bias associated with the parameter $\lambda$, we have only considered face weights. Natural questions are whether there is a similar formalism for edge weights and whether face {\it and} edge weights can be combined. We will see that the answers to both questions are positive. In a sense, this should not be surprising since the face and edge most general weightings are separately complete, meaning that the edge weights can be fully absorbed in face weights or vice-versa. However in the case we call inhomogeneous bias, the edge weights enter through a further generalization of $\lambda$-determinants.

In addition to the vertical bias $\sqrt\lambda$, one can add a horizontal bias $\sqrt\mu$ to all horizontal dimers. This however brings nothing really new since, up to a global factor, the partition function will depend on the ratio $\lambda/\mu$. More interesting is to consider inhomogeneous bias, by which the vertical and horizontal bias depends on which row and column the dimer is located. In the notations of Section 2, this corresponds to take the weights $\alpha_{k,\ell},\gamma_{k,\ell}$ resp. $\beta_{k,\ell},\delta_{k,\ell}$ to depend on $\ell$ resp. $k$ only. This system of inhomogeneous bias is precisely the situation analyzed by Di Francesco \cite{DF13} (although it was not stated in these terms). Adapting his notations to the present framework, we denote by $\sqrt{\lambda_a}$ the vertical bias assigned to a vertical dimer in row $a$ (i.e. the common value of all $\alpha_{k,a}=\gamma_{k,a}$), and similarly by $\sqrt{\mu_b}$ the horizontal bias assigned to a horizontal dimer in column $b$ (common value of all $\beta_{b,\ell}=\delta_{b,\ell}$). For the Aztec graph of order $n$, the labels $a$ and $b$ take integer values from $-(n-1)$ to $n-1$. Let us also denote by $T_n(P,Q|\text{\boldmath$\lambda,\mu$})$ the partition function for perfect matchings of the Aztec graph of order $n$, with a general face weighting $(P,Q)$ and inhomogeneous bias, where $\text{\boldmath$\lambda$} = (\lambda_{-(n-1)},\ldots,\lambda_{n-1})$ and $\text{\boldmath$\mu$} = (\mu_{-(n-1)},\ldots,\mu_{n-1})$ are finite sequences of length $2n-1$. Because the numbers of vertical dimers in any row and of horizontal dimers in any column are even, the partition function is polynomial in the $\lambda_a$ and $\mu_b$.

In view of the general octahedral recurrence (\ref{octa}), the recurrence (\ref{octaPQ}) generalizes to
\bea
T_n(P,Q|\text{\boldmath$\lambda,\mu$}) \, T_{n-2}(P_\mathrm{C},Q_\mathrm{C}|\text{\boldmath$\lambda$}_{\mathrm{CC}},\text{\boldmath$\mu$}_{\mathrm{CC}}) \egal \mu_\ast \, T_{n-1}(P_{\mathrm{UL}},Q_{\mathrm{UL}}|\text{\boldmath$\lambda$}_\mathrm{C},\text{\boldmath$\mu$}_\mathrm{L}) \, T_{n-1}(P_{\mathrm{LR}},Q_{\mathrm{LR}}|\text{\boldmath$\lambda$}_\mathrm{C},\text{\boldmath$\mu$}_\mathrm{R}) \nonumber\\
\noalign{\smallskip}
&& \hspace{-2cm} + \: \lambda_\ast \, T_{n-1}(P_{\mathrm{LL}},Q_{\mathrm{LL}}|\text{\boldmath$\lambda$}_\mathrm{C},\text{\boldmath$\mu$}_\mathrm{C}) \, T_{n-1}(P_{\mathrm{UR}},Q_{\mathrm{UR}}|\text{\boldmath$\lambda$}_\mathrm{R},\text{\boldmath$\mu$}_\mathrm{C}),
\label{octaPQlamu}
\eea
where $\lambda_\ast$ and $\mu_\ast$ are the central entries of $\text{\boldmath$\lambda$}$ and $\text{\boldmath$\mu$}$ ($\lambda_0$ and $\mu_0$ in the previous equation). Also the left, central and right truncations of $\text{\boldmath$\lambda$}$, and similarly for $\text{\boldmath$\mu$}$, are defined by 
\be
\text{\boldmath$\lambda$}_\mathrm{L} = (\lambda_{-(n-1)},\ldots,\lambda_{n-3}), \quad \text{\boldmath$\lambda$}_\mathrm{C} = (\lambda_{-(n-2)},\ldots,\lambda_{n-2}), \quad \text{\boldmath$\lambda$}_\mathrm{R} = (\lambda_{-(n-3)},\ldots,\lambda_{n-1}),
\ee
and $\text{\boldmath$\lambda$}_{\mathrm{CC}}$ and $\text{\boldmath$\mu$}_{\mathrm{CC}}$ refer to a double central restriction. The previous recurrence allows one to compute recursively all partitions functions for $n \ge 2$ from the initial conditions,
\be
T_0(p_{11},-|-,-) = p_{11}, \qquad T_1\left(\SmallMatrix{p_{11} & p_{12} \\ p_{21} & p_{22}},q_{11}\big|(\lambda_0),(\mu_0)\right) = \frac{\mu_0 \, p_{11}\,p_{22} + \lambda_0 \, p_{12}\,p_{21}}{q_{11}}.
\label{initPQlamu}
\ee
For $n=2$, we obtain the following expression generalizing (\ref{T2}),
\bea
&& \hspace{-7mm} T_2(P,Q|(\lambda_{-1},\lambda_0,\lambda_1),(\mu_{-1},\mu_0,\mu_1)) = \frac 1{p_{22}} \Big[\mu_0 \, \frac{\mu_{-1} \, p_{11}\,p_{22} + \lambda_0 \, p_{12}\,p_{21}}{q_{11}} \cdot \frac{\mu_1 \, p_{22}\,p_{33} + \lambda_0 \, p_{23}\,p_{32}}{q_{22}} \nonumber\\ 
\noalign{\smallskip}
&& \hspace{4cm} + \; \lambda_0 \, \frac{\mu_0 \, p_{21}\,p_{32} + \lambda_{-1} \, p_{22}\,p_{31}}{q_{21}} \cdot \frac{\mu_0 \, p_{12}\,p_{23} + \lambda_1 \, p_{13}\,p_{22}}{q_{12}}\Big] \nonumber\\
\noalign{\smallskip}
\egal \mu_{-1}\mu_0\mu_1 \,\frac{p_{11}\,p_{22}\,p_{33}}{q_{11}\,q_{22}} + \lambda_0\mu_0\mu_1 \, \frac{p_{12}\,p_{21}\,p_{33}}{q_{11}\,q_{22}} + \lambda_0\mu_{-1}\mu_0\,\frac{p_{11}\,p_{23}\,p_{32}}{q_{11}\,q_{22}} + \lambda_{-1}\lambda_0\mu_0 \, \frac{p_{12}\,p_{23}\,p_{31}}{q_{12}\,q_{21}} \nonumber\\
\noalign{\medskip}
&& \hspace{2mm} + \: \lambda_0\lambda_1\mu_0 \,\frac{p_{13}\,p_{21}\,p_{32}}{q_{12}\,q_{21}}  + \lambda_{-1}\lambda_0\lambda_1 \, \frac{p_{13}\,p_{22}\,p_{31}}{q_{12}\,q_{21}} + \lambda_0\mu_0^2 \, \frac{p_{12}\,p_{21}\,p_{23}\,p_{32}}{p_{22}\,q_{12}\,q_{21}} + \lambda_0^2\mu_0 \,\frac{p_{12}\,p_{21}\,p_{23}\,p_{32}}{p_{22}\,q_{11}\,q_{22}},
\eea
where the eight terms correspond, in the same order, to the eight dimer configurations pictured in Section 3.1, below (\ref{initPQ}). One may check that this gives the eight configurations the correct bias; the first term for instance corresponds to the all horizontal dimer configuration, and has indeed one pair of horizontal dimers in each column $b=-1,0,1$.

The recurrence (\ref{octaPQlamu}) and the initial conditions (\ref{initPQlamu}) for $Q=1$ have been considered in \cite{DF13} as yet a further multiparameter generalization of the $\lambda$-Desnanot-Jacobi recurrence. The solution  was viewed as a generalized, inhomogeneous $\lambda$-determinant, and denoted by $\det_{\text{\boldmath$\lambda,\mu$}}$.

The similarity with a \ldet follows from the recurrence itself, which implies that the $\text{(\boldmath$\lambda,\mu$)}$-determinant of a general square matrix $A$ can be computed by using a modified condensation algorithm. The modification concerns the way connected $2 \times 2$ minors are computed: the value of a minor will depend on its position inside the matrix $A$ that contains it. Let $A$ be a matrix size $n+1$ (we will soon apply the algorithm to $A=P$ of size $n+1$); we label the diagonals of $A$ by an integer $a$, ranging from $-n$ in the lower left corner to $n$ in the upper right corner, so that $a=0$ corresponds to the main diagonal ($i=j$). Similarly we label the antidiagonals of $A$ by an integer $b$, with value $-n$ in the upper left corner and value $n$ in the lower right corner, $b=0$ being the label of the main antidiagonal ($i+j=n+2$). Then the value of a connected $2 \times 2$ minor of a matrix $X$ is (in the condensation algorithm, $X$ will be one of the matrices $A_k$ produced at intermediate steps)
\be
\left|\:
\begin{matrix}
x_{i,j} & x_{i,j+1} \\
x_{i+1,j} & x_{i+1,j+1}
\end{matrix}\:
\right|_{\text{\boldmath$\lambda,\mu$}} = \mu_b \, x_{i,j} \,x_{i+1,j+1} + \lambda_a \, x_{i,j+1} \, x_{i+1,j},
\ee
if the elements $x_{i,j},x_{i+1,j+1}$ belong to the diagonal $a$ in $X$ and $x_{i,j+1},x_{i+1,j}$ belong to the antidiagonal $b$. For such a minor, the values of $a$ and $b$ are between $-(s-1)$ and $s-1$ if $X$ has size $s$.
It follows that in the condensation algorithm recalled in Section 1, the equation (\ref{cond}) is to be replaced by the following one (recall that $A_k$ has size $n+2-k$),
\be
(A_k)_{i,j} = \big[\mu_{i+j-(n+3-k)} \,(A_{k-1})_{i,j} \, (A_{k-1})_{i+1,j+1} + \lambda_{j-i} \, (A_{k-1})_{i+1,j} \, (A_{k-1})_{i,j+1}\big]/(A_{k-2})_{i+1,j+1}.
\label{reclm}
\ee
Starting from $A_0 = (1)_{1 \le i,j \le n+2}$ and $A_1=A$ of size $n+1$, the recursive calculation of the sequence $(A_k)$ terminates with $A_{n+1} = \det_{\text{\boldmath$\lambda,\mu$}} A$.

The following gives a combinatorial interpretation of the inhomogeneous determinants introduced in \cite{DF13}. 

\begin{theo}
The partition function for perfect matchings of the Aztec graph of order $n$, with face weighting $(P,1)$, extra bias $\sqrt{\lambda_a}$ on vertical dimers in row $a$ and $\sqrt{\mu_b}$ on horizontal dimers in column $b$, is given by $T_n(P,1|\text{\boldmath$\lambda,\mu$}) = \det_{\text{\boldmath$\lambda,\mu$}} P$.
\end{theo}

\noindent {\it Proof.} We only have to show that the generalized recurrence (\ref{reclm}) for $A=P$ leads to a Desnanot-Jacobi-type identity of the form (\ref{octaPQlamu}) with $Q=1$. Let us observe, from (\ref{reclm}), that the first matrix produced, $A_2$, of size $n$, depends on all parameters in $\text{\boldmath$\lambda$} = (\lambda_{-(n-1)},\ldots,\lambda_{n-1})$ and $\text{\boldmath$\mu$} = (\mu_{-(n-1)},\ldots,\mu_{n-1})$, and the same is true of the subsequent matrices $A_{k \ge 2}$, in particular of $A_n$ (size 2) and $A_{n+1} = \det_{\text{\boldmath$\lambda,\mu$}} P$. 

Let us now focus on the specific entry $(A_n)_{1,1}$. Following the recurrence backwards, we see that its value only depends on the UL restrictions of $A_{n-1},\, A_{n-2},\, \ldots,\, A_2,\,A_1$, where the UL restriction means in each case omitting the last row and last column. As a consequence, $(A_n)_{1,1}$ only depends on the parameters in $\text{\boldmath$\lambda$}_\mathrm{C}$ and $\text{\boldmath$\mu$}_\mathrm{L}$ (let the indices $i,j$ of $A_2$ vary between 1 and $n-1$) and is actually equal to $\det_{\text{\boldmath$\lambda$}_\mathrm{C},\text{\boldmath$\mu$}_\mathrm{L}} P_{\mathrm{UL}}$. The same argument applies to the three other entries $A_n$, namely $(A_n)_{1,2}, \, (A_n)_{2,1},\,(A_n)_{2,2}$, with the appropriate restrictions UR,\,LL, \,LR, and also to the central entry $(A_{n-1})_{2,2}$ of $A_{n-1}$, for which the central restriction C is used (one omits the first and last rows and columns).

The relation (\ref{reclm}) for $k=n+1$ (and $i=j=1$) then implies the following relation,
\be
\textstyle \det_{\text{\boldmath$\lambda,\mu$}} P \cdot \det_{\text{\boldmath$\lambda$}_\mathrm{CC},\text{\boldmath$\mu$}_\mathrm{CC}} P_{\mathrm C} = \mu_0 \, \det_{\text{\boldmath$\lambda$}_\mathrm{C},\text{\boldmath$\mu$}_\mathrm{L}} P_{\mathrm{UL}} \cdot \det_{\text{\boldmath$\lambda$}_\mathrm{C},\text{\boldmath$\mu$}_\mathrm{R}} P_{\mathrm{LR}} + \lambda_0 \, \det_{\text{\boldmath$\lambda$}_\mathrm{L},\text{\boldmath$\mu$}_\mathrm{C}} P_{\mathrm{LL}} \cdot \det_{\text{\boldmath$\lambda$}_\mathrm{R},\text{\boldmath$\mu$}_\mathrm{C}} P_{\mathrm{UR}},
\ee
identical to the recurrence satisfied by $T_n(P,1|\text{\boldmath$\lambda,\mu$})$. It follows that $T_n(P,1|\text{\boldmath$\lambda,\mu$}) = \det_{\text{\boldmath$\lambda,\mu$}} P$ for all $n$ since, from (\ref{initPQlamu}), their values coincide for $n=0$ and $n=1$. \cqfd

\medskip
Inhomogeneous bias allows one to focus on vertical or horizontal dimers in special rows or columns. For instance one may ask for the number $V_{n,k}$ of perfect matchings of the Aztec graph of order $n$ which have exactly $k$ pairs of vertical dimers in the zeroth (central) row. To compute its generating function $G_n(\lambda_0) = \sum_{k=0}^n  V_{n,k} \, \lambda_0^k$, one simply sets all $\lambda_a,\mu_b$ to 1 except $\lambda_0$, takes $P=1$, and evaluates
\be
G_n(\lambda_0) = \textstyle{\det_{\text{\boldmath$\lambda,\mu$}} (p_{ij}=1)_{1 \le i,j \le n+1}}, \qquad \text{\boldmath$\lambda$} = (1,\ldots,1,\lambda_0,1,\ldots,1), \: \text{\boldmath$\mu$} = (1,\ldots,1).
\label{Ginhom}
\ee
They can be easily computed for finite $n$. The first few are listed in the following.

\begin{coro}
For the first values of $n$, the generating functions $G_n(x)$ are given by 
\begin{subequations}
\bea
G_1(x) \egal 1 + x, \\
G_2(x) \egal 1 + 6x + x^2, \\
G_3(x) \egal 1 + 47 x + 15 x^2 + x^3, \\
G_4(x) \egal 1 + 572 x + 390 x^2 + 60 x^3 + x^4, \\
G_5(x) \egal 1 + 9\,197 x + 17\,010 x^2 + 5\,970 x^3 + 589 x^4 + x^5, \\
G_6(x) \egal 1 + 173\,058 x + 1\,118\,191 x^2 + 661\,532 x^3 + 135\,151 x^4 + 9\,218 x^5 + x^6.
\eea
\end{subequations}
\end{coro}

\medskip
These polynomials seem to have a number of intriguing properties: (1) all their roots are real negative, (2) the roots of $G_n(x)$ and $G_{n+1}(x)$ are interlaced, (3) every $G_n(x)$ has a unique root which is, in norm, much larger that the others and which seems to be asymptotic to $-G_{n-1}'(0)$. For instance, the two largest, in norm, roots of $G_7(x)$ are $-173\,060.196$ and $-9.9974$, to be compared with $-G_6'(0)=-173\,058$. It follows that the coefficients $G_{n-1}(x)\big|_x$ and $G_n(x)\big|_{x^{n-1}}$ are close. Indeed we also observe: (4) the identity 
\be
G_n(x)\big|_{x^{n-1}} = G_{n-1}(x)\big|_x + 4n-3
\ee 
appears to be exact, but remains to be properly understood.

Similarly to what we did at the end of Section 3, we can absorb the bias $\sqrt{\lambda_0}$ on the vertical edges of the central row in the redefinition of the matrices $P \to P_{\lambda_0}$ and $Q \to Q_{\lambda_0}$. Using the general Robbins-Rumsey formula (\ref{genRR}), we can express the $\text{(\boldmath$\lambda,\mu$)}$-determinant in (\ref{Ginhom}) as a single sum over alternating sign matrices. The result is the following,
\be
G_n(\lambda_0) = \sum_{B \in {\rm ASM}_{n+1}} \lambda_0^{P_0(B)} \, \Big(\frac{1+\lambda_0}2\Big)^{N_-^0(B)} \, 2^{N_-(B)},
\ee
where $P_0(B)$ and $N_-^0(B)$ are the restrictions of $P(B)$ and $N_-(B)$ to the main diagonal (thus $P_0(B)$ is the number of zeros on the main diagonal of $B$ which have non-zero entries to the right and below, and such that the first non-zero entry in both directions is a one). For $n=2$, the identity matrix contributes a factor 1, each of the other five permutation matrices a factor $\lambda_0$, whereas the seventh alternating sign matrix brings a contribution $\lambda_0(1 + \lambda_0)$. For general $n$ however, this formula does not seem to help much to understand the structure of the polynomials $G_n(x)$.


\vskip 0.5truecm
\noindent 
{\bf \large Acknowledgements}

\medskip
\noindent
We would like to thank Pierre Bieliavsky, Alexei Borodin, Philippe Di Francesco, Maurice Duits, Christian Hagendorf, James Propp and Michael Somos for valuable discussions and encouragement. This work was supported by the Fonds de la Recherche Scientifique\,--\,FNRS and the Fonds Wetenschappelijk Onderzoek\,--Vlaanderen (FWO) under EOS project no 30889451. J-FdK and PR are respectively Aspirant Fellow, under the grant FC 38477, and Senior Research Associate of FRS-FNRS (Belgian Fund for Scientific Research); NR is supported by a Belgian FRIA grant.


\vskip 0.5truecm
\appendix \noindent 
{\bf \large Appendix A. On elliptic curves related to biased two-periodic Aztec diamonds}
\addcontentsline{toc}{subsection}{Appendix A. On elliptic curves related to biased two-periodic Aztec diamonds}
\setcounter{section}{1}
\setcounter{equation}{0}

\medskip
\noindent
The material presented in this Appendix owes much to exchanges with Alexei Borodin and Maurice Duits, and with Michael Somos for the last part.

In Section 3.2, we have discussed the problem of computing the partition function for tilings of two-periodic Azted diamonds, for which every vertical domino receives an extra weight $\sqrt\lambda$. The result, given in (\ref{tnabl}), can be written in terms of the first $n$ terms of a sequence $(r_k)_{k \in \Z}$, defined recursively by
\be
r_{k+1} \, r_{k-1} = \frac{\lambda + r_k^2}{1 + \lambda \, r_k^2} , \qquad r_0 = 1, \; r_1=t.
\label{recr}
\ee
We noted that when $\lambda$ and $t$ satisfy certain polynomial conditions, the sequence $(r_k)$ is periodic with a certain periodicity $p \ge 3$, that is, satisfies $r_{k+p} = r_k$ for all $k$.

The surprising observation is that these polynomials are precisely those found in \cite{BD23} to characterize when a certain point $P$ of an elliptic curve is a torsion point, that is, when $P$ has finite order $p$ for the Abelian addition law on the elliptic curve. When $P$ is a torsion point, it generates a periodic flow on the elliptic curve, given by $P_k = P_{k-1} + P$ for some initial point $P_0$. That flow encodes a sequence of Wiener-Hopf factorizations for a product of two-by-two matrices. 

More specifically, the elliptic curve considered in \cite{BD23} is given by (the two parameters $\alpha$ and $a$ used in \cite{BD23} are related to ours by $\alpha \leftrightarrow t$ and $a^2 \leftrightarrow \lambda$),
\be
{\cal E}_1 \;:\;\;y^2 = x^2 + \frac{4x\big(x-\lambda\big)\big(x-1/\lambda\big)}{\big(\sqrt\lambda + 1/\sqrt\lambda\big)^2\,\big(t + 1/t\big)^2}.
\ee
The specific point $P$ generating the flow is $P=(\frac1\lambda,\frac1\lambda)$. Thus $P$ acts on the elliptic curve by translations, $(x,y) \to \sigma(x,y) = (x,y) + P$, with
\be
\sigma(x,y) = \Big(\frac{(\lambda-x)(y-x)}{(1-\lambda\,x)(x+y)} \,, \, \frac{\lambda \big(x^2+y(\lambda-\lambda^{-1})-1\big)(y-x)}{(1-\lambda \, x)^2(x+y)}\Big).
\label{add}
\ee
The iteration of this transformation defines the flow, namely $P_k \equiv (x_k,y_k) = \sigma^k(P_0)$ for a given initial point $P_0$; in the context of \cite{BD23}, $P_0 = \big(\!-\!1,\frac{t^2-1}{t^2+1}\big)$. Because $P_0$ and $P$ are respectively on the negative and positive component of ${\cal E}_1$, all $P_k$ are on the negative component, $x_k < 0$ for all $k$.

Although the recurrence (\ref{recr}) satisfied by the sequence $(r_k)$ and the discrete flow $P_k = P_0 + k \cdot P$ on ${\cal E}_1$ do not have any apparent connection, the striking fact is the following,

\medskip \noindent
{\it The sequences $(r_k)_{k \ge 0}$ and $(P_k)_{k \ge 0}$ are simultaneously periodic with the same period, namely, in each case, the conditions ensuring their $p$-periodicity are identical.}

\medskip
This strongly suggests that the recurrence relation for the sequence $(r_k)$ is intimately connected to the elliptic curve ${\cal E}_1$. Below we indeed exhibit a direct connection with ${\cal E}_1$, but surprisingly, the sequence $(r_k)$ is more naturally connected to a translational flow on a different elliptic curve ${\cal E}_2$, which however shares the same periodicities as the flow on ${\cal E}_1$.

\medskip
\noindent
{\bf Element \#1.} On general grounds, a relation with the elliptic curve ${\cal E}_1$ used in \cite{BD23} is expected, based on the following chains of connections. The recurrence (\ref{recr}) comes directly from the condensation algorithm applied to the computation of $\detl Q^{-1}$, where $Q$ is the matrix with coefficients $a$ and $b$ alternating on rows and columns, see (\ref{matQ}). As shown in Section 3.5, the condensation algorithm and the shuffling algorithm (or the urban renewal) applied to the $q$-faces work in the same way, reducing, at each step, the order of the Aztec graph by one unit at the price of a redefinition of the face weights. Finally, these successive redefinitions of the face weights and the translational flow on the elliptic curve ${\cal E}_1$ recalled above, were shown to be identical \cite{CD23}.

\bigskip \noindent
{\bf Element \#2.} One can exhibit a direct connection between the two sequences $(r_k)$ and $(P_k)$ as follows. By direct calculation, one finds from (\ref{add}) that $(x_k,y_k) = \sigma(x_{k-1},y_{k-1})$ are related in such a way that the following identity holds,
\be
\frac{1 - \lambda \, x_k}{\lambda - x_k} \cdot \frac{y_k - x_k}{x_k + y_k} = \frac 1{x_{k-1}}.
\label{inv}
\ee
Using (\ref{add}) once more, the terms $x_k$ satisfy a second-order recurrence relation,
\be
x_{k+1} = \frac{\lambda - x_k}{1 - \lambda \,x_k} \cdot \frac{y_k - x_k}{x_k + y_k} = \Big(\frac{\lambda - x_k}{1 - \lambda\,x_k}\Big)^2 \: \frac1{x_{k-1}}.
\ee
Multiplying both sides by $-1$ and taking the positive square root (all $x_k$ are negative), we obtain 
\be
\sqrt{-x_{k+1}} = \frac{\lambda + (-x_k)}{1 + \lambda \, (-x_k)} \; \frac1{\sqrt{-x_{k-1}}},
\ee
which shows that $r_k>0$ and $\sqrt{-x_k}$ satisfy the same recurrence relation. The initial conditions match since $x_0=-1$ implies $r_0=1$, whereas $x_1=-t^2$, which may be computed from (\ref{add}) for $(x,y)=(x_0,y_0)$, yields $r_1=t$. Therefore both sequences coincide, $x_k = -r_k^2$ for all $k \ge 0$.

The relation (\ref{inv}) can be solved for $y_k$ in terms of $x_k$ and $x_{k-1}$, themselves expressible in terms of the $r_k$'s. The explicit expressions read
\be
x_k = -r_k^2, \qquad
y_k = r_k^2 \: \frac{(\lambda + r^2_k) - (1 + \lambda \, r^2_k) \, r^2_{k-1}}{(\lambda + r^2_k) + (1 + \lambda \, r^2_k) \, r^2_{k-1}} = r_k^2 \: \frac{r_{k+1} - r_{k-1}}{r_{k+1} + r_{k-1}}.
\label{rkell}
\ee
Because the terms $r_k$ are positive, the two sequences $(P_k)$ and $(r_k)$ are simultaneously periodic with the same period.

\bigskip \noindent
{\bf Element \#3.} It turns out that the recurrence (\ref{recr}) belongs to a larger class of recurrence relations that have a fairly long history. Because it is of direct relevance for what follows, we will more specifically refer to \cite{BR05} (see also \cite{Ho07}).

\begin{figure}[t]
\begin{center}
\includegraphics[angle=0,width=65mm]{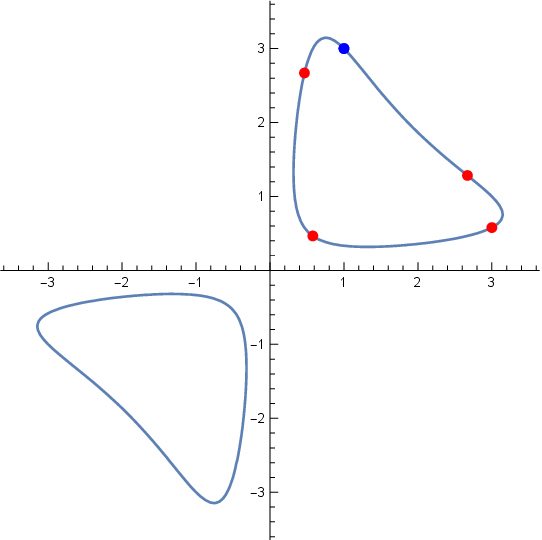}
\hspace{1cm}
\includegraphics[angle=0,width=65mm]{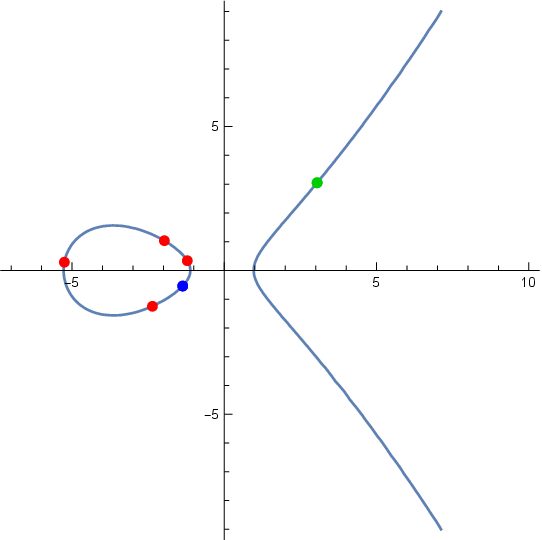}
\end{center}
\caption{Left: Real section of the biquadratic curve (\ref{biq}) for $\lambda=2$ and $t=3$ ($K=5$). The red dots represent the points $P_1,P_2,P_3,P_4$ obtained from $P_0=(1,3)$, in blue, by the flow described in the text. Right: The same flow represented on the isomorphic cubic curve ${\cal E}_2'$; $P_0$ is the blue dot on the bounded component, whereas the green dot on the unbounded component is the point $P$ in terms of which the flow can be viewed as a translational flow.}
\label{fig9}
\end{figure}

The basic but crucial observation is that the recurrence (\ref{recr}) is such that the map $\mu \,:\,(r_k,r_{k+1}) \to (r_{k+1},r_{k+2})$ preserves the following biquadratic curve ${\cal E}_2$ (see Prop. 2.5 in \cite{Ho07} for a more general statement),
\be
{\cal E}_2 \;:\;\; X^2 \, Y^2 + \lambda^{-1} \, (X^2 + Y^2) + 1 = K \,XY,
\label{biq}
\ee
where $K$ is a constant. For the initial conditions (\ref{recr}), namely $r_0=1$ and $r_1=t$, it is given by $K = (1 + \frac 1\lambda)(t + \frac 1t)$. The real section of ${\cal E}_2$ has two bounded symmetric components related by $(X,Y) \leftrightarrow (-X,-Y)$, see Figure~\ref{fig9}.

In a sense, ${\cal E}_2$ is already present in the relation (\ref{rkell}) because the fact that the point $(x_k,y_k)$ lies on the elliptic curve ${\cal E}_1$ implies that $(r_{k-1},r_k)$ satisfies the biquadratic relation (\ref{biq}). However the map $(r_{k-1},r_k) \to (x_k,y_k)$ given in (\ref{rkell}) is not birational (in particular, on the reals, its image only yields the negative component of ${\cal E}_1$).

The biquadratic curve ${\cal E}_2$ has genus 1 and is therefore an elliptic curve. The flow on ${\cal E}_2$ obtained by applying iteratively the map $\mu$ on the initial point $(r_0,r_1)=(1,t)$ is geometrically easy to describe \cite{BR05}. Because ${\cal E}_2$ is symmetric under $X \leftrightarrow Y$, the two points $P_k = (r_k,r_{k+1})$ and $P^*_{k+1}=(r_{k+2},r_{k+1})$, the diagonally reflected image of $P_{k+1}$, belong to ${\cal E}_2$ and are the only two intersection points of ${\cal E}_2$ with the horizontal line passing through $P_k$ (for $Y=r_{k+1}$, (\ref{biq}) is quadratic in $X$ with roots equal to $r_k$ and $r_{k+2}$). Thus $P_{k+1}$ is found in three steps: draw a horizontal line through $P_k$, find the other intersection with ${\cal E}_2$, and diagonally reflect it. For $\lambda$ and $t$ positive, the flow remains on the positive component, see Figure~\ref{fig9}. Moreover, similarly to the map $\sigma$ discussed above, the flow induced by $\mu$ can be seen as the iterated addition on ${\cal E}_2$ of a specific point $P$ \cite{Ho07}. 


Since ${\cal E}_1$ and ${\cal E}_2$ are both elliptic, the natural question is whether they are isomorphic. It turns out that the answer is negative. To see this, we first follow \cite{BR05} where a sequence of birational transformations is defined that brings the biquadratic curve to an elliptic curve in standard, cubic form. We refer to \cite{BR05} for the details and merely quote the final result. 

Let $u$ be the smallest positive root of $X^4 + (2 \lambda^{-1} - K) X^2 + 1=0$. Then the biquadratic curve ${\cal E}_2$ is birationally equivalent to the following elliptic curve ${\cal E}'_2$ written in standard form as,
\be
{\cal E}'_2 \;:\;\; \textstyle \frac{4(K+2\lambda^{-1}-2u^2)}q \, y^2 = 4 x^3 + b_2 x^2 + 2 b_4 x + b_6,
\label{stand}
\ee
where
\bea
&& \gamma = 2u^2 + \frac2\lambda - K, \qquad p = - \frac{4u}{2 + \lambda K - 2\lambda u^2}, \qquad q = p(\gamma + 4 p u + 4p^2),\\
&& b_2 = -\frac 4q (\gamma + 8pu + 12 p^2), \qquad b_4 = \frac 8q (u + 3p), \qquad b_6 = -\frac {16}q.
\eea
The point $P$ on ${\cal E}'_2$ defining the translational flow is given by $P = \big(\frac1{p+u},\frac1{p+u}\big)$ and lies on the diagonal.

To check whether the elliptic curves ${\cal E}_1$ and ${\cal E}_2'$ are isomorphic, we merely compute  their $j$-invariant. With respect to the standard form given in (\ref{stand}), into which ${\cal E}_1$ can easily be recast, the invariant is defined by \cite{Si09}
\be
j = \frac{(b_2^2 - 24 b_4)^3}{9b_2b_4b_6 - \frac 14 b_2^3 \, b_6 + \frac 14 b_2^2 \, b_4^2 - 8b_4^3 - 27b_6^2}.
\ee
We find surprisingly close but definitely different values in terms of the two parameters $\lambda$ and $K$,
\bea
j({\cal E}_1) \egal \frac{\left(16 \lambda ^4-16 \lambda ^2+\lambda ^4 K^4-8 \lambda ^4 K^2-8 \lambda ^2 K^2+16\right)^3}{\lambda ^8 \big[(K+2)^2 - 4\lambda^{-2}\big] \, \big[(K-2)^2 - 4\lambda^{-2}\big]} , \\
\noalign{\medskip}
j({\cal E}_2') \egal \frac{\left(16 \lambda ^4+224 \lambda ^2+\lambda ^4 K^4-8 \lambda ^4 K^2-8
\lambda ^2 K^2+16\right)^3}{\lambda ^{10} \big[(K+2)^2 - 4\lambda^{-2}\big]^2 \, \big[(K-2)^2 - 4\lambda^{-2}\big]^2}.
\eea
It follows that the two curves are not isomorphic over $\C$ and therefore also not over $\R$. Yet the translational flows $P_k = P_0 + k \cdot P$ defined in terms of $P_0$ and $P$, and respectively different for ${\cal E}_1$ and ${\cal E}_2'$, are simultaneously periodic with the same period. Although this has been proved above, somewhat indirectly since the argument was made on the sequence $(-r_k^2)$, it remains to be properly understood.

\bigskip \noindent
{\bf Element \#4.} After submitting the manuscript, Michael Somos informed us of his observation that the two sequences $(a_k)$ and $(b_k)$ introduced in (\ref{akbk}) are each generalized Somos-4 sequences, i.e. satisfy a quadratic recurrence relation of order 4,
\be
s_k s_{k-4} = \alpha \, s_{k-1}s_{k-3} + \beta \, s_{k-2}^2.
\ee
The initial conditions $s_0,s_1,s_2,s_3$ are different for the $a_k$ and $b_k$, but the coefficients $\alpha$ and $\beta$ are identical for both sequences, and easily related to $a_0=1,\,a_1=a^{-1},\,b_0=1$ and $b_1=b^{-1}$,
\begin{subequations}
\bea
\alpha \egal \Big[\frac{a_0b_1}{a_1b_0} + \frac{a_1b_0}{a_0b_1} + \lambda \frac{a_0a_1}{b_0b_1} + \lambda \frac{b_0b_1}{a_0a_1}\Big]^2 = (1+\lambda)^2 \,(t + t^{-1})^2,\\
\noalign{\medskip}
\beta \egal -\alpha + (1-\lambda^2)^2 = (1 + \lambda)^2 \big[(1-\lambda)^2 - (t + t^{-1})^2\big],
\eea
with $t = \frac ab$.
\end{subequations}

Somos sequences are at the heart of the {\it elliptic realm}, and central in topics like elliptic divisibility and the Laurent phenomenon (see for instance the item A006720 on the On-line Encyclopedia of Integer Sequences \cite{OEIS} for a rich list of references). The intriguing observation made by Somos adds a layer of ellipticity to our problem, and in fact shed a new light on what we had already observed, as summarized above. Indeed there is a well-documented relation between Somos-4 sequences and elliptic curves, in particular translational flows on elliptic curves. In short, a Somos-4 sequence can be associated with a translational flow $P_0 + k \cdot P$, and vice-versa. 

More precisely, let $(x_k,y_k)$ be the coordinates of the points $P_0+k \cdot P$ on an elliptic curve $\cal E$, for $P_0=(x_0,y_0)$ and $P=(\xbar x,\xbar y)$ non-singular points on $\cal E$. Then the sequence defined by
\be
\frac{s_k}{s_{k-1}} = (\xbar x - x_{k-1}) \, \frac{s_{k-1}}{s_{k-2}}, \qquad {\rm for\ }k \ge 1,
\ee
with $s_{-1}$ and $s_0$ arbitrary, is a Somos-4 sequence \cite{Sw03}.

Applying this to the elliptic flow on ${\cal E}_1$ discussed in \cite{BD23} and recalled above, for which $\xbar x = \frac1\lambda$, we readily obtain that the terms $\tilde s_k = \lambda^{k(k+1)/2} s_k$ form an equivalent Somos-4 sequence and satisfy 
\be
\frac{\tilde s_k}{\tilde s_{k-1}} = (1 - \lambda \, x_{k-1}) \, \frac{\tilde s_{k-1}}{\tilde s_{k-2}} = (1 + \lambda \, r^2_{k-1}) \, \frac{\tilde s_{k-1}}{\tilde s_{k-2}}, \qquad {\rm for\ }k \ge 1,
\ee
where the second equality follows from $x^{}_{k-1}=-r^2_{k-1}$, see (\ref{rkell}). Comparing with the recurrence satisfied by $a_k$, see (\ref{recak}), confirms that $a_k = \tilde s_k$ is a Somos-4 sequence. The same holds for the sequence $(b_k)$. This clearly reinforces the tie with \cite{BD23} since $(a_k)$ and $(b_k)$ are Somos-4 sequences that are associated in a natural way with the translational flow on ${\cal E}_1$ considered in \cite{BD23}. 


\end{document}